\documentclass[aps,rmp,reprint,amsmath,amssymb,graphicx,longbibliography]{revtex4-1}

\usepackage{bm}

\usepackage{graphicx}
\usepackage{mdframed}
\usepackage{enumitem}

\begin{document}

\title{Colloquium: Quantum skyrmionics}

\author{Hector Ochoa}
\altaffiliation[Current address: ]{Department of Physics, Columbia University, New York, NY 10027, USA}
\author{Yaroslav Tserkovnyak} 
\affiliation{Department of Physics and Astronomy, University of California, Los Angeles, California 90095, USA}

\date{\today{}}

\begin{abstract}
{Skyrmions are topological solitons that emerge in many physical contexts. In magnetism, they appear as textures of the spin-density field stabilized by different competing interactions and characterized by a topological charge that counts the number of times the order parameter wraps the sphere. They can behave as classical objects, when the spin texture varies slowly on the scale of the microscopic lattice of the magnet. However, the fast development of experimental tools to create and stabilize skyrmions in thin magnetic films has lead to a rich variety of textures, sometimes of atomistic sizes. In this article, we discuss, in a pedagogical manner, how to introduce quantum interference in the translational dynamics of skyrmion textures, starting from the micromagnetic equations of motion for a classical soliton. We study how the nontrivial topology of the spin texture manifests in the semiclassical regime, when the microscopic lattice potential is treated quantum-mechanically, but the external driving forces are taken as smooth classical perturbations. We highlight close relations to the fields of noncommutative quantum mechanics, Chern-Simons theories, and the quantum Hall effect.

}
\end{abstract}


\maketitle

\tableofcontents{}

\section{Introduction}

Solitons in continuous media are nonlinear excitations with well-defined energy that behave as particles. Sometimes, these excitations are robust against decay to the ground state due to \textit{topological} constraints imposed by the boundary conditions. Skyrmions are examples of topological solitons in systems described by a certain class of nonlinear $\sigma$-models. These objects owe their name to T. H. R. Skyrme, who proposed topologically stable configurations of the pion field describing the effective interaction between nucleons as candidates for hadronic matter \cite{skyrme1962}. This type of textures are also present in condensed matter when the orbital and spin degrees of freedom are mixed; notorious examples are superfluid $^3$He \cite{Shankar1977,Volovik1977} or Bose-Einstein condensates with ferromagnetic order \cite{Usama2001,Ueda2014}.

The definition was later broadened to include whirling configurations of a directional order parameter in a planar system (i.e., effectively in two spatial dimensions) characterized by an integer index of the form \cite{A.Belavin1975}\begin{align}
\label{eq:Q}
\mathcal{Q}\equiv\frac{1}{4\pi} \int d^2\mathbf{r}\,\,\bm{n}\cdot\left(\partial_x\bm{n}\times\partial_y\bm{n}\right).
\end{align}
Here $\bm{n}(\mathbf{r})$ represents the unit vector along the order parameter, for example, the saturated spin density in a planar magnet. The skyrmion charge $\mathcal{Q}$ labels the number of times $\bm{n}(\mathbf{r})$ winds the unit sphere $S^2$. An example of this type of texture is shown in Fig.~\ref{fig:textures}(a). At the center of the skyrmion, which we denote by $\bm{R}$ from now on, the order parameter is reversed with respect to the uniform ground state. Mathematically, we say that the skyrmion charge classifies topologically distinct configurations of the order parameter according to the second homotopy group of the sphere, $\pi_2(S^2)=\mathbb{Z}$, provided that far away from $\bm{R}$ all the vectors are aligned uniformly. Textures with different skyrmion charge cannot be continuously deformed into each other, which in practice means that the system has to overcome a high-energy barrier determined by microscopic details of the system.

This type of low dimensional or \textit{baby} skyrmions characterized by the topological index in Eq.~\eqref{eq:Q} appear in many areas of condensed matter physics. They have been discussed, for example, in the context of spinor condensates \cite{Ho1998,Mizushima2002} and chiral superfluids\footnote{The situation in superfluid $^3$He deserves additional clarification. The order-parameter manifold in the A phase is SO(3), corresponding to rotations of the tetrad formed by $\bm{\hat{l}}$, the orbital angular momentum of the Cooper pair, and orthogonal unit vectors $\bm{\hat{\Delta}_{1,2}}$ parametrizing the superfluid complex-vector order parameter, $\bm{\hat{\Delta}}=\bm{\hat{\Delta}}_1+i\bm{\hat{\Delta}}_2$. Rotations along $\bm{\hat{l}}=\bm{\hat{\Delta}}_1\times\bm{\hat{\Delta}}_2$ change the phase of the order parameter, while rotations of $\bm{\hat{l}}$ produce spin textures. Shankar skyrmions  \cite{Shankar1977} are associated with winding of $S^3$, provided that SO(3) is topologically equivalent to a hypersphere with antipodal points identified as the same. \textit{Baby} skyrmions are in fact coreless $4\pi$-vortex lines in the phase of the order parameter \cite{Anderson1977}, where $\bm{\hat{l}}$ produces a texture with $|\mathcal{Q}|=1$ in order to avoid the singularity. This relation between the topological charge of the spin texture and the vorticity of the mass superflow is known as the Mermin-Ho relation \cite{Mermin-Ho}.} like $^3$He-A \cite{Anderson1977,Salomaa1987} or triplet superconductors \cite{Knigavko1999,Li2009}, where they can form a regular lattice akin to Abrikosov vortices \cite{Abrikosov1957}. This skyrmion lattice resembles the blue phase of cholesteric liquid crystals \cite{Wright1989,deGennes_book}. Nevertheless, magnetism in solid state systems is the context where these objects have garnered most attention recently.

\subsection{Skyrmions in solid state}

\begin{figure}
\includegraphics[width=1\columnwidth]{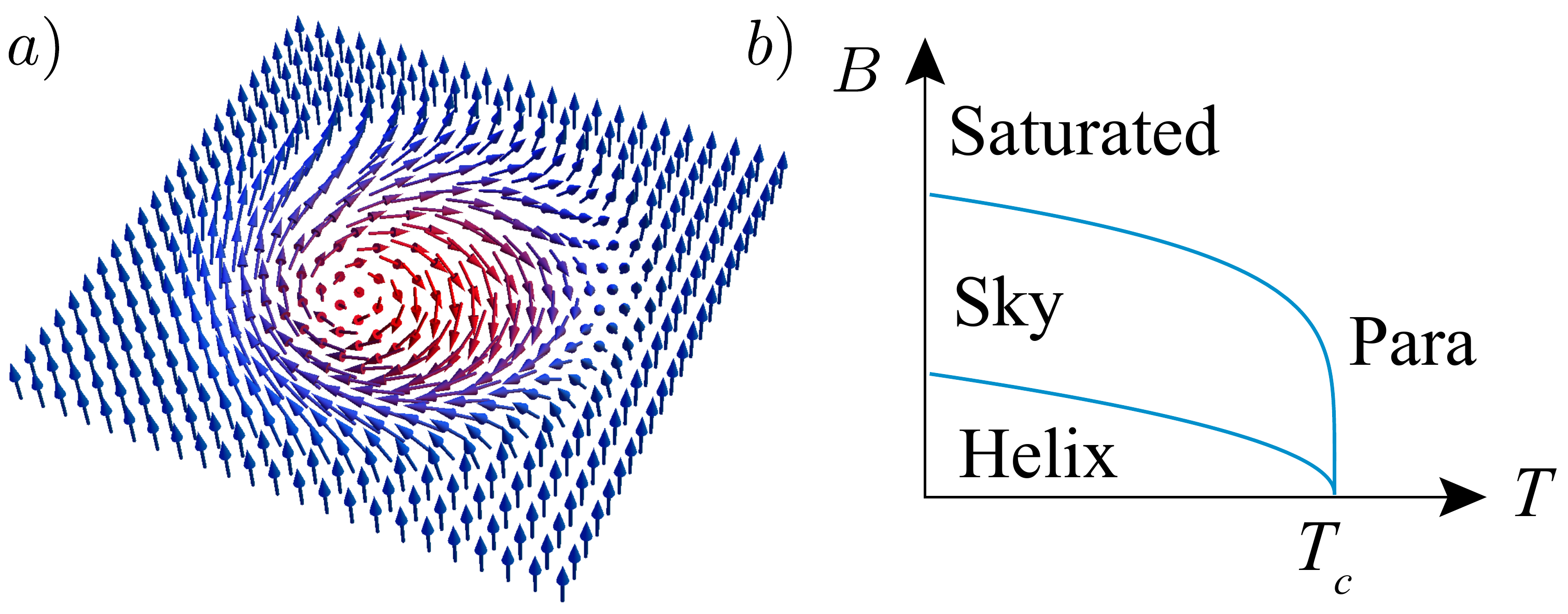}
\caption{a) Bloch-type skyrmion stabilized by interfacial Dzyaloshinskii-Moriya (DM) interactions in helimagnets. Colors indicate the out-of-plane polarization (blue corresponds to up, red to down). Interfacial DM interactions, usually ascribed to $C_{nv}$ point groups of magnetic heterostructures, tend to stabilize N\'eel-type skyrmions. b) Phenomenological phase diagram of thin helimagnets like, e.g., Cu$_2$OSeO$_3$ \cite{Seki2012}. At intermediate fields, skyrmions on top of the uniformly ordered state nucleate to form a regular lattice. The skyrmion lattice becomes unstable with respect to helical order at lower fields.}
\label{fig:textures}
\end{figure}

One of the first examples of magnetic skyrmions appears in the physics of the quantum Hall effect \cite{DasSarmaPinczuk}. When a two-dimensional electron gas is subjected to a strong magnetic field, eventually the system (here disorder and temperature plays a major role) evolves into an incompressible liquid characterized by a quantized Hall response. The kinetic energy is quenched by the strong magnetic field, with the single-particle spectrum consisting of discrete Landau levels. Many-body effects are therefore relevant and usually manifest in the cleanest samples. Even in the limit of negligible $g$-factor (as is the case of GaAs due to relativistic corrections, for example) the ground state can develop a spontaneous spin ordering that minimizes the energy cost of Coulomb exchange interactions. Skyrmions in these quantum Hall ferromagnets carry nontrivial quantum numbers identical to those of Laughlin quasiparticles at the same filling factor \cite{YangSondhi}. Their energy and size are determined by the competition between the Zeeman energy and Coulomb repulsion: the former tries to minimize the texture while the latter tries to expand it. It is accepted that skyrmions are the lowest charged excitations in $\nu=1$ quantum Hall ferromagnets (i.e., with one electron per magnetic flux quantum) in materials with small $g$-factor \cite{Sondhi1993,Schmeller1995}. Slightly away from $\nu=1$, the additional charge is arranged in a skyrmion lattice \cite{Brey1995,Bayot1996}.

Helimagnets are another family of materials where skyrmion solitons are stable, in this case from the competition between the Zeeman energy, various anisotropy terms, and relativistic Dzyaloshinskii-Moriya (DM) interactions \cite{Dzyaloshinskii1957,Dzyaloshinsky1958,Moriya1956}. The simplest magnetic free-energy functional describing this class of materials reads\begin{align}
\label{eq:H}
H=\int d\mathbf{r}\,\left[\frac{\mathcal{A}}{2}\left(\partial_i\bm{n}\right)^2+\mathcal{D}\,\bm{n}\cdot\left(\bm{\nabla}\times\bm{n}\right)-\gamma\,\bm{B}\cdot\bm{s}\right].
\end{align}
Here $\mathcal{A}$ represents the magnetic stiffness and $\mathcal{D}$ measures the strength of the DM interaction compatible, in this case, with $D_{n}$ symmetry (with the $n$-fold axis along the out-of-plane direction $\bm{\hat{z}}$ of the film). The last term is a Zeeman coupling with an external field $\bm{B}=B\bm{\hat{z}}$, where $\gamma\equiv g\mu_B/\hbar$ is the gyromagnetic ratio, $g$ and $\mu_B$ being the $g$-factor and Bohr magneton, respectively, and $\bm{s}\left(\mathbf{r}\right)=s\,\bm{n}\left(\mathbf{r}\right)$, $s$ being the saturated spin density.

Since the DM interaction is linear in the gradient of the order parameter, the magnet gains energy by allowing spatial modulations of the spin-density field. When $B=0$, the ground state consists in a helix with wavelength $\lambda_{\textrm{helix}}\sim \mathcal{A}/\mathcal{D}$. At fields larger than $B\sim \mathcal{D}^2/\gamma s\mathcal{A}$, however, the spin-density field is fully polarized. Metastable soliton solutions with nontrivial skyrmion charge exist on top of the field-polarized background. Its characteristic radius scales as $\mathcal{R}\sim \sqrt{\mathcal{A}/\gamma sB}$ \cite{Bogdanov1994}, whereas the sign of $\mathcal{D}$ define the chirality of the texture (the sense of rotation of in-plane spins). These textures can be induced by current pulses \cite{Romming2013}, spin-orbit torques \cite{Jiang2015}, vortex beams \cite{Fujita2017}, or local annealing \cite{Koshibae2014}.

In thin films, the field-polarized ground state becomes unstable against the formation of a skyrmion lattice at intermediate fields \cite{Bogdanov1989,Roessler2006,Tewari2006,Binz2006a}, giving rise to the schematic phase diagram represented in Fig.~\ref{fig:textures}(b). In bulk systems, however, the skyrmion lattice competes with the more favorable conical order, and it is only stabilized by critical fluctuations close to the Curie temperature \cite{Muhlbauer2009,Buhrandt2013}. Interestingly, this transition corresponds to the Landau-Brazovskii paradigm of weak crystallization \cite{Brazovskii1975,Brazovskii1987}, as it has been confirmed experimentally \cite{Janoschek2013}. The skyrmion lattice has been observed in various helimagnets using different imaging techniques, including metallic materials like MnSi or FeGe \cite{Muhlbauer2009,Li2013,Yu2011,Wilhelm2011} or multiferroic insulators like Cu$_2$OSeO$_3$ \cite{Seki2012,Adams2012}. In the former case, the formation of the lattice can also be detected in magnetotransport through the topological Hall effect \cite{Neubauer2009,Kanazawa2011,Li2013,Gallagher2017}, resulting from the emergent magnetic field experienced by electrons when their spin follows adiabatically the skyrmion texture.


\subsection{Skyrmion dynamics vs. cyclotron motion}

This colloquium concerns the dynamics of isolated skyrmions in low-dimensional (planar) magnets, when these objects appear as metastable soliton solutions in the background of a spin-polarized medium. The usual starting point is a continuum, classical description of magnetic interactions (like, e.g, the one provided by Eq.~\ref{eq:H}), in which the microscopic lattice structure of the material is ignored. We are going to assume that the magnetic film behaves effectively as a two-dimensional system, meaning that its thickness is much smaller than the exchange lengths and, therefore, the magnetization remains uniform along the $z$-axis. In the case of helimagnets, for example, the thickness should be smaller than the helix pitch, e.g., $\lambda_{\textrm{helix}}\sim 50$ nm in Cu$_2$OSeO$_3$ \cite{Seki2012}. We are going to focus on electrically insulating magnets, where the underlying microscopic degrees of freedom can be represented by localized magnetic moment on a lattice.

The skyrmion solution breaks spontaneously the translational invariance of the continuum theory. Translations of the skyrmion texture correspond then to a soft mode of the magnetization dynamics. We are going to consider here the limit of slow (adiabatic) dynamics, in which the skyrmion moves in a rigid fashion. The skyrmion center $\bm{R}$ will be promoted to a collective dynamical variable, whose evolution is generated by the Poisson bracket \cite{Papanicolaou1991,Tchernyshyov2015}
\begin{align}
\label{eq:G3}
\left\{R_i,R_j\right\}=\frac{\epsilon_{ij}}{4\pi s\mathcal{Q}},
\end{align}
where $\epsilon_{ij}$ is the Levi-Civita symbol.

The skyrmion position resembles the guiding center of electrons subjected to a magnetic field $\bm{B}=\pm B\bm{\hat{z}}$, $\{R_x,R_y\}=\pm c/eB$, where in that case $c$ and $e$ would correspond to the speed of light and the (minus) electron charge, respectively. The most intriguing aspects of the skyrmion dynamics are rooted in this analogy. The Poisson bracket in Eq.~\eqref{eq:G3} gives rise to a Magnus force in the equation of motion, which could be interpreted as a sort of electromagnetic Lorentz force. Its origin is completely different, though: the Magnus force is effectively created by the surrounding medium in response to the nontrivial topology of the skyrmion texture \cite{Nikiforov1983}, without electromagnetic fields coupled to it. The trajectories of skyrmions are therefore deflected with respect to the driving force \cite{Jiang2016,Litzius2016}, where the latter could be induced either by current-induced torques \cite{Jonietz2010,Yu2012}, thermal gradients \cite{Kong2013,Lin2014}, or even the application of electric fields \cite{White2014}.

Differences between these two physical scenarios become more evident when the role of the microscopic lattice is considered. In electronic systems, e.g., in a semiconductor quantum well, the original orbital bands split into sub-band states that develop into the quantization of cyclotron motion, i.e., the Landau levels that we mentioned before. The energy spectrum can be derived from a set of Harper equations \cite{Harper1955} usually expressed in a basis of atomic orbitals, either if we start from an effective mass Hamiltonian or directly from a tight-binding description. In the latter approach, the hopping integrals are modified in minimal coupling by means of the Peierls substitutions, like in the Hofstadter model \cite{Hofstadter1976}. The justification for this approach is that even for the highest magnetic fields that can be achieved in a lab, the associated length scale $\ell_B=\sqrt{\hbar c /eB}$ characterizing the radius of the semiclassical cyclotron orbits\footnote{When promoted to quantum mechanical operators, the electron guiding centers satisfy the commutator algebra $[R_x,R_y]=\pm i\ell_B^2$.} are much longer than the typical distance $a$ between atoms in a solid. For example, in a zinc-blende heterostructure, we would need to apply fields of the order of $B\approx2500$ T in order to have magnetic lengths of $\ell_B\sim 6$ \AA, the lattice constant of GaAs.

\begin{figure}
\includegraphics[width=1\columnwidth]{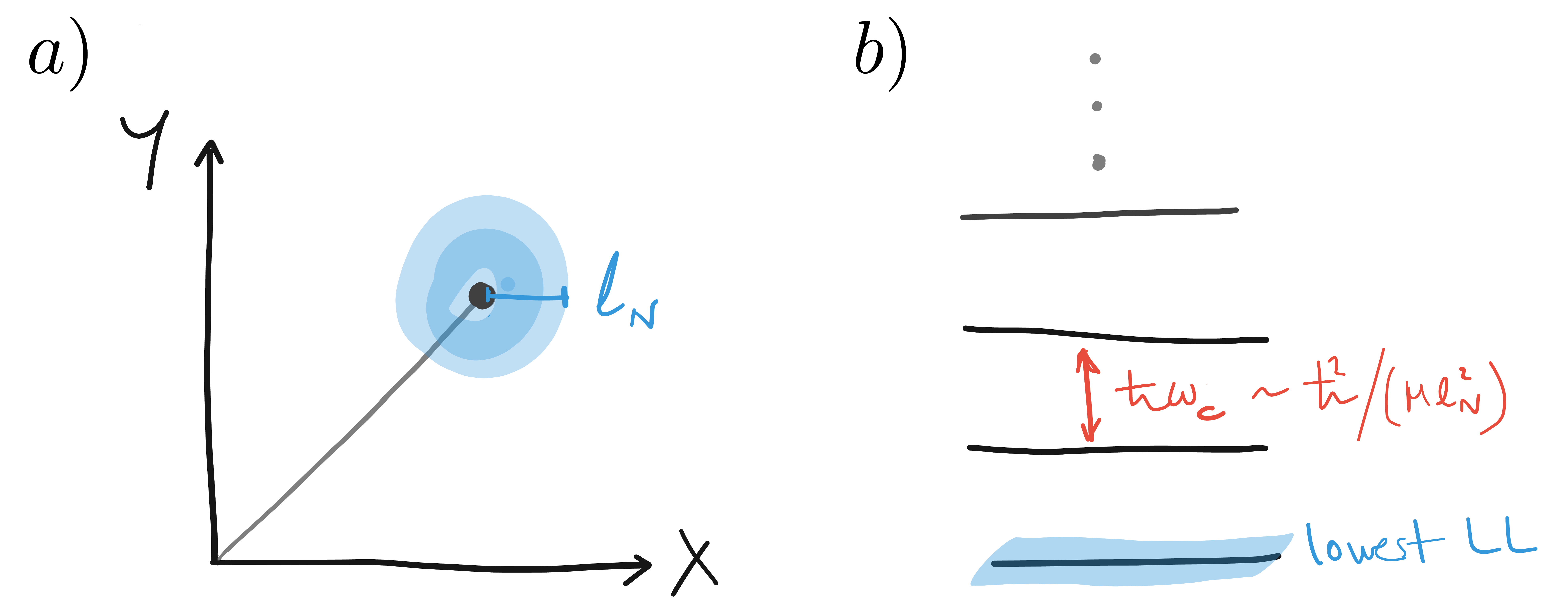}
\caption{a) A rigid skyrmion is identified with a coherent state in the phase space of collective coordinates. These coordinates can be identified with the center of the topological charge. The \textit{magnetic length} defined in Eq.~\eqref{eq:magnetic_length} characterizes the extension of quantum fluctuations. b) The spectrum of quantum skyrmions is organized in \textit{Landau levels}, where the cyclotron gap is proportional to the inverse of the skyrmion mass $M$, the latter resulting from dynamical deformations of the texture. In the limit of rigid textures ($M=0$), the Hilbert space is truncated to the lowest Landau level.}
\label{fig:qhe}
\end{figure}

In the case of skyrmions, the analog to this length scale follows from Eq.~\eqref{eq:G3} and can be related to the extension of quantum fluctuations in the phase space of collective coordinates, as we represent in Fig.~\ref{fig:qhe}(a). By promoting the Poisson bracket in Eq.~\eqref{eq:G3} to a quantum commutator we obtain\begin{align}
\label{eq:magnetic_length}
\ell_{N}\equiv\sqrt{\frac{A_c}{2\pi N}},\,\, \text{with}\,\, N=2S|\mathcal{Q}|\,\,\text{an integer}.
 \end{align}
This definition follows from the quantization of the spin density as $s=\hbar S/A_c$, where $S$ is the length of quantum spin operators defined in a lattice with unit-cell area $A_c\sim a^2$. From Eq.~\eqref{eq:magnetic_length}, we already see that the present problem is in the opposite limit, $\ell_N\leq a$, as compared to electrons in a semiconductor. In this case, we should express the Harper equations in a basis of \textit{Landau levels} associated with a fictitious magnetic field, reminiscent of the special kinematics of the classical soliton.

\subsection{Quantum corrections to skyrmion dynamics}

As it is clear from the previous discussion, we expect a new regime of skyrmion dynamics with no analogue in electronic systems when the microscopic lattice starts to play a role. Lattice effects become relevant when the characteristic size of the skyrmion, $\mathcal{R}$, starts to be comparable with the lattice spacing, $a$ \cite{Cai2012}. In that situation, quantum fluctuations characterized by the $\textit{magnetic length}$ in Eq.~\eqref{eq:magnetic_length} cannot be disregarded. One immediate consequence is the possibility of nucleation/annihilation of skyrmions \cite{Diaz2016,Derras-Chouk2018}. These processes consist of tunneling events on the atomic scale and their amplitude decreases exponentially with the number of spins involved and their length, $S\mathcal{R}^2/a^2\sim(\mathcal{R}/\ell_N)^2$. Our immediate interest, in contrast, is to elucidate how the nontrivial topology of the texture is manifested in the semiclassical regime characterized by\begin{align}
\label{eq:inequalities}
\ell_N<a\lesssim\mathcal{R}\ll\text{rest of length scales},
\end{align}
when the microscopic lattice is introduced as a weak quantum potential breaking the translational invariance of the classical theory \cite{Galkin2007,Takashima2016}, but keeping the number of skyrmions approximately unchanged. In other words, we are introducing quantum fluctuations on the scale of the lattice around the classical, continuum spin texture, but restricting the dynamics to the collective translational mode. Recently, a full spin-wave expansion revealed that quantum fluctuations stabilize\footnote{For noncollinear textures, quantum fluctuations generically reduce the energy of the ground state with respect to the saddle point of the corresponding classical model; see also \onlinecite{Ivanov2007}.} skyrmions \cite{Roldan2015}; those considerations, however, are beyond the scope of our theory since we are not concerned about internal modes of the skyrmion dynamics.

The first of the inequalities in \eqref{eq:inequalities} is ensured by considering large spin numbers, $S$, which can be controlled by modifying the thickness of the magnetic film, so long as the system remains effectively two dimensional. In order to have small, almost atomic-size skyrmions in helimagnets, very large magnetic fields must be applied, of the order of $B\sim JS/g\mu_B$ according to our previous estimates, where $J\approx\mathcal{A}/S^2$ is the microscopic exchange coupling. In practice, anisotropy terms not included in Eq.~\eqref{eq:H} but definitively present in low-dimensional systems can help to stabilize smaller skyrmions. Different phases with skyrmion radius comparable to the lattice spacing have been also proposed in frustrated magnets \cite{Okubo2012,Leonov2015,Lin2016}. Moreover, skyrmions with extensions of just a few nanometers have been observed in one atom-thick Fe layers on Ir surfaces \cite{Heinze2011,Romming2013}, stabilized, possibly, by interfacial DM interactions.\footnote{The stability can be further enhanced by other terms like, e.g., ring-exchange interactions, which have been suggested to be relevant in this system.} The dynamics of these much smaller objects are more difficult to detect in neutron scattering or electron microscopy, but their motion, on the other hand, can give rise to new transport phenomena.

The fact that $\ell_N$ is the shortest length scale in our approach may lead one to think that quantum fluctuations do not play a significant role. However, already in the limit of rigid textures considered here, we may encounter difficulties in invoking a classical to quantum correspondence for some symmetry operations. 
In this limit, the kinetic energy of skyrmions is totally quenched and the quantum dynamics are effectively constrained to the lowest \textit{Landau level}, as sketched in Fig.~\ref{fig:qhe}(b). This truncation of the Hilbert space introduces subtle effects in the skyrmion quantum numbers, more prominently in the angular-momentum spectrum. The way to regularize the quantum theory is by including additional degrees of freedom, corresponding to distortions of the texture that lend some inertia to the skyrmion motion \cite{Makhfudz2012,Psaroudaki2016}.

A more immediate consequence, though, is the lifting of the macroscopic degeneracy of the lowest Landau levels due to quantum interference. As we stressed before, the analogy with the dynamics of charged particles in a strong magnetic field ultimately manifests the special kinematics of spins in ferromagnetic insulators, for which the energy-stress tensor is ill-defined \cite{Haldane1986}. Nevertheless, and contrary to the case of, e.g., domain walls \cite{Yan2013}, the linear momentum of skyrmions is well defined \cite{Papanicolaou1991,Tchernyshyov2015}. {The translation of this to the quantum realm is} that the energy spectrum of skyrmions should be arranged in bands, with quantum states labeled by a well-defined quasimomentum. Naively, the only way to reconcile this observation with the analog to cyclotron motion is if the latter preserves the discrete translational symmetry of the lattice, i.e., if the extension $2\pi (\ell_N)^2$ of quantum fluctuations in phase space is commensurate with $A_c$. But this is precisely what Eq.~\eqref{eq:magnetic_length} tells us. The initially featureless Landau level splits then into $N$ dispersive bands \cite{Galkin2007,Takashima2016}. These bands are characterized by nonzero Berry curvatures \cite{Berry1984}, which can be interpreted as the quantum descendant of the classical Magnus force. The nontrivial topology of the skyrmion bands anticipates the existence of edge modes localized at the physical terminations of the system, just like in the quantum Hall effect \cite{Halperin1982}. The number and chirality of the edge modes, dictated by the Chern numbers of the skyrmion bands, depend crucially on $S$.

All these features of the skyrmion spectrum are manifested in the transport coefficients dominated by their translational dynamics. Here we are going to focus on thermal transport, for these measurements are a powerful technique for the study of nonequilibrium phenomena in magnetic insulators and can reveal the presence of exotic quasiparticles \cite{Hirschberger2015a,Kasahara,Lee2015,Hirschberger2015}. Systems hosting skyrmionic quasiparticles naturally display a thermal Hall effect, regardless of the particular symmetry of the lattice and as a consequence of the nontrivial topology of the magnetization texture \cite{KimSoo2016}. In addition to that, we find that details of the skyrmion spectrum, like Dirac points, avoided crossings, etc., play a major role in the Hall response of the system when the temperature is comparable with the skyrmion bandwidth, $T\lesssim t$. The skyrmion bandwidth $t$ is controlled by the Fourier components of the lattice potential, corresponding to a fraction of the energy of the classical texture that decays only algebraically with $\mathcal{R}/a$. As $S$ increases, the gaps in the skyrmion spectrum decrease and, therefore, these details are expected to become less important. We find, however, that differences between integer and half-integer spins persist even in the semiclassical regime, reflecting the importance of quantum interference in the skyrmion dynamics when the size of the texture is comparable with the microscopic lattice. These parity effects can be traced to the number and chirality of skyrmion edge states carrying the energy flow.



\subsection{Structure of this article}

The manuscript is divided in three main sections. In Sec.~\ref{sec:sky_quantization}, we introduce quantum fluctuations in the dynamics of skyrmion textures starting from the classical equations of motion. The first two subsections are devoted to understand the origin of Eq.~\eqref{eq:G3}. Supplementary details on the symplectic structure of the skyrmion dynamics are provided in Appendices~\ref{sec:A}~and~\ref{sec:B}. Readers familiarized with micromagnetics (Sec.~\ref{sec:micromagnetics}) and collective coordinates (Sec.~\ref{sec:coordinates}) can jump directly to Sec.~\ref{sec:quantization}, where we quantize the rigid motion of skyrmions and derive the expression for the \textit{magnetic length} in Eq.~\eqref{eq:magnetic_length}. Issues related to the representation of wave functions and the prescription for operator ordering in this truncated Hilbert space are saved for Appendix~\ref{sec:C}. Appendix~\ref{sec:anomaly} discusses in more detail the anomaly in the angular-momentum spectrum associated with the reduction of degrees of freedom.

Equipped with this formalism, we study in Sec.~\ref{sec:lattice} the problem of nanoscale skyrmions subjected to a periodic potential. We evaluate first in Sec.~\ref{sec:potential} the amplitude of the umklapp processes generated by the microscopic lattice, which decays (algebraically) with the size of the skyrmion texture measured in units of the lattice spacing. Section~\ref{sec:translations} deals with the algebra of skyrmion translations and the resemblance with magnetic lattices. In Sec.~\ref{sec:bands}, we compute the resulting skyrmion bands in a simplified geometry, their Berry curvatures, and the appearance of edge states localized at the physical terminations of the system.

In Sec.~\ref{sec:semiclassic}, we study how an ensemble of nanoscale skyrmions respond to a nonequilibrium bias, assuming that the driving forces vary slowly on the atomic scale. Section~\ref{sec:wp} discusses the semiclassical equations of motion describing the dynamics of skyrmion wave packets. We introduce skyrmion transport currents in Sec.~\ref{sec:transport}; some technical details are reserved for Appendix~\ref{sec:currents}. We show in Sec.~\ref{sec:thermal} that the thermal conductivity driven by skyrmions possesses in general two different contributions: the usual dissipative term coming from the deviation from the equilibrium distribution function, and a transverse nondissipative term related to the flow of quasiparticles and energy at the boundaries of the system. This thermal Hall effect depends on the number of skyrmion modes and is therefore sensitive to the value of $S$. Finally, we summarize our main findings in Sec.~\ref{sec:outlook}, pointing at experimental and theoretical prospects.



\section{Skyrmion quantization}
\label{sec:sky_quantization}

\begin{figure}
\includegraphics[width=0.6\columnwidth]{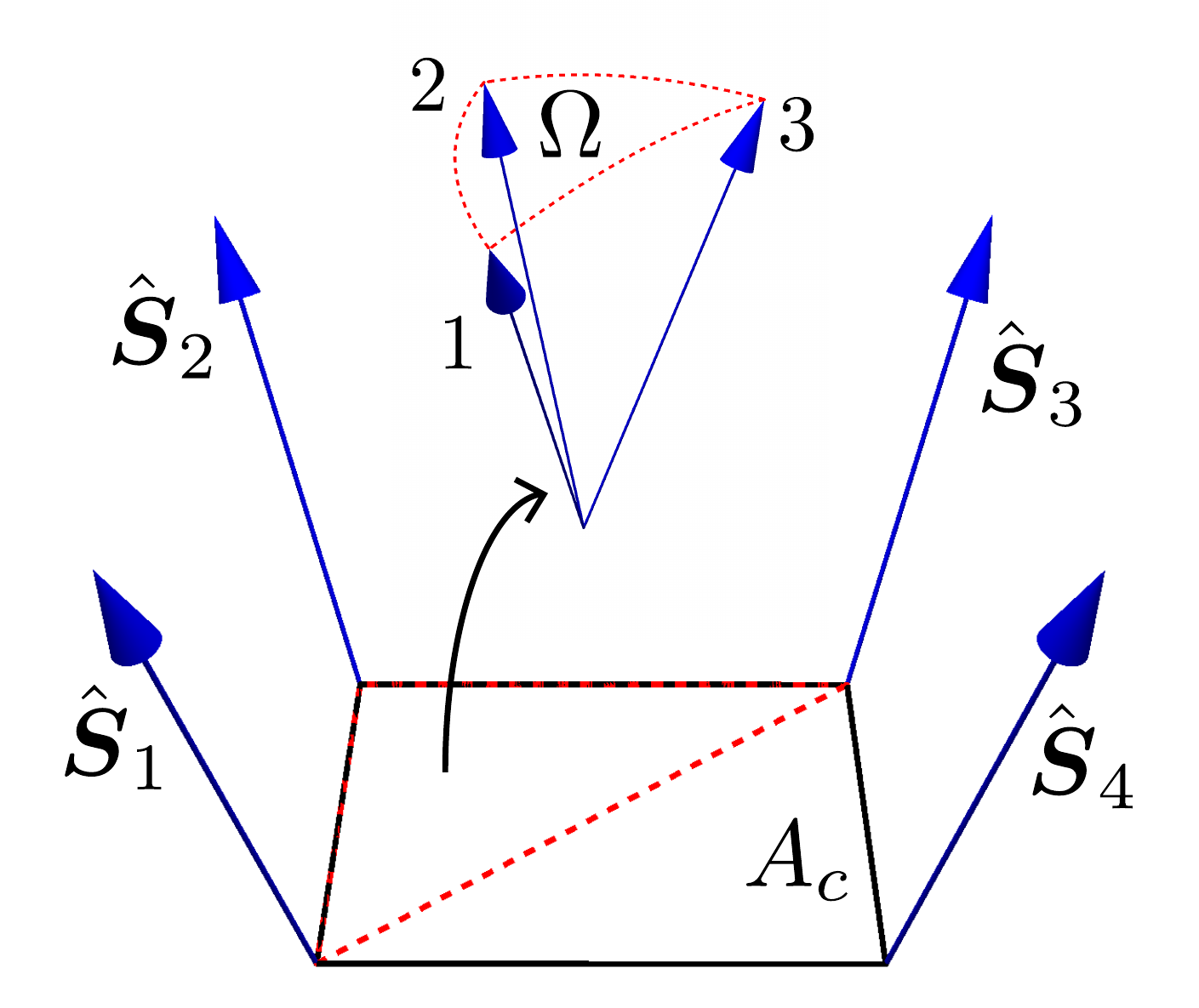}
\caption{Coarse-graining of the skyrmion charge by triangulation of the spin lattice. The topological density can be estimated from the solid angle $\Omega$ subtended by triads of arrows representing the macroscopic state of the magnet.}
\label{fig:textures_bis}
\end{figure}

Magnetism is intrinsically a quantum-mechanical phenomenon \cite{vanVleck}. 
The magnetic response of many solids is the ultimate manifestation of the internal angular momentum of their constituents, the spin of the electrons. Nevertheless, emergent excitations on top of a symmetry-broken ground state can be regarded as classical, in particular at temperatures much lower than the ordering transition, when short-ranged fluctuations are inconsequential. In the field of micromagnetics \cite{Brown1963}, the magnetization dynamics is assumed to be dominated by these classical, hydrodynamical modes that vary smoothly on the scale of the microscopic lattice \cite{Halperin1969}. The dynamics of magnetic textures is often discussed within this framework.

The basic idea for our quantization procedure is that these hydrodynamical modes can be built from a semiclassically coarse-grained spin-density operator
\begin{align}
\label{eq:quantum_s}
\hat{\bm{s}}\left(\mathbf{r}\right)=\hbar\sum_i\hat{\bm{S}}_i\,\delta\left(\mathbf{r}-\mathbf{R}_i\right).
\end{align}
In electrically insulating magnets, $\hat{\bm{S}}_i=(\hat{S}_i^x,\hat{S}_i^y,\hat{S}_i^z)$ is a vector of spin operators defined in lattice positions $\mathbf{R}_i$, satisfying the usual commutation relations $[\hat{S}_i^{\alpha},\hat{S}_j^{\beta}]=i\,\epsilon_{\alpha\beta\gamma}\,S_i^{\gamma}\,\delta_{ij}$; in this last expression $\epsilon_{\alpha\beta\gamma}$ and $\delta_{ij}$ correspond to the Levi-Civita and Kronecker-delta symbols, respectively, and summation over repeated indices is assumed. The saturated spin-density field can be formally defined as the expectation value of this operator in the spin-coherent representation \cite{Klauder1979} of the macroscopic state of the magnet, $\bm{s}\left(\mathbf{r}\right)\approx\langle \Psi_{\textrm{sc}}| \hat{\bm{s}}(\mathbf{r}) |\Psi_{\textrm{sc}}\rangle$ (see Appendix~\ref{sec:A}). The following regularization of the Dirac delta in Eq.~\eqref{eq:quantum_s} is implicit in this construction:\begin{align}
\label{eq:regularization}
\delta \left( \mathbf{R}_i-\mathbf{R}_j\right)\longrightarrow A_c^{-1}\,\delta_{ij},
\end{align}
In principle, $A_c$ does not correspond necessarily to the size of the microscopic cell, but rather to the short-wavelength cutoff for the continuum description in the $XY$ plane of the film. Hence, $\hat{\bm{S}}_i$ must be interpreted as a quantum \textit{macrospin} more than the true microscopic degree of freedom of the magnet. As we mentioned before, we are going to assume that the magnetization remains uniform along the $z$-axis and the argument of the spin-density field is just a 2-dimensional vector, $\mathbf{r}=(x,y)$. The number of layers along the $z$-axis is incorporated in $S$, the spin quantum number.

In a similar manner, the skyrmion charge defined in Eq.~\eqref{eq:Q} can be coarse-grained by triangulation of the lattice, as illustrated in Fig.~\ref{fig:textures_bis}. The topological density is given by the averaged solid angle $\Omega$ (divided by $4\pi$) subtended by groups of 3 spins; the solid angle can be computed from Berg's formula \cite{Berg1981}
\begin{align}
e^{\frac{i\Omega}{2}}=\frac{1+\bm{n}_1\cdot\bm{n}_2+\bm{n}_1\cdot\bm{n}_3+\bm{n}_2\cdot\bm{n}_3+i\bm{n}_1\cdot\left(\bm{n}_2\times\bm{n}_3\right)}{\sqrt{2\left(1+\bm{n}_1\cdot\bm{n}_2\right)\left(1+\bm{n}_1\cdot\bm{n}_3\right)\left(1+\bm{n}_2\cdot\bm{n}_3\right)}}.
\end{align}
Based on this heuristic construction, the goal of this section is to \textit{requantize} the dynamics of the magnet, now constrained to the phase space of rigid skyrmion textures. We are going to apply a symplectic reduction \cite{Faddeev1988} instead of imposing second class constraints like in Dirac's method \cite{Dirac1964}. The path-integral formulation of the quantum motion of a skyrmion can be found in~\onlinecite{Psaroudaki2016}. 

\subsection{Micromagnetics}
\label{sec:micromagnetics}

In a planar magnet at temperatures well below $T_c$, the magnetization density saturates at some fixed value. Macroscopic variations in this magnitude are strongly penalized and short-ranged fluctuations can be safely omitted, so a continuum description is valid. The dynamics of the spin-density field is governed by the Landau-Lifshitz equation \cite{Landau1,Landau2},\begin{align}
\dot{\bm{s}}\left(\mathbf{r}\right)=\bm{s}\left(\mathbf{r}\right)\times\bm{h}_{\textrm{eff}}\left(\mathbf{r}\right),
\label{eq:LL}
\end{align}
where $\bm{h}_{\textrm{eff}}$ is the thermodynamic conjugate force to $\bm{s}$, $\bm{h}_{\textrm{eff}}\equiv -\delta H/\delta\bm{s}$. This equation describes the transverse magnetization dynamics preserving the free energy $H$. 
The coupling with microscopic degrees of freedom introduces dissipation, phenomenologically described by Gilbert damping \cite{Gilbert2004}.

Since $\bm{s}\left(\mathbf{r}\right)$ can be understood as an average of the quantum spin-density operator defined in Eq.~\eqref{eq:quantum_s}, the classical magnetization dynamics should be expressed in terms of a Liouville-like equation associated with the evolution of this operator in the Heisenberg picture, where the phenomenological free energy, which is a functional of the coarse-grained field only, replaces the microscopic spin Hamiltonian. Indeed, the Landau-Lifshitz equation can be recast as \cite{Dzyaloshinskii1980} \begin{align}
\label{eq:Hamilton_eq}
\dot{\bm{s}}\left(\mathbf{r}\right)=\left\{\bm{s}\left(\mathbf{r}\right),H\right\}=\int d^2\mathbf{r}'\left\{\bm{s}\left(\mathbf{r}\right),s_{\alpha}\left(\mathbf{r}'\right)\right\}\frac{\delta H}{\delta s_{\alpha}\left(\mathbf{r}'\right)},
\end{align}
with the Poisson brackets given by\begin{align}
\label{eq:s_algebra}
\left\{s_{\alpha}\left(\mathbf{r}\right),s_{\beta}\left(\mathbf{r}'\right)\right\}=\epsilon_{\alpha\beta\gamma}\,s_{\gamma}\left(\mathbf{r}\right)\delta\left(\mathbf{r}-\mathbf{r}'\right).
\end{align}
The quantum-commutation relations for the spin-density operator defined in Eq.~\eqref{eq:quantum_s} are recovered form this last expression upon the usual identification,\begin{align}
\label{eq:identification}
& \{\,,\,\}\longrightarrow  -\frac{i}{\hbar}\,[\,,\,].
\end{align}



The quantization rules for the soliton dynamics follow from this relation once canonical variables are identified. This is not straightforward, though. The difficulty comes from a kinematic constraint: since $s=|\bm{s}|$ remains constant during the evolution, there are more elements in the algebra of Eq.~\eqref{eq:s_algebra} than dynamical variables. The phase space is then arranged in \textit{symplectic manifolds} \cite{Morrison1998} corresponding to spheres in the 3-dimensional space of coordinates $\bm{s}=(s_x,s_y,s_z)$. A generalization of Darboux's theorem guarantees that canonical variables can be defined, but only locally in general. In fact, there is not a global parametrization covering the whole symplectic manifold, making canonical variables globally ill-defined for generic textures. For example, if we try to project these spheres onto a plane of generalized coordinates there is always a point $\bm{n}_0$ (the north pole in the usual stereographic projection) that is singular. Everywhere except at that point, the mapping is bijective, and the Landau-Lifshitz equation can be obtained from the Lagrangian \begin{align}
\label{eq:L_main}
L\left[\bm{n}\right]=\int d^2\mathbf{r}\,\, \bm{a}\left[\bm{n}\left(\mathbf{r}\right)\right]\cdot\dot{\bm{s}}\left(\mathbf{r}\right)-H\left[\bm{n}\right].
\end{align}
As we explain in detail in Appendix~\ref{sec:A}, the Euler-Lagrange equations derived from Eq.~\eqref{eq:L_main} reduces to Eq.~\eqref{eq:LL} as long as  $\bm{\nabla}_{\bm{n}}\times \bm{a}=-\bm{n}$, i.e., $\bm{a}[\bm{n}]$ corresponds to the gauge field created by a monopole at the center of the sphere,\begin{align}
\label{eq:a}
\bm{a}\left[\bm{n}\right]=\frac{\bm{n}_0\times\bm{n}}{1-\bm{n}_0\cdot\bm{n}},
\end{align}
where $\bm{n}_0$ represents the direction of the Dirac string connecting the source with a infinitely distant monopole of opposite charge. The election of canonical variables as well as the choice of the gauge ($\bm{n}_0$) in Eq.~\eqref{eq:a} is not unique and depends on the parametrization of the order parameter. In spherical coordinates, for example, we can write $\bm{n}=(\sin\theta \cos\phi,\sin\theta\sin\phi,\cos\theta)$, where $\phi$ and $s\cos\theta$ form a canonical pair, $\{\phi(\mathbf{r}),s\cos\theta(\mathbf{r}')\}=\delta(\mathbf{r}-\mathbf{r}')$. The \textit{kinetic} part in the Lagrangian of Eq.~\eqref{eq:L_main} reduces to the familiar Wess-Zumino term, $L_{\textrm{WZ}}=s\int d^2\mathbf{r}\,\, \dot{\phi}\left(\cos\theta\pm1\right)$ \cite{Wess1971}, where the sign $\pm$ applies to the Dirac string intersecting the north/south pole, $\bm{n}_0=\pm\bm{z}$.

This Lagrangian formulation can be extended to the quantum limit by means of the appropriate path integral \cite{Braun1996,Klauder1979}, from which we have\begin{align}
\label{eq:S}
s=\frac{\hbar S}{A_c}.
\end{align}
The field in Eq.~\eqref{eq:a} can be identified then with the Berry-phase connection of the spin-coherent state \cite{Kovner1989}, where the phase ambiguity in the definition of the latter provides the apparent gauge freedom. Here it is worth emphasizing that the quantization condition in Eq.~\eqref{eq:S} is independent of the specific coarse-graining procedure and stems from the single-valueness of the semiclassical (macroscopic) state of the magnet (i.e., gauge invariance). Consider a slow (adiabatic) change of the order parameter along a closed path on the sphere, for which at each point of the sample $\phi(\mathbf{r})\rightarrow\phi(\mathbf{r})+2\pi$ in a time interval $\left[0,T\right]$. This is equivalent to a gauge transformation of the associated spin coherent state, as discussed in Appendix~\ref{sec:A}. The quantum amplitude of this process is
\begin{align}
e^{i\frac{\delta\mathcal{S}}{\hbar}}= e^{\frac{i}{\hbar}\int_0^{T} dt\, L_{\textrm{WZ}}}=e^{-\frac{i s}{\hbar} \int d^2\bm{r}\,\Omega\left(\bm{r}\right)},
\end{align}
where $\Omega(\mathbf{r})$ is the solid angle subtended by $\bm{n}(\mathbf{r})$ during the adiabatic evolution. There is an ambiguity in the definition of $\Omega(\mathbf{r})$ associated with the intersection of the Dirac string, $\Omega(\mathbf{r})=2\pi\times[\pm 1-\cos\theta(\mathbf{r})]$. Therefore, the phase of the quantum action is ambiguous unless the saturated spin-density $s$ multiplied by the area of the magnetic film is an integer or half-integer multiple of $\hbar$, which is equivalent to the condition expressed in Eq.~\eqref{eq:S}.

\subsection{Collective coordinates}
\label{sec:coordinates}

The spin-density field can be parametrized, in principle, by an infinite number of generalized coordinates. Nevertheless, the dynamics of stable textures is often assumed to be dominated by just a few slow modes with long relaxation times \cite{Tretiakov2008}. When the system is translationally invariant, $\partial_i H\approx0$, the displacements of the skyrmion as a rigid texture cost no energy. The collective coordinate, $\bm{R}=\left(X,Y\right)$, corresponding to the center of the skyrmion charge \cite{Papanicolaou1991,Moutafis2009,Kravchuk2018},\begin{align}
\label{eq:sky_center}
\bm{R}\equiv\frac{\int d^2\mathbf{r}\,\,\mathbf{r}\,\,\bm{n}\cdot\left(\partial_x\bm{n}\times\partial_y\bm{n}\right)}{\int d^2\mathbf{r}\,\,\bm{n}\cdot\left(\partial_x\bm{n}\times\partial_y\bm{n}\right)},
\end{align}
evolves on time scales much larger than the rest of coordinates parametrizing the texture.\footnote{Hereafter the magnon spectrum is assumed to be gapped due to an external field.}

In the low-frequency limit, the skyrmion dynamics can be approximated by $\dot{\bm{s}}\approx -\dot{R}_i\,\partial_i\bm{s}$, in which the texture is assumed to move in a rigid fashion. The Hamiltonian dynamics are generated by the Poisson bracket in Eq.~\eqref{eq:G3}, which can be derived from the algebra in Eq.~\eqref{eq:s_algebra} constrained to the phase space of rigid skyrmion textures, see Appendix~\ref{sec:B}. The equation of motion, $\dot{\bm{R}}=\left\{\bm{R},V(\bm{R})\right\}$, corresponds to the Thiele equation \cite{Thiele1973}\begin{align}
\label{eq:classical}
4\pi s\mathcal{Q}\,\dot{\bm{R}}\times\mathbf{z}=\bm{F},
\end{align}
where $\bm{F}\equiv -\partial V/\partial{\bm{R}}$ is the generalized force (e.g., by a confining potential that breaks the translational symmetry). In this expression $V(\bm{R})\equiv H[\bm{n}_{\textrm{sk}}(\mathbf{r}-\bm{R})]$ must be taken as the free-energy functional evaluated with the skyrmion solution.

The variable canonically conjugate to $\bm{R}$ is $\bm{\Pi}\equiv4\pi s \mathcal{Q}\,\bm{R}\times\mathbf{z}$, $\{R_i,\Pi_j\}=\delta_{ij}$; this is indeed the generator of translations of the rigid texture,\begin{align}
\{\Pi_i,\bm{s}\}=4\pi s\mathcal{Q}\,\epsilon_{ij}\{R_j,\bm{s}\}=\epsilon_{ij}\epsilon_{jk}\frac{\partial \bm{s}}{\partial R_k}\approx\partial_i\bm{s}.
\end{align}
However, the algebra of translations is not closed,\begin{align}
\label{eq:generators}
\{\Pi_i,\Pi_j\}=4\pi s\mathcal{Q}\,\epsilon_{ij},
\end{align}
 resulting from the geometrical nature of the kinetic term in the Lagrangian of Eq.~\eqref{eq:L_main}. This is ultimately related to the appearance of a Magnus force in the left-hand side of Eq.~\eqref{eq:classical} \cite{Thouless1996,Watanabe2014}. It is worth noticing at this point that despite the formal resemblance with the case of a charged particle in the presence of a magnetic field, there is not a true gauge field acting here and, therefore, the generators of translations do not lose their physical meaning. This is reflecting, in the last instance, the fact that magnetic skyrmions are local excitations, involving only a finite number of reversed spins around its core, in contrast, for example, to domain walls.

Gapped modes of the magnetization dynamics generate an inertial term in the effective Lagrangian for the skyrmion center \cite{Makhfudz2012,Buttner2015,Lin2017,Psaroudaki2016},\begin{align}
\label{eq:Leff}
\mathcal{L}_{\textrm{eff}}=2\pi s\mathcal{Q}|\dot{\bm{R}}\wedge\bm{R}|+\frac{M}{2}|\dot{\bm{R}}|^2-V\left(\mathbf{R}\right),
\end{align}
where $|\dot{\bm{R}}\wedge\bm{R}|\equiv \mathbf{z}\cdot(\dot{\bm{R}}\times\bm{R})$ and $\times$ stands for the usual vectorial product. In the Hamiltonian formalism, the dynamics in the limit of rigid textures ($M=0$) is recovered by imposing the second class constraint $\bm{P}=0$, where $\bm{P}=M\dot{\bm{R}}$ is the kinetic momentum of massive skyrmions (see Appendix~\ref{sec:B}). The mass $M$ accounts for deformations of the moving skyrmion compared with the static solution. Their effects are enhanced in confined geometries, when translational symmetry is broken. In that situation one must be careful in the definition of collective coordinates. Equation~\eqref{eq:sky_center} ensures that the skyrmion displacement coincides with a traveling-wave mode \cite{Kravchuk2018}. All other degrees of freedom can in principle be accounted for by integrating them out with respect to the skyrmion center in the path integral formalism \cite{Psaroudaki2016}.

\subsection{Phase-space quantization}
\label{sec:quantization}

Once the dynamics of the skyrmion has been reduced to a symplectic manifold parametrized by collective coordinates $\bm{R}=(X,Y)$, we can introduce quantum fluctuations just by promoting these variables to operators, $\bm{R}\longrightarrow\hat{\bm{R}}=(\hat{X},\hat{Y})$, while preserving the canonical structure of the classical dynamics as prescribed by Eq.~\eqref{eq:identification}; from Eqs.~\eqref{eq:G3}~and~\eqref{eq:S}, we have\begin{align}
\label{eq:G4}
i\hbar\left\{R_i,R_j\right\}\longrightarrow [\hat{R}_i,\hat{R}_j]=\frac{iA_c}{4\pi S\mathcal{Q}}\,\epsilon_{ij}\equiv\pm\, i\,\ell_{N}^2\epsilon_{ij},
\end{align}
where the upper/lower sign applies to positive/negative $\mathcal{Q}$ hereafter. In the last expression we have recast the commutator in terms of the \textit{magnetic length} $\ell_N$ defined in Eq.~\eqref{eq:magnetic_length}. 
This is thus a measure of the extension of the quantum fluctuations in the space of skyrmion positions, as depicted in Fig.~\ref{fig:qhe}(a). The limit $\ell_N\ll\sqrt{A_c}\sim a$ amounts to the classical limit of large spins, $S\rightarrow\infty$.

This length scale defines also the high-energy cut-off for our quantization procedure, while the skyrmion mass serves as a control parameter for the hybridization with gapped modes of the magnetization dynamics. More specifically, the spectrum of the quantum Hamiltonian derived from Eq.~\eqref{eq:Leff} (omitting $V(\mathbf{R})$ for the moment) is organized in Landau levels separated by energy gaps $E_{\textrm{gap}}=\hbar^2/M\ell_N^2$, as illustrated in Fig.~\ref{fig:qhe}(b). The Hilbert space of the reduced theory is, therefore, truncated, corresponding to the projection onto the lowest Landau level $|LL=0\rangle$ created by the field of the Dirac monopole. The dimension of the truncated Hilbert space is indeed\footnote{This is just the degeneracy of the lowest Landau level, corresponding to the number of fictitious flux quanta crossing the system. Note that the area covered by the semiclassical cyclotron orbits is $2\pi\ell_{N}^2$, thus the unit cell of the lattice encloses $N=2S|\mathcal{Q}|$ flux quanta.} 
\begin{align}
\label{eq:dimension}
\frac{A_c\times N_c}{2\pi\left(\ell_{N}\right)^2}=N\times N_c=2S\left|\mathcal{Q}\right|N_c,
\end{align}
where $N_c$ is the number of unit cells in the lattice. 

The truncation of the Hilbert space is non-perturbative, as reflected by the divergence of the Landau gap in the limit of $M=0$. An universal subtraction (i.e., a renormalization of the ground-state energy) removes the infinities in the energy spectrum of the truncated theory. The effect on the angular-momentum spectrum is more subtle, though. In the constrained theory, the $z$-component of the angular momentum corresponds to $L_z=|\bm{R}\wedge\bm{\Pi}|/2=-2\pi s\mathcal{Q}|\bm{R}|^2$, which is the generator of rotations of rigid skyrmion textures,\begin{align}
\left\{L_z,\bm{s}\right\}=-4\pi s\mathcal{Q}\,R_i\left\{R_i,\bm{s}\right\}\approx\epsilon_{ij}\,R_i\,\partial_j\bm{s}.
\end{align}
Promoting $L_z$ to a quantum-mechanical operator following the same prescription as for Eq.~\eqref{eq:G4} leads to \begin{align}
\label{eq:Lz}
L_z\longrightarrow \hat{L}_z=\mp\,\hbar\left(\hat{a}^{\dagger}\hat{a}+\frac{1}{2}\right),
\end{align}
where $\hat{a}^{\dagger}=(\hat{X}\mp i\hat{Y})/\sqrt{2}\ell_N$, $\hat{a}=(\hat{X}\pm i\hat{Y})/\sqrt{2}\ell_N$ are ladder operators satisfying the usual boson algebra, $[\hat{a},\hat{a}^{\dagger}]=1$. As we explain in Appendix~\ref{sec:C}, we can use the coherent states associated with these operators to construct wave functions in the truncated Hilbert space. In this so-called holomorphic representation, $z=(x\mp i y)/\sqrt{2}\ell_N$ plays the role of position variable and therefore operators must be expressed in anti-normal order before acting on the truncated wave functions. This procedure is analogous to the Hilbert space reduction in the quantum Hall effect, when all the observables are projected onto the lowest Landau level \cite{Girvin1984}.

Although a skyrmion can be interpreted as a coherent superposition of magnon bound states and, therefore, it is a local, boson excitation,\footnote{In quantum Hall ferromagnets, however, nontrivial quantum numbers arise from the incompressibility of the spin-polarized electron liquid along with the mixed internal-rotational symmetry of the texture \cite{Nayak1996}.} the associated angular-momentum spectrum is quantized in half-integers according to Eq.~\eqref{eq:Lz}.\footnote{Note that this quantization rule resembles the kinetic angular momentum of a charged particle orbiting around a tube that encloses a (superconducting) quantum of magnetic flux, $\Phi_0=h/2e$ \cite{Wilczek1982}.} This apparent paradox stems from a quantum anomaly \cite{anomalies_book} in the truncated Hilbert space of rigid skyrmions, resulting from the suppression of hard modes of the magnetization dynamics. As we mentioned earlier, the completion of the theory is provided by Eq.~\eqref{eq:Leff}, for which the angular-momentum spectrum takes integer values, see Appendix~\ref{sec:anomaly}. Thus, the inclusion of a skyrmion mass in the quantum theory reconciles the anomalous behavior of the angular momentum with the rotational symmetry of the classical theory.

\section{Skyrmions on the lattice}
\label{sec:lattice}

We expect quantum effects to become relevant when the characteristic size of the skyrmion texture is not too large compared to the underlying lattice spacing. In that case, the free energy of the texture cannot be taken as translationally invariant anymore, i.e., translations cost certain amount of energy that defines the potential created by the microscopic lattice, $V\left(\bm{R}\right)$. Since $V\left(\bm{R}+\mathbf{R}_i\right)=V\left(\bm{R}\right)$ (neglecting boundary effects for the moment), the lattice potential admits a Fourier expansion of the form
\begin{align}
\label{eq:Vlat}
V\left(\bm{R}\right)=\sum_{\left\{\mathbf{G}\right\}}V_{\mathbf{G}}\, e^{i\mathbf{G}\cdot\bm{R}} \longrightarrow \hat{H}=\sum_{\left\{\mathbf{G}\right\}}V_{\mathbf{G}}\,\hat{T}\left(\mathbf{G}\right).
\end{align}
Here the sum is extended to the vectors in the reciprocal lattice and $V_{\mathbf{G}}=\left(V_{-\mathbf{G}}\right)^*$ are the Fourier components (or harmonics) of the expansion.

Having promoted the skyrmion coordinates to quantum operators, the Hamiltonian can be expressed in terms of translation operators $\hat{T}$ in the phase space of rigid skyrmion textures. These operators resemble the algebra of translations in the presence of a magnetic field \cite{Zak1964}, reminiscent of the symplectic structure of the micromagnetic dynamics. The flux of this field is commensurate with the lattice, as we anticipated, preserving the discrete translational symmetry. 
The Hamiltonian can be easily diagonalized using algebraical methods, as we are going to explain in this section. Akin to the Magnus force in the classical dynamics, the skyrmion bands are characterized by Berry curvatures, the latter associated with the accumulation of geometrical phases by the skyrmion Bloch states in their coherent evolution in reciprocal space. 

\subsection{Lattice potential}
\label{sec:potential}

The first task is to evaluate the Fourier components $V_{\mathbf{G}}$. For a continuum texture, these are formally defined as\begin{align}
V_{\mathbf{G}}\equiv\frac{1}{A}\int d^2\bm{R}\, \,H[\bm{n}_{\textrm{sk}}(\bm{R})]\, e^{-i\mathbf{G}\cdot\bm{R}},
\label{eq:V_def}
\end{align}
where $A=N_c\times A_c$ is the area of the magnetic film and the integration is performed over the position of the skyrmion $\bm{R}$. 
The Hamiltonian corresponds to the free-energy functional evaluated with the classical soliton solution,\begin{align}
H[\bm{n}_{\textrm{sk}}(\bm{R})] & =\int d^2\mathbf{r}\,\mathcal{H}_{\textrm{sky}}\left(\mathbf{r}-\bm{R}\right)
\nonumber\\
& \approx A_c\sum_{i}\mathcal{H}_{\textrm{sky}}\left(\mathbf{R}_i-\bm{R}\right),
\label{eq:H_approx}
\end{align}
where $\mathcal{H}_{\textrm{sky}}\left(\mathbf{r}-\bm{R}\right)$ represents the coarse-grained free-energy density and the last sum is extended to positions on the lattice. By plugging Eq.~\eqref{eq:H_approx} into Eq.~\eqref{eq:V_def} we obtain
\begin{align}
V_{\mathbf{G}} & \approx\frac{A_c}{A}\sum_{i}\int d^2\bm{R}\,\,\mathcal{H}_{\textrm{sky}}\left(\mathbf{R}_i-\bm{R}\right) e^{-i\mathbf{G}\cdot\bm{R}}
\nonumber\\
&\approx \int d^2\mathbf{r}\,\mathcal{H}_{\textrm{sky}}\left(\mathbf{r}\right)e^{i\mathbf{G}\cdot\mathbf{r}}.
\label{eq:V_approx}
\end{align}
In the second line of this equation we have changed the integration variable, $\bm{R}\rightarrow \mathbf{r}\equiv \mathbf{R}_i-\bm{R}$ (assuming that the system is very large, so boundary effects are neglected for the moment), in such a way that the positions in the lattice only enter through exponentials; those are summed up as $\sum_ie^{-i\mathbf{G}\cdot\mathbf{R}_i}=N_c$ for vectors $\mathbf{G}$ of the reciprocal lattice.

The final result in Eq.~\eqref{eq:V_approx} is very convenient in order to estimate the strength of the Fourier components in terms of the parameters defining the classical texture. The $\mathbf{G}=0$ component, for example, is just the energy of the skyrmion solution, $V_{\mathbf{G}=0}\equiv\varepsilon_0$. Higher harmonics with $|\mathbf{G}|\neq 0$ incorporate the effect of the lattice, removing the degeneracy of the classical solution by umklapp processes. These effects are controlled by the size $\mathcal{R}$ of the skyrmion texture with respect to the lattice spacing $a\sim \sqrt{A_c}$, and can be ignored in the limit $\mathcal{R}\gg a$. For simplicity, we can assume that $\mathcal{H}_{\textrm{sky}}\left(\mathbf{r}\right)$ varies only within a region of radius $\mathcal{R}$. 
The integral in Eq.~\eqref{eq:V_approx} can be approximated by\begin{align}
V_{\left|\mathbf{G}\right|}\approx2\pi\int_0^{\mathcal{R}} dr\,r\,\bar{\mathcal{H}}_{\textrm{sky}}\left(r\right)J_0\left(\left|\mathbf{G}\right|r\right),
\end{align}
where we have assumed an axially symmetric skyrmion solution. Here $J_i(x)$ are Bessel functions of the first class. The strength of the harmonics decreases algebraically with $\mathcal{R} |\mathbf{G}|$, i.e., with the ratio between the size of the skyrmion texture and the lattice spacing. Finally, we can approximate $\mathcal{H}_{\textrm{sky}}\left(r\right)$ by $\varepsilon_0/\pi \mathcal{R}^2$ within the radius of a classical solution to obtain\begin{align}
\label{eq:approx}
V_{\left|\mathbf{G}\right|}\approx\frac{2\,\varepsilon_0}{\mathcal{R}\left|\mathbf{G}\right|}\,J_1\left(\mathcal{R}\left|\mathbf{G}\right|\right).
\end{align}

\subsection{The group of translations}
\label{sec:translations}

The classical potential created by the lattice can be written as in the left-hand side of Eq.~\eqref{eq:Vlat}, where the Fourier coefficients are just numbers estimated from Eq.~\eqref{eq:approx}. Now we promote the collective coordinates to quantum-mechanical operators:
\begin{align}
e^{i\mathbf{G}\cdot\bm{R}}\longrightarrow e^{i\mathbf{G}\cdot\hat{\bm{R}}}\equiv\hat{T}\left(\mathbf{G}\right).
\end{align}
Here we have applied the usual Weyl ordering prescription \cite{Weyl1927} in the noncommutative plane of skyrmion coordinates \cite{Ezawa2003}. The (unitary) Weyl-Wigner operators $\hat{T}\left(\mathbf{G}\right)$ form a \textit{projective} or \textit{ray} group \cite{Zak1964}, akin to the group of magnetic translations in a crystal, satisfying the algebra\begin{align}
\label{eq:algebra}
\hat{T}\left(\mathbf{G}\right)\hat{T}\left(\mathbf{G}'\right)=\exp\left(\frac{\pm i\left|\mathbf{G}\wedge\mathbf{G}'\right| A_c}{4\pi N}\right)\,T\left(\mathbf{G}+\mathbf{G}'\right),
\end{align}
where the upper/lower sign applies to positive/negative $\mathcal{Q}$ hereafter. Equation~\eqref{eq:algebra} follows immediately from the Baker-Campbell-Hausdorff formula\footnote{$e^{\hat{X}}e^{\hat{Y}}=e^{\hat{X}+\hat{Y}+[\hat{X},\hat{Y}]/2}$} and the commutation relations between the skyrmion coordinates, Eq.~\eqref{eq:G4}.

\begin{figure}
\includegraphics[width=1\columnwidth]{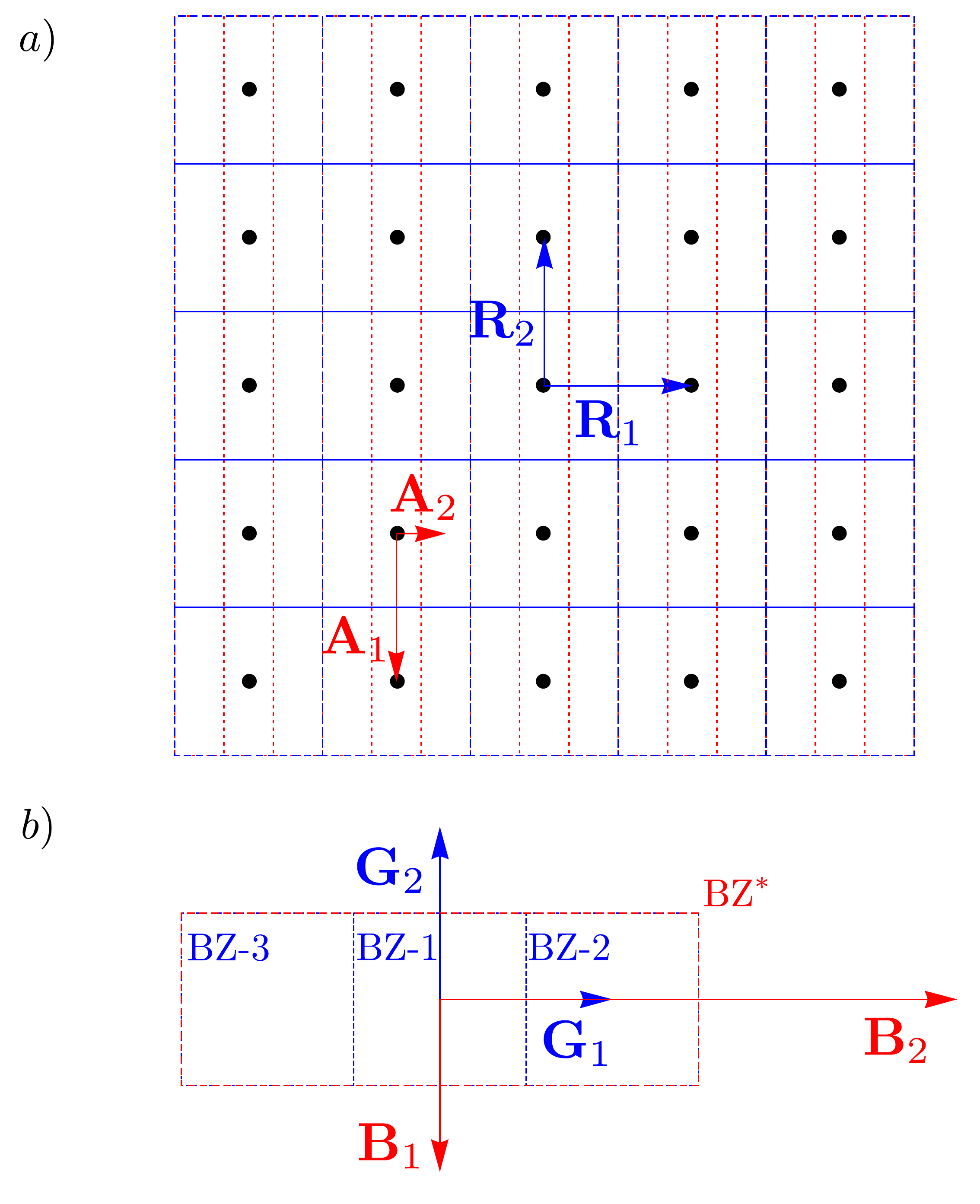}
\caption{$S=3/2$ spins ($N=3$) in a square lattice. $a)$ Lattice in real space. The dots represent the location of the spin operators in the microscopic lattice, highlighted in blue. The von Neumann lattice is represented in red. The primitive vectors $\mathbf{A}_{1,2}$ correspond to $\mathcal{Q}=-1$ skyrmions. $b)$ First Brillouin zone BZ$^*$ of the von-Neumann lattice, consisting of $N=3$ copies of the Brillouin zone BZ of the spin lattice.}
\label{fig:lattice}
\end{figure}

Let us write the vectors in the reciprocal lattice as $\mathbf{G}\equiv\mathbf{G}_{n,m}=n\,\mathbf{G}_1+m\,\mathbf{G}_2$, where $n$, $m$ are integers and \begin{subequations}\begin{align}
& \mathbf{G}_1=2\pi\,\frac{\mathbf{R}_2\times\hat{\mathbf{z}}}{\left|\mathbf{R}_1\wedge\mathbf{R}_2\right|},\\
& \mathbf{G}_2=2\pi\,\frac{\hat{\mathbf{z}}\times\mathbf{R}_1}{\left|\mathbf{R}_1\wedge\mathbf{R}_2\right|}.
\end{align}\end{subequations}
Here $\mathbf{R}_{1(2)}$ are the primitive vectors of the spin lattice, see Fig.~\ref{fig:lattice}. From Eq.~\eqref{eq:algebra}, we can express the translation operators as\begin{align}
\label{eq:operator}
\hat{T}_{n,m}\equiv\hat{T}\left(\mathbf{G}_{n,m}\right)=e^{\mp\frac{ i nm\pi}{N}}\,\left(\hat{T}_1\right)^n\left(\hat{T}_2\right)^m,
\end{align}
where we have introduced the generators\begin{subequations}\begin{align}
& \hat{T}_{1}\equiv e^{i\mathbf{G}_{1}\cdot\hat{\bm{R}}},\\
& \hat{T}_{2}\equiv e^{i\mathbf{G}_{2}\cdot\hat{\bm{R}}}.
\end{align}\end{subequations}These operators act on the skyrmion positions as\begin{subequations}
\label{eq:actions}
\begin{align}
\label{eq:action1}
& \hat{T}_1^{\dagger}\,\hat{\bm{R}}\,\,\hat{T}_1=\hat{\bm{R}}\pm\frac{\mathbf{R}_2}{N},\\
\label{eq:action2}
& \hat{T}_2^{\dagger}\,\hat{\bm{R}}\,\,\hat{T}_2=\hat{\bm{R}}\mp\frac{\mathbf{R}_1}{N}.
\end{align}\end{subequations}

We introduce now Bloch states, formally defined as the simultaneous eigenstates of mutually commuting translation operators. Notice that $\hat{T}_1$ and $\hat{T}_2$ do not commute, but since $|\mathbf{G}_1\wedge\mathbf{G}_2|=4\pi^2/A_c$, it follows from Eq.~\eqref{eq:algebra} that either $\hat{T}_{N,0}=(\hat{T}_1)^N$ or $\hat{T}_{0,N}=(\hat{T}_2)^N$ commutes with any other element of the ray group. Notice also that $(\hat{T}_{1})^N$, $(\hat{T}_{2})^N$ are the generators of translations in the spin lattice (in blue in Fig.~\ref{fig:lattice}a), as inferred from Eqs.~\eqref{eq:actions}. Having written the generators in the order of Eq.~\eqref{eq:operator}, it is convenient to introduce the basis of simultaneous eigenstates of $(\hat{T}_1)^{N}$ and $\hat{T}_2$.\footnote{In other words, we have implicitly chosen a representation of wave functions in which the projection along $\mathbf{G}_2$ plays the role of \textit{positions}.} These must be interpreted as the generators of translations in a fictitious \textit{magnetic lattice} spanned by primitive vectors $\mathbf{A}_1=\pm \mathbf{R}_2$, $\mathbf{A}_2=\mp\mathbf{R}_1/N$. The unit cell of this lattice has area $A_c^*=A_c/N=2\pi(\ell_{N})^2$, enclosing a single flux quantum of the fictitious field. In fact, this is the von Neumann lattice \cite{vonNeumann} associated with the centers of coherent states forming a complete subset in the phase space of collective coordinates (see Appendix~\ref{sec:C}). Figure~\ref{fig:lattice} shows the von Neumann lattice associated with a square spin lattice with $N=3$ ($S=3/2$, $\mathcal{Q}=-1$).

The successive application of $(\hat{T}_1)^{N}$ and $\hat{T}_2$ generates two distinct cyclic subgroups (assuming periodic boundary conditions for the wave functions, as in the band theory of solids; see, e.g., \onlinecite{Dresselhaus}). These subgroups are abelian and therefore have only 1-dimensional irreducible representations, forming a set of phase factors (the characters) of the form $e^{i\mathbf{k}\cdot\mathbf{A}_{1,2}}$. The crystal momentum $\mathbf{k}$ labels the representation and is restricted to the first Brillouin zone of the von Neumann lattice, denoted by BZ$^*$ in Fig.~\ref{fig:lattice}. Thus, we have\begin{subequations}
\begin{align}
\label{eq:Bloch1a}
\left(\hat{T}_1\right)^{N}\left|\mathbf{k}\right\rangle & =e^{i\mathbf{k}\cdot\mathbf{A}_1}\left|\mathbf{k}\right\rangle,\\
\label{eq:Bloch1b}
 \hat{T}_2\left|\mathbf{k}\right\rangle & =e^{i\mathbf{k}\cdot\mathbf{A}_2}\left|\mathbf{k}\right\rangle.
\end{align}\end{subequations}
Moreover, the number of irreducible representations of each cyclic subgroup equals the number of elements and, therefore, the total number of Bloch states is given by the number of cells in the von Neumann lattice, $N_c^*=N\times N_c$. This is indeed the dimension of the truncated Hilbert space, Eq.~\eqref{eq:dimension}.\footnote{In fact, the set of Bloch states $\left\{\left|\mathbf{k}\right\rangle\right\}$ can be understood as the dual (Fourier transform) of a complete and orthogonal (and therefore delocalized) set defined in the von Neumann lattice (see Eq.~\eqref{eq:von_Neumann} and the subsequent discussion in Appendix~\ref{sec:C}).} The spectral decomposition of the identity reads in this basis (in the continuum limit)\begin{align}
\label{eq:spectral}
\hat{1}=\int_{\textrm{BZ}^*}\frac{d\mathbf{k}}{\left(2\pi\right)^2}\,\left|\mathbf{k}\right\rangle\left\langle\mathbf{k}\right\rangle,
\end{align}
where the Bloch states are normalized as $\langle\mathbf{k}|\mathbf{k}'\rangle=(2\pi)^2\,\delta(\mathbf{k}-\mathbf{k}')$. Notice also that they satisfy the periodic conditions $|\mathbf{k}\rangle\equiv|\mathbf{k}+\mathbf{B}\rangle$, where $\mathbf{B}$ is a vector of the reciprocal von Neumann lattice spanned by\begin{subequations}\begin{align}
& \mathbf{B}_1=2\pi\,\frac{\mathbf{A}_2\times\hat{\mathbf{z}}}{\left|\mathbf{A}_1\wedge\mathbf{A}_2\right|}=\pm\mathbf{G}_2,\\
& \mathbf{B}_2=2\pi\,\frac{\hat{\mathbf{z}}\times\mathbf{A}_1}{\left|\mathbf{A}_1\wedge\mathbf{A}_2\right|}=\mp N\mathbf{G}_1.
\end{align}\end{subequations}

The subspace of eigenstates of $(\hat{T}_1)^{N}$ with eigenvalue $e^{i\mathbf{k}\cdot\mathbf{A}_1}$ is $N$-fold degenerate since
\begin{align}
\left(\hat{T}_1\right)^{N}\,\left(\hat{T}_1\right)^n \left|\mathbf{k}\right\rangle=e^{i\mathbf{k}\cdot\mathbf{A}_1}\left(\hat{T}_1\right)^n\,\left|\mathbf{k}\right\rangle,
\end{align}
where $n=0,1,\,...\,\,N-1$, but $(\hat{T}_1)^n\,|\mathbf{k}\rangle\neq |\mathbf{k}\rangle$. Notice that $(\hat{T}_1)^n\,|\mathbf{k}\rangle$ has indeed a different eigenvalue with $\hat{T}_2$:\begin{align}
 \hat{T}_2\,\left(\hat{T}_1\right)^n \left|\mathbf{k}\right\rangle & =e^{\mp i\frac{2\pi n}{N}}\, \left(\hat{T}_1\right)^n\,\hat{T}_2 \left|\mathbf{k}\right\rangle\nonumber\\
& =e^{i\mathbf{k}\cdot\mathbf{A}_2\mp\frac{i2\pi n}{N}}\left(\hat{T}_1\right)^n \left|\mathbf{k}\right\rangle,
\end{align}
where we have used Eq.~\eqref{eq:algebra}. The result in the second line of this equation implies that $(\hat{T}_1)^n |\mathbf{k}\rangle\propto|\mathbf{k}\mp n\,\mathbf{B}_2/N\rangle\equiv |\mathbf{k}+n\,\mathbf{G}_1\rangle$. There is of course a phase (gauge) freedom in the definition of these states, but it is convenient to use\begin{align}
\label{eq:gauge}
\left(\hat{T}_1\right)^n \left|\mathbf{k}\right\rangle=e^{\frac{i n \mathbf{k}\cdot\mathbf{A}_1}{N}} \left|\mathbf{k}+n\,\mathbf{G}_1\right\rangle.
\end{align}

\subsection{Skyrmion bands}
\label{sec:bands}

The complete set $\left\{\left|\mathbf{k}\right\rangle\right\}$ provides a suitable basis to try to the diagonalize the Hamiltonian. Using the spectral decomposition in Eq.~\eqref{eq:spectral} we can write \begin{align}
\label{eq:H2}
\hat{H}=\sum_{\mathbf{k}_1,\mathbf{k}_2}\sum_{n,m} V_{n,m}\,\left\langle\mathbf{k}_1\right|\hat{T}_{n,m}\left|\mathbf{k}_2\right\rangle\,\left|\mathbf{k}_1\right\rangle\left\langle\mathbf{k}_2\right|,
\end{align}
where $\sum_{\mathbf{k}}\equiv\int_{\text{BZ}^*} d\mathbf{k}/(2\pi)^2$ and we have introduced the notation $V_{n,m}\equiv V_{\mathbf{G}_{n,m}}$. The matrix elements can be easily computed from the action of the translation operators on the Bloch states; with our previous gauge choice, we have\begin{align}
\label{eq:irr}
\hat{T}_{n,m}\left|\mathbf{k}\right\rangle=e^{\frac{in\mathbf{k}\cdot\mathbf{A}_1}{N}+im\mathbf{k}\cdot\mathbf{A}_2\mp \frac{inm\pi}{N}}\left|\mathbf{k}+n\,\mathbf{G}_1\right\rangle.
\end{align}
The harmonics multiple of $\mathbf{G}_1$ couple Bloch states belonging to different copies of the first Brillouin zone of the original spin lattice (BZ). There are $N$ copies of BZ, denoted by BZ$_q$, with $q=1,2\,...\,N$, as illustrated in Fig.~\ref{fig:lattice}(b). The matrix elements of the translation operators open gaps at the BZ edges and the single-particle spectrum splits into $N$ bands. It is convenient to fold these copies onto the original BZ$_1$ $\equiv$ BZ and introduce new quantum numbers $q$. This is implemented in 2 steps:\begin{enumerate}[label=(\roman*)]
\item We split the integrations over BZ$^*$ in Eq.~\eqref{eq:H2} into the $N$ copies of BZ:
\begin{align*}
\int_{\text{BZ}^*}\frac{d\mathbf{k}}{\left(2\pi\right)^2}\longrightarrow \sum_{q=1}^{N}\int_{\text{BZ}_q}\frac{d\mathbf{k}_q}{\left(2\pi\right)^2}.
\end{align*}
\item We introduce a new quantum number, i.e., we identify $\left|\mathbf{k}_q\right\rangle\longrightarrow \left|\mathbf{k},q\right\rangle$, since we can write\begin{align*}
\mathbf{k}_q=\mathbf{k}+\left(q-1\right)\mathbf{G}_1,
\end{align*}
and $\mathbf{k}$ is now restricted to BZ$_1$ $\equiv$ BZ.
\end{enumerate}
We can introduce then a $N$-component ket $\Psi_{\mathbf{k}}^{\dagger}=\left(\left|\mathbf{k},1\right\rangle,\left|\mathbf{k},2\right\rangle\,...\,\left|\mathbf{k},N\right\rangle\right)$, so the Hamiltonian reads\begin{align}
\hat{H}=\int_{\text{BZ}}\frac{d\mathbf{k}}{\left(2\pi\right)^2}\Psi_{\mathbf{k}}^{\dagger}\,\mathcal{H}_{\mathbf{k}}\,\Psi_{\mathbf{k}},
\end{align}
where the matrix elements of $\mathcal{H}_{\mathbf{k}}$ are just \begin{align}
\label{eq:matrix_elements}
& \left(\mathcal{H}_{\mathbf{k}}\right)_{\alpha\beta}=\sum_{p,m\in\mathbb{Z}}\left(-1\right)^{mp}e^{\mp\frac{im2\pi}{N}\left(\frac{\alpha+\beta}{2}-1\right)}\nonumber\\
& \times V_{\alpha-\beta+pN,m}\,e^{\pm\frac{i\left(\alpha-\beta\right)\mathbf{k}\cdot\mathbf{R}_2}{N}\mp\frac{im\mathbf{k}\cdot\mathbf{R}_1}{N}\pm ip\mathbf{k}\cdot\mathbf{R}_2}.
\end{align}
The eigenvalue problem of this matrix reduces to $N$ coupled Harper equations \cite{Harper1955}.


\begin{figure}
\includegraphics[width=1\columnwidth]{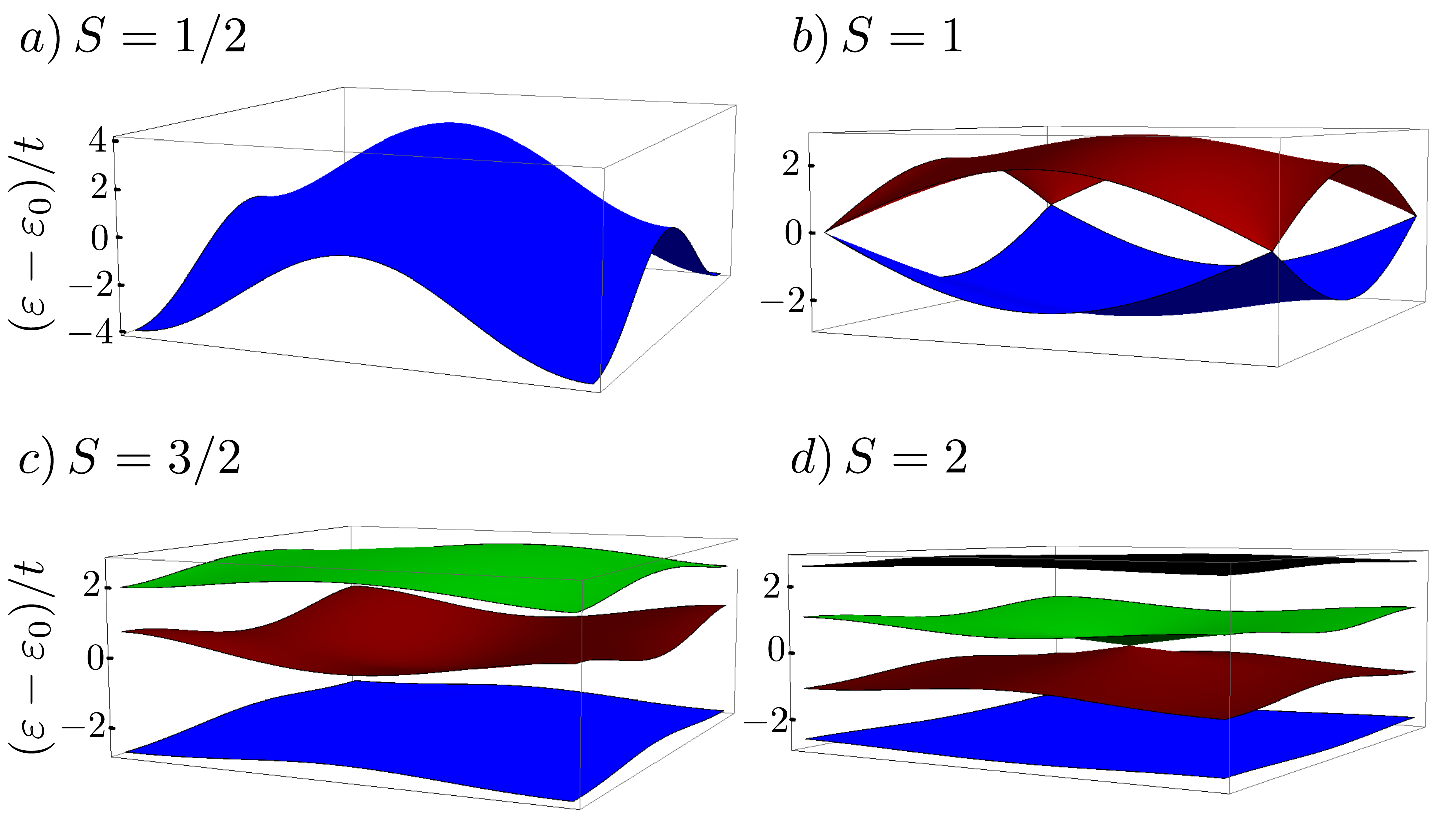}
\caption{Skyrmion bands ($|\mathcal{Q}|=1$) in the square lattice for different spin numbers. Only the first harmonics of the lattice potential (parametrized by $t$) are included.}
\label{fig:bands}
\end{figure}

\begin{figure*}[t!]
\includegraphics[width=1\textwidth]{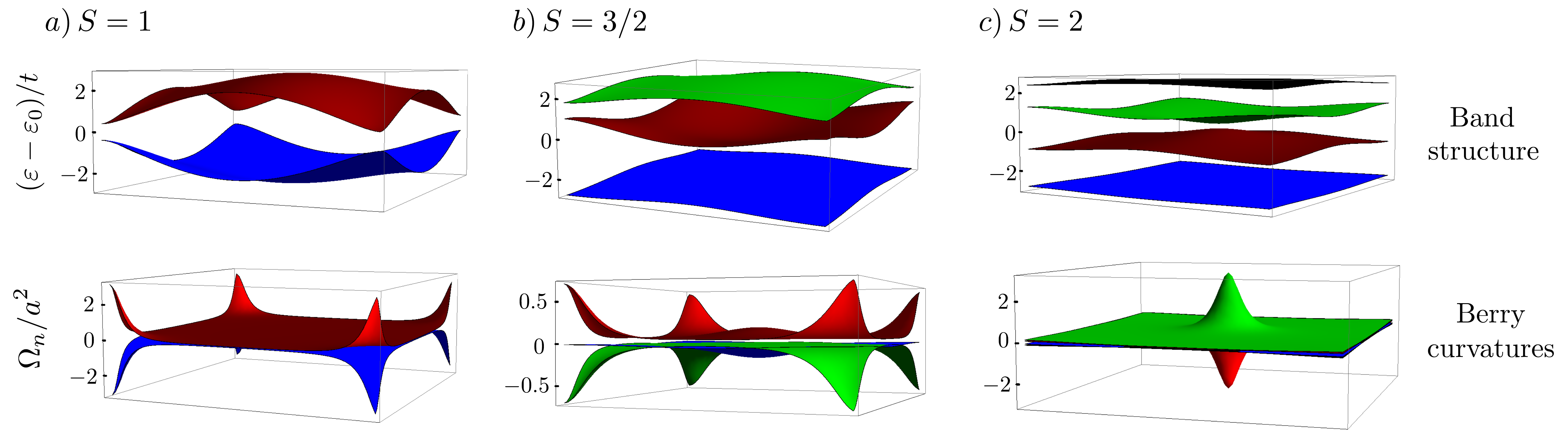}
\caption{Band structure and Berry curvatures of skyrmions ($\mathcal{Q}=-1$) in the square lattice for different spin numbers. We have included first ($t$) and second harmonics ($t'=-0.1t$) of the lattice potential.}
\label{fig:berry}
\end{figure*}

As an illustration, we compute the skyrmion bands in the square lattice, $\mathbf{R}_1=\left(a,0\right)$, $\mathbf{R}_2=\left(0,a\right)$. For the moment, we consider only the first harmonics of the lattice potential, \begin{align}
V_{\pm 1,0}=V_{0,\pm 1}\equiv t\approx\frac{a\,\varepsilon_0}{\pi \mathcal{R}}\,J_1\left(\frac{2\pi \mathcal{R}}{a}\right).
\end{align}
Figure~\ref{fig:bands} shows the skyrmion spectrum for different spin numbers. As $S$ increases, the bands become less dispersive. Interestingly, the middle bands with energies around $\varepsilon_0$ develop Dirac points when $S$ is integral. 
In the reduced zone scheme of Fig.~\ref{fig:bands}, the Dirac points appear at the inequivalent BZ corners $\left(\pm \pi/a,\pi/a\right)$ when $S$ is odd, and at the zone center when $S$ is even. The Dirac points always appear in pairs\footnote{Points connected by $\mathbf{G}_1=(2\pi/a,0)$ are inequivalent since the periodic boundary conditions are defined on BZ$^*$, not on BZ.} in the (extended) Brillouin zone of the von Neumann lattice, BZ$^*$, as prescribed by a doubling theorem \cite{Nielsen1981}. These $2S$ inequivalent Dirac points stem from a chiral symmetry\footnote{Notice, in particular, that the skyrmion spectrum for integer $S$ is symmetric with respect to the energy of the classical texture.} \cite{Wen1989} specific to the square lattice. This symmetry and subsequently the Dirac crossings are removed by the inclusion of second harmonics of the form\begin{align}
V_{\pm 1,\pm 1}\equiv t'\approx\frac{a\,\varepsilon_0}{\sqrt{2}\pi \mathcal{R}}\,J_1\left(\frac{2\pi\sqrt{2} \mathcal{R}}{a}\right),
\end{align}
as is shown in the calculations of Fig.~\ref{fig:bands}. 

\subsubsection{Band topology}

The $N-$component Bloch eigenstate of a given band acquires a geometrical phase when evolving adiabatically in reciprocal space along a path $\mathcal{C}$, $\gamma_n(\mathcal{C})=\int_{\mathcal{C}}d\mathbf{k}\cdot\bm{\mathcal{A}}_n(\mathbf{k})$, where $\bm{\mathcal{A}}_n(\mathbf{k})=i\langle \Psi_{\mathbf{k},n} | \nabla_{\mathbf{k}} |\Psi_{\mathbf{k},n}\rangle$ is the Berry connection \cite{Berry1984}. Here $|\Psi_{\mathbf{k},n}\rangle$ diagonalizes the matrix Hamiltonian $\mathcal{H}_{\mathbf{k}}$ with eigenvalue $\varepsilon_{\mathbf{k}}^{n}$, $n$ labels the band. The Berry connection depends on the gauge choice, already set in Eq.~\eqref{eq:gauge}, so it is convenient to introduce the Berry curvature
\begin{align}
\bm{\Omega}_n(\mathbf{k})=\nabla_{\mathbf{k}}\times\bm{\mathcal{A}}_n(\mathbf{k})=\Omega_n(\mathbf{k})\,\bm{\hat{z}}.
\end{align}
Figure~\ref{fig:berry} shows the Berry curvatures deduced from the previous calculations in the square lattice. We see how this function is strongly picked around avoided crossings, particularly for integer $S$ around the gapped Dirac points controlled by the strength of the second harmonics, $t'$.

The Berry curvature of the non-degenerate band $n$ integrated over a compact surface (in this case, the Brillouin zone BZ$^*$ of the von Neumann lattice) defines the integer-valued index \cite{Thouless1982,Avron1983}\begin{align}
\label{eq:Chern}
C_n=\int_{\text{BZ}^*}\frac{d\mathbf{k}}{2\pi}\,\Omega_n\left(\mathbf{k}\right)\in \mathbb{Z}.
\end{align} 
This is the Chern number, which provides us with a topological classification of the skyrmion bands viewed as smooth mappings between BZ$^*$ and the Hilbert space of Bloch states. The first example of nontrivial skyrmion bands is $S=1$. Each inequivalent (gapped) Dirac point contributes with $\pm 1/2$, where the sign depend on both $t'$ and the skyrmion charge. In the case of $S=3/2$, the highest and lowest energy bands have Chern number $C_{1,3}=\text{sign} (\mathcal{Q})$, and the remaining band $C_2=-2\,\text{sign} (\mathcal{Q})$. This sequence of Chern numbers can be generalized to higher $S$, as summarized in Tab.~\ref{tab:Chern}.

\subsubsection{Edge states}

A consequence of the nontrivial topology of the skyrmion bands is the appearance of counter-propagating chiral modes localized at the boundaries of a confined geometry, like in the quantum Hall effect \cite{Halperin1982}. A bulk property, the Chern number of the skyrmion bands, dictates the number and chirality of the edge modes. This is the so-called bulk-boundary correspondence \cite{Hatsugai1993}: $|\nu_i|$ modes localized at the boundary with a \textit{trivial} vacuum appear within the energy gap between band $i$ and $i+1$, where $\nu_i=\sum_{j\leq i}C_j$; the sign of $\nu_i$ determines the chirality, i.e., the direction of propagation. The fact that the Chern numbers are only integer-valued when the Berry curvature is integrated over the whole Brillouin zone of the von Neumann lattice anticipates the special nature of the wave functions near the edges. Like the operator-ordering issues and the anomaly in the angular-momentum spectrum, this is a consequence of the truncation of the Hilbert space.

\begin{figure*}[t!]
\includegraphics[width=1\textwidth]{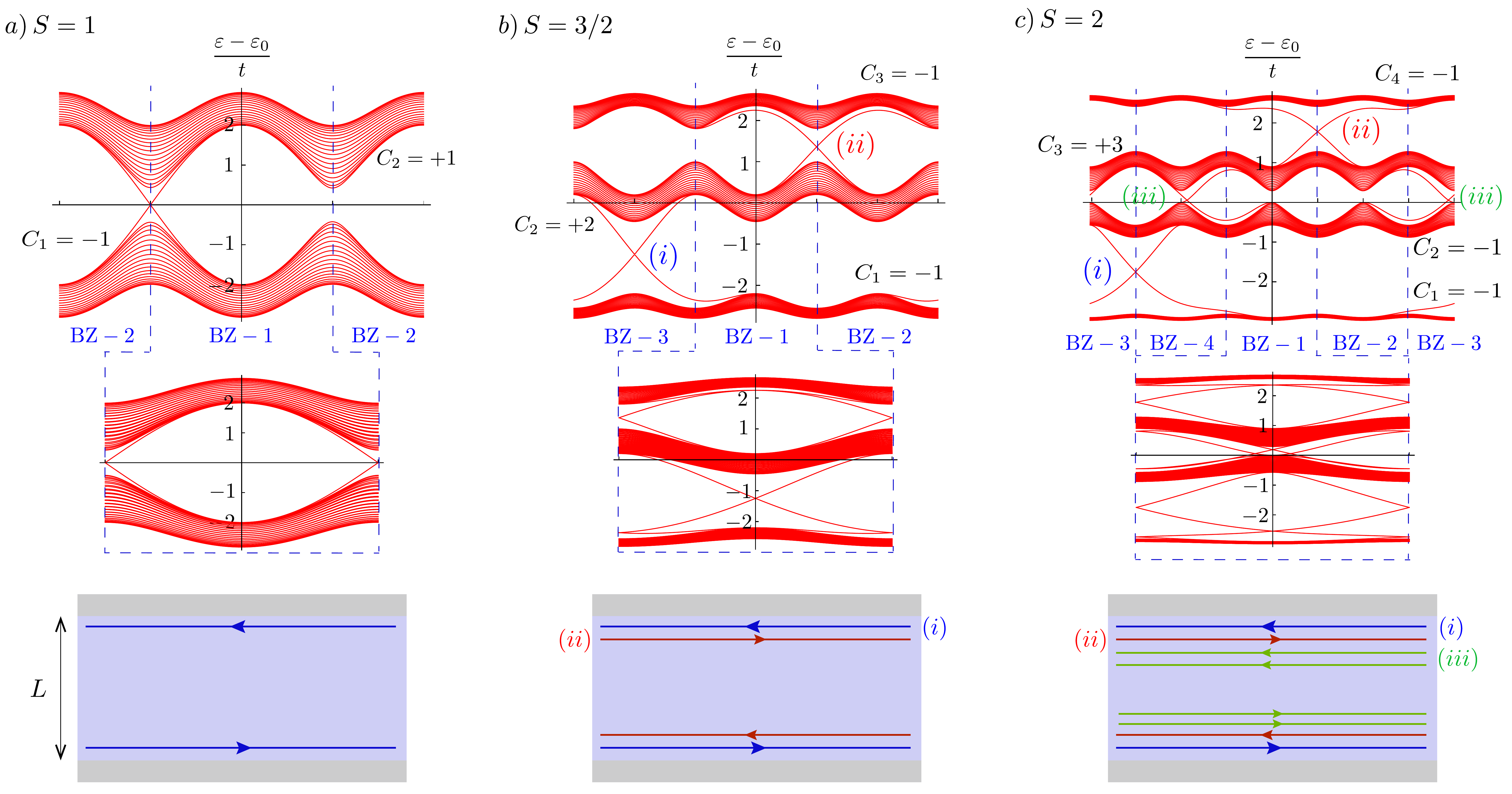}
\caption{Band structure of skyrmions ($\mathcal{Q}=-1$) in a rectangular strip with $L=20\,a$ in the hard-wall approximation for different spin quantum numbers. We have included second harmonics of the lattice potential ($t'=-0.1t$ in all cases). The bands are represented both in the extended zone (top panel) and reduced zone schemes. The bottom panels show the chirality of the edge modes within the projected bulk gaps.}
\label{fig:edge}
\end{figure*}

In order to illustrate the case, let us consider the problem in the presence of a confining potential along the direction defined by $\mathbf{G}_2$, the \textit{position} variable in our representation of wave functions. The total Hamiltonian reads $\hat{H}=\hat{H}_0+\hat{V}$, where $\hat{H}_0$ represents now the lattice Hamiltonian in Eq.~\eqref{eq:H2}. The matrix elements of $\hat{V}$ in the basis of eigenstates of the position operator projected along $\mathbf{G}_2$ are given by the confining potential $V(y)=\langle y|\hat{V}|y\rangle$. Since the translational symmetry along this direction is broken, it is convenient to introduce the set\begin{align}
\label{eq:new_set}
\left| k_x,q, \alpha \right\rangle = \int_{\mathbf{G}_2} \frac{dk_y}{2\pi}\, e^{\pm\frac{i\left(q-1\right)\mathbf{k}\cdot\mathbf{R}_2}{N}} e^{\pm i\alpha\mathbf{k}\cdot\mathbf{R}_2} e^{i\theta_{k_y}}\left|\mathbf{k},q\right\rangle,
\end{align}
where $k_y$ represents the projection of $\mathbf{k}$ along $\mathbf{G}_2$ and the integration is upon a period in reciprocal space. Here $\theta_{k_y}$ is a $k_y$-dependent phase, not specified yet. The upper (lower) sign corresponds to $\mathcal{Q}=+1$ ($\mathcal{Q}=-1$).

\begin{table}
\begin{tabular}{|c||c|c|}
\hline
Chern $\#$ & Middle bands $\left(\varepsilon_{\mathbf{k}}^n\sim \varepsilon_0\right)$ & Other bands  \\
\hline\hline
Half-integer $S$  & $C_{S+1/2}=\mp\left(2S-1\right)$ & $C_n=\pm 1$\\
\hline
Integer $S$ & $\begin{array}{c} C_{S+1}=\mp\left[S-1-\text{sign}\left(t'\right) S\right] \\ C_S=\mp\left[S-1+\text{sign}\left(t'\right) S\right] \end{array}$ & $C_n=\pm 1$\\ 
\hline
\end{tabular}
\caption{Chern numbers of the skyrmion bands in the square lattice up to second harmonics $t'$. The upper (lower) sign corresponds to positive (negative) skyrmion charge.}
\label{tab:Chern}
\end{table}

By construction, this set forms a complete, orthogonal basis; the orthogonality in the index $\alpha$ follows from the periodicity of the Bloch states along the direction of $\mathbf{G}_2$ in reciprocal space. This new quantum number is related to the position of the center of the wave function, $\psi_{k_x,q,\alpha}\left(y\right)\equiv\left\langle y| k_x,q,\alpha\right\rangle$, as inferred from the action of the translation operators on these states; specifically, we have\begin{align}
\left(\hat{T}_1\right)^N\left|k_x,q,\alpha\right\rangle=\left| k_x,q, \alpha+1 \right\rangle.
\end{align}
We can use this set to expand the Hamiltonian as in Eq.~\eqref{eq:H2}. For a given $k_x$, the lattice Hamiltonian $\hat{H}_0$ can be written as a block-matrix of the form\begin{align}
\label{eq:chain2}
\mathcal{H}_{k_x}=\left(\begin{array}{ccccc}
... & ... & ... &... &...\\
... & \mathcal{H}_{0} & \mathcal{V}_{-1} & \mathcal{V}_{-2} & ...\\
... & \mathcal{V}_{1} & \mathcal{H}_0 & \mathcal{V}_{-1} &...\\
... &\mathcal{V}_{2} &\mathcal{V}_{1} & \mathcal{H}_0 & ... \\
... & ... & ... &... &...
\end{array}\right),
\end{align}
each block acting on the subspace spanned by $\Psi_{k_x,\alpha}^{\dagger}= (| k_x,1,\alpha\rangle, | k_x,2,\alpha\rangle\,...\, | k_x,N,\alpha\rangle)$. The matrix elements of the diagonal and off-diagonal blocks are given, respectively, by\begin{subequations}\begin{align}
& \left(\mathcal{H}_0\right)_{\alpha\beta}=\sum_m V_{\alpha-\beta,\, m}\, e^{\mp\frac{imk_x\left|\mathbf{R}_1\right|}{N}\mp\frac{im2\pi}{N}\left(\frac{\alpha+\beta}{2}-1\right)},\\
& \left(\mathcal{V}_p\right)_{\alpha\beta}=\sum_{p,m} \left(-1\right)^{mp} V_{\alpha-\beta+pN,\, m}\, e^{\mp \frac{im}{N}\left[k_x\left|\mathbf{R}_1\right|+\pi\left(\alpha+\beta-2\right)\right]},
\end{align}\end{subequations}

We can also expand the confining potential in this basis. In principle, the periodicity of Bloch states along the $\mathbf{G}_2$-direction guarantees the existence of a gauge $\theta_{ky}$ for which the wave function is localized (exponentially decaying) around $y_{\alpha}\equiv\pm\alpha\, \mathbf{G}_2\cdot\mathbf{R}_2/|\mathbf{G}_2|$ \cite{Kohn1959}. With this choice, $\hat{V}$ can be taken as approximately diagonal in $\alpha$. This corresponds to a tight-binding approximation in which the confining potential is replaced by a set of on-site energies in the basis of wave functions localized along $y=y_{\alpha}$. However, a generic confining potential mixes states with different $q$ number, i.e., lying on different copies of the Brillouin zone BZ of the physical lattice. 
Nevertheless, the set introduced in Eq.~\eqref{eq:new_set} is still a good basis to diagonalize the problem in the hard-wall approximation,\begin{align}
\label{eq:hard-wall}
V\left(y\right)=\begin{cases}
\infty & \text{if}\,\,\left|y\right|\geq \frac{L}{2},\\
0 & \text{if}\,\,\left|y\right|< \frac{L}{2}.
\end{cases}
\end{align}
This problem corresponds to a finite strip of width $L$ in which the wave function is imposed to go to 0 at the boundaries. The spectrum is deduced then from the solution to a finite set of Harper equations corresponding to the eigenvalue problem of the matrix Hamiltonian introduced in Eq.~\eqref{eq:chain2}, now truncated to a finite number of blocks corresponding to the number of unit cells within $L$. The hybridizations along the BZ edges can be neglected since $V=0$ in the interior of the strip, but the calculation only makes sense if extended to the entire BZ$^*$. The bands can be represented either in the extended zone BZ$^*$ or in the reduced zone BZ after folding, see Fig.~\ref{fig:edge}. The projection of the ``bulk" bands is repeated at each copy of BZ in the extended zone scheme. The number and chirality of the edge modes follow the prescription of the bulk-boundary correspondence.

The fact that the wave functions of the edge states must be expressed in an overcomplete basis (the set introduced in Eq.~\eqref{eq:new_set} repeated over $N$ copies of the original Brillouin zone of the lattice) is a consequence of the Hilbert space truncation. It is not present when additional degrees of freedom are taken into account. 
Generically, a potential that breaks the translational symmetry and is sharp in the scale of the lattice spacing is going to hybridize the skyrmion translations with other modes of the magnetization dynamics \cite{Psaroudaki2016,Makhfudz2012}. In that situation, the skyrmion edge states cannot be identified with the rigid translation of the texture anymore. The truncated Hilbert space does not reflect these additional degrees of freedom, what is manifested as the reduction of the spectral weight of these modes.

\section{Semiclassical transport theory}
\label{sec:semiclassic}

We consider now classical perturbations to the skyrmion dynamics. Specifically, the driving forces acting on the skyrmions are assumed to vary in length scales much larger than $\ell_N$ and at frequencies much slower than $t/\hbar$. 
A reactive force $\bm{F}$ can incorporated via a potential energy of the form $V\left(\bm{r}\right)=-\bm{F}\cdot\bm{r}$, where $\bm{r}$ represents the expectation value of the skyrmion position in certain quantum state. Since $V\left(\bm{r}\right)$ is weak, the quantum numbers of Bloch states in a given band $n$, which are not longer conserved, will evolve adiabatically within an iso-energetic surface defined by the equation $\varepsilon_{\bm{k}}^n+V=\text{cte}$; during the evolution we have then $\dot{\bm{k}}\cdot\partial \varepsilon_{\bm{k}}^n/\partial\bm{k}-\bm{F}\cdot\dot{\bm{r}}=0$, where $\bm{k}$ represents the averaged quasimomentum of the skyrmion state expanded in the eigenstate basis $|\Psi_{\mathbf{k},n}\rangle$. If we identify $\dot{\bm{r}}$ with the group velocity, then this equation reduces to $\hbar \dot{\bm{k}}=\bm{F}$. This is also true if we add to the velocity an \textit{anomalous} term orthogonal to $\dot{\bm{k}}$ reflecting the multicomponent nature of the wave function. The expression of a Bloch state with crystal momentum $\mathbf{k}$ in terms of the instantaneous Bloch states with $\mathbf{k}=\mathbf{k}_0+t \bm{F}/\hbar$ reads \cite{Thouless1983,Chang1995a}
\begin{align}
\left|\Psi_{\mathbf{k}_0,n}\right\rangle\approx \left|\Psi_{\mathbf{k},n}\right\rangle-\sum_{n'\neq n}\frac{\left\langle \Psi_{\mathbf{k},n} \right|i\hbar\frac{d}{dt}\left| \Psi_{\mathbf{k},n} \right\rangle}{\varepsilon_{\mathbf{k}}^n- \varepsilon_{\mathbf{k}}^{n'}}\,\left|\Psi_{\mathbf{k},n}\right\rangle,
\end{align}
where we have omitted an irrelevant dynamical phase factor. Then we can identify \cite{Xiao2010}
\begin{align}
\label{eq:semi-classical_0}
\dot{\bm{r}}\equiv \left\langle \Psi_{\bm{k},n} \left|\frac{1}{\hbar}\frac{\partial \mathcal{H}_{\mathbf{k}}}{\partial\mathbf{k}}\right| \Psi_{\bm{k},n} \right\rangle\approx\frac{1}{\hbar}\frac{\partial \varepsilon_{\bm{k}}^n}{\partial\bm{k}}-\dot{\bm{k}}\times\bm{\Omega}_n\left(\bm{k}\right),
\end{align}
where $\bm{\Omega}_n\left(\mathbf{k}\right)=\Omega_n\left(\bm{k}\right)\bm{\hat{z}}$ is the Berry curvature of band $n$ introduced in Sec.~\ref{sec:lattice}. By plugging $\hbar \dot{\bm{k}}=\bm{F}$ into this last equation we finally arrive at the following semiclassical equation of motion:
\begin{align}
\label{eq:semi-classical}
\dot{\bm{r}}=\frac{1}{\hbar}\frac{\partial \varepsilon_{\bm{k}}^n}{\partial\bm{k}}-\frac{\bm{F}}{\hbar}\times\bm{\Omega}_n\left(\bm{k}\right).
\end{align}

In the limit of large spin numbers, the skyrmion bands become less dispersive, as it can be verified from the calculations in Sec.~\ref{sec:lattice}. As long as the skyrmion texture is comparable to the lattice spacing and the gaps remain open, the Chern number of the lowest/highest energy bands remain fixed to $\pm 1$, where the chirality is defined by the sign of the skyrmion charge. Thus, the Berry curvature of these bands tends to an uniform value $\Omega_n\approx \pm A_c/(2\pi N)=A_c/(4\pi S\mathcal{Q})$; in the classical limit $S\rightarrow\infty$ we have then
\begin{subequations}
\begin{align}
\frac{\partial \varepsilon_{\bm{k}}^n}{\partial\bm{k}}\longrightarrow0,\,\,\,\,\frac{\Omega_n\left(\bm{k}\right)}{\hbar}\longrightarrow \frac{1}{4\pi s\mathcal{Q}}.
\end{align}\end{subequations}
The semiclassical dynamics expressed in Eq.~\eqref{eq:semi-classical} reduces to the classical equation of motion~\eqref{eq:classical} if we identify $\bm{r}$ with the collective coordinate $\bm{R}$.

The collective dynamics of skyrmions can be analyzed now in the framework of a master equation describing the evolution of their distribution function, $f_n(\bm{r},\bm{k})$, from which hydrodynamic currents can be coarse-grained. The distribution function expresses the probability of finding a skyrmion in a given quantum state (labelled by $n$ and $\bm{k}$) within an infinitesimal volume defined around position $\bm{r}$. The drift in the semiclassical phase space caused by the driving force $\bm{F}$ is captured by Eq.~\eqref{eq:semi-classical}, where the Berry curvature incorporates the quantum mechanical corrections arising from the accumulation of Berry phases in reciprocal space, reminiscence of the nontrivial topology of the classical skyrmion texture. However, there is still a limitation in this equation, for it neglects the spatial extension of the skyrmion semiclassical state. In the presence of statistical forces, i.e., gradients of thermodynamical intensive variables (e.g., the temperature $T$), the systems is not longer homogenous and the reactive torques exerted by gapped modes of the magnetization dynamics \cite{Schutte2014} make the skyrmions to rotate \cite{Jonietz2010,Seki2012b,Mochizuki2015}. This \textit{self-rotation} of the skyrmions leads to a divergence-free contribution to the local currents, which must be subtracted in the definition of linear response coefficients.

\subsection{Dynamics of skyrmion wave packets}
\label{sec:wp}

The previous heuristic derivation can be formalized by means of algebraical methods \cite{Wilkinson1996} or as a WKB expansion \cite{Sokoloff,Nenciu}, although Eq.~\eqref{eq:semi-classical} is often regarded as the equation of motion for the center of a wave packet \cite{Chang1995b,Sundaram1999}, which is the perspective that we are going to adopt from now on. Let us consider then a wave packet $|\Psi_n\rangle$ expanded in the set of eigenvectors $|\Psi_{\mathbf{k},n}\rangle$ of the unperturbed lattice Hamiltonian, $\hat{H}_0$, according to certain distribution in crystal momenta. The distribution is centered at $\bm{k}$, defining the quasimomentum of the wave packet, and is narrow in the scale of the Brillouin zone by construction. Therefore, the wave packet is spread over several unit cells of the spin lattice; the expectation value of the position operator, $\bm{r}=\langle \Psi_n|\hat{\bm{R}}|\Psi_n\rangle$, is identified with the center of the wave packet in real space. These variables evolve according to a variational principle dictated by the semiclassical Lagrangian \cite{Chang1995b,Sundaram1999}\begin{align}
\label{eq:L_0}
\mathcal{L}\left(\bm{r},\dot{\bm{r}};\bm{k},\dot{\bm{k}}\right) & \equiv\left\langle \Psi_n\left|i\hbar\frac{d}{dt} -\hat{H}\right|\Psi_n\right\rangle
\\
& \approx\hbar\,\bm{\mathcal{A}}_n\left(\bm{k}\right)\cdot\dot{\bm{k}}+\hbar\,\bm{k}\cdot\dot{\bm{r}}-\varepsilon_{\bm{k}}^n+\bm{F}\cdot\bm{r},
\nonumber
\end{align}
where we have replaced the lattice Hamiltonian by the operator $\hat{H}=\hat{H}_0-\bm{F}\cdot\hat{\bm{R}}$ and neglected a total time derivative. The variational principle assumes that $\mathcal{L}$ can be evaluated at the wave-packet center $(\bm{r},\bm{k})$ for smooth enough perturbations and narrow enough distributions in reciprocal space. Combining the Euler-Lagrange equations derived from Eq.~\eqref{eq:L_0} leads to Eq.~\eqref{eq:semi-classical}.

This Lagrangian, however, does not take into account the structure of the wave packet in real space, which is required to properly describe the dynamics of its center when the system is not in thermal equilibrium. The inclusion of inhomogeneous forces arising from the exchange of linear and angular momenta with other degrees of freedom can be introduced as a fictitious gauge field, as we describe in Appendix~\ref{sec:currents}. There two important consequences for the skyrmion semiclassical dynamics: First of all, there is a Zeeman-like correction to the energy of the wave packet of the form $\bm{L}_n\left(\bm{k}\right)\cdot\bm{B}(\bm{r},t)$, where $\bm{B}(\bm{r},t)$ is a fictitious magnetic field associated with the angular-momentum density transferred from the magnetization dynamics to the skyrmion ensemble; $\bm{L}_n(\bm{k})$ is an intrinsic quantum property related to the Berry curvature, the orbital moment of the skyrmion band \cite{Chang1995b,Sundaram1999}, which is a measure of the wave-packet self-rotation. Secondly, the inclusion of these forces modify the density of skyrmion states in the semiclassical phase space. This change gives rise to an additional term in the transport currents, which can be associated with the circulation of skyrmions at the physical terminations of the system.

\subsection{Local vs. transport currents}
\label{sec:transport}

The main assumption of linear-response theory is that hydrodynamic deviations reach a state of local equilibrium in a short, microscopic time scale $\tau$. Nonequilibrium deviations of the skyrmion distribution can be described by a Boltzmann equation in the relaxation-time approximation,
\begin{align}
\label{eq:Boltzman}
\dot{f}_n=\left(\partial_t+\dot{\bm{r}}\cdot\bm{\nabla}_{\bm{r}}+\dot{\bm{k}}\cdot\bm{\nabla}_{\bm{k}}\right)f_n=-\frac{f_n-f_n^0}{\tau},
\end{align}
where $f_n^0$ is the local equilibrium (Bose-Einstein) distribution function, corresponding to the occupation number when the thermodynamic parameters equal their local values. The relaxation process is fast and thermodynamically irreversible in general, so $\tau$ is related to the energy dissipated by the skyrmion dynamics due to the coupling with other microscopic degrees of freedom. Here we are going to consider only dissipative processes conserving the number of skyrmions (i.e., the total topological charge) assumed that these are well-defined quasiparticles of the magnetic system.

The particle and energy currents are determined only by the local values of the number of skyrmions and the energy density, their gradients, and by the gradients of the thermodynamical parameters. A proper coarse-graining of the local currents corresponds to \cite{Xiao2006} \begin{widetext}\begin{subequations}
\begin{align}
 \bm{j}\left(\bm{r}\right) & =\sum_n\int_{\text{BZ}}\frac{d\bm{k}}{\left(2\pi\right)^2}\,f_n\left(\bm{r},\bm{k}\right)\dot{\bm{r}}
+\bm{\nabla}\times\sum_n \int_{\text{BZ}}\frac{d\bm{k}}{\left(2\pi\right)^2}\,f_n\left(\bm{r},\bm{k}\right) \bm{L}_{n}\left(\bm{k}\right),\\
 \bm{j}^{\varepsilon}\left(\bm{r}\right) & = \sum_n\int_{\text{BZ}}\frac{d\bm{k}}{\left(2\pi\right)^2}\,f_n\left(\bm{r},\bm{k}\right)\left(\varepsilon_{\mathbf{k}}^n-\mu\right)\dot{\bm{r}}
+ \bm{\nabla}\times\sum_n \int_{\text{BZ}}\frac{d\bm{k}}{\left(2\pi\right)^2}\,f_n\left(\bm{r},\bm{k}\right)\bm{L}_{n}\left(\bm{k}\right)\left(\varepsilon_{\mathbf{k}}^n-\mu\right),
\end{align}\end{subequations}
\end{widetext}
Here we are considering a gas of skyrmions on top of the collinear order, where the skyrmion density is controlled by the external magnetic field acting, effectively, as the chemical potential $\mu$. This is a well defined thermodynamical quantity as long as the number of skyrmions is conserved. 

In a quantum mechanical system when time-reversal symmetry is explicitly broken, the local currents are not necessarily 0 in equilibrium. There may be nonzero circulating currents related to the intrinsic angular momentum of the skyrmions. These currents can be written as $\bm{j}_{\bm{m}}(\bm{r})=\bm{\nabla}\times\bm{m}(\bm{r})$, $\bm{j}_{\bm{m}}^{\varepsilon}(\bm{r})=\bm{\nabla}\times\bm{m}_{\varepsilon}(\bm{r})$, where $\bm{m}$ ($\bm{m}_{\varepsilon}$) may be interpreted as the (thermal) orbital magnetization of the skyrmion ensemble in response to the fictitious field $\bm{B}$, i.e., the angular velocity of the skyrmion ensemble and the associated energy flow in response to a reactive torque exerted by gapped modes of the magnetization dynamics. In linear response, these velocities can be computed even in thermal equilibrium, when the thermodynamic parameters ($\mu$ and $T$) take constant values. The derivation can be found in Appendix~\ref{sec:currents}; the final result for $\bm{m}$ reads\begin{widetext}
\begin{align}
\label{eq:am_part}
 \bm{m}\left(\bm{r}\right)=& \sum_n\int_{\text{BZ}}\frac{d\bm{k}}{\left(2\pi\right)^2}\,\bm{L}_n\left(\bm{k}\right)f_{n}\left(\bm{r},\bm{k}\right)
-\sum_n\int_{\text{BZ}}\frac{d\bm{k}}{\left(2\pi\right)^2}\frac{\bm{\Omega}_n\left(\bm{k}\right)}{\hbar}
\int_{\varepsilon_{\bm{k}}^n-\mu}^{\infty}d\varepsilon\, f_{n}\left(\varepsilon\right).
\end{align}
The generalization for the energy flow, in the absence of external forces, reads just
\begin{align}
\label{eq:am_thermal}
 \bm{m}_{\varepsilon}\left(\bm{r}\right)= & \sum_n\int_{\text{BZ}}\frac{d\bm{k}}{\left(2\pi\right)^2}\,\bm{L}_n\left(\bm{k}\right)f_{n}\left(\bm{r},\bm{k}\right)\,\left(\varepsilon_{\bm{k}}^n-\mu\right)
 -\sum_n\int_{\text{BZ}}\frac{d\bm{k}}{\left(2\pi\right)^2}\frac{\bm{\Omega}_n\left(\bm{k}\right)}{\hbar}
\int_{\varepsilon_{\bm{k}}^n-\mu}^{\infty}d\varepsilon\, f_{n}\left(\varepsilon\right)\varepsilon.
\end{align}
The first terms in Eqs.~\eqref{eq:am_part}~and~\eqref{eq:am_thermal} are just the average over the orbital moments of the skyrmion bands. The second term, arising from the correction to the density of quantum states in the semiclassical phase space, is the contribution from the current circulating at the edges, as we see next.

Let us assume for a moment that the dynamics of the skyrmion wave packet close to the boundaries is well described by Eq.~\eqref{eq:semi-classical}, where the force $\bm{F}$ is induced by a confining potential, $\bm{F}=-\bm{\nabla}_{\perp} V$ (here $\perp$ denotes the gradient along the normal to the boundary), such that $V\rightarrow0$ ($V\rightarrow \infty$) in the interior (exterior) of the sample. The anomalous velocity induces a motion of the wave-packet along the edge, $\dot{\bm{r}}=\bm{\nabla}_{\perp}V\times\bm{\Omega}_n$. After integrating over all the states, we have for the skyrmion current in thermal equilibrium \cite{Matsumoto2011,Matsumoto2011a,Buttiker1988}
\begin{align}
\label{eq:edge_current}
\bm{I}_{\textrm{edge}}=\int_{\textrm{in}}^{\textrm{out}} dr_{\perp}\,\sum_n\int_{\text{BZ}}\frac{d\bm{k}}{\left(2\pi\right)^2}\,\frac{\bm{\nabla}_{\perp}V\left(\bm{r}\right)\times\bm{\Omega}_n\left(\bm{k}\right)}{\hbar}\,f_n\left(\varepsilon_{\bm{k}}^n+V\left(\bm{r}\right)-\mu\right)=\sum_n\int_{\text{BZ}}\frac{d\bm{k}}{\left(2\pi\right)^2}\frac{\bm{\hat{n}}\times\bm{\Omega}_n\left(\bm{k}\right)}{\hbar}
\int_{\varepsilon_{\bm{k}}^n-\mu}^{\infty}d\varepsilon\, f_{n}\left(\varepsilon\right),
\end{align}\end{widetext}
where $\bm{\hat{n}}$ is the normal (exterior) to the edge. The last result assumes that the confining potential varies slowly in the scale of the lattice. Note, however, that in the final result the confining potential does not appear explicitly, and actually $\bm{\hat{n}}\times\bm{I}_{\textrm{edge}}$ is just the second term in Eq.~\eqref{eq:am_part}. This current, although localized near the boundaries, is a bulk property of the system, and we expect this result to hold even in the case of a hard-wall potential, Eq.~\eqref{eq:hard-wall}.

In thermal equilibrium, the velocity fields are uniform and therefore the angular-momentum currents vanish in the interior of the sample. However, there will be currents at the edges of the sample, as we just saw. In the presence of thermal gradients, there will be bound currents in the interior of the sample as well due to the inhomogeneous angular velocity of the skyrmion ensemble, but these cannot be detected in a transport experiment \cite{Cooper1997}. Hence, the true transport currents, from which the linear-response coefficients are inferred, must be defined from the local currents after subtracting the divergence-free angular momentum components: $\bm{J}(\bm{r})=\bm{j}(\bm{r})-\bm{j}_{\bm{m}}(\bm{r})$ for the skyrmion current, and similarly for the energy current, $\bm{J}^{\varepsilon}(\bm{r})=\bm{j}^{\varepsilon}(\bm{r})-\bm{j}_{\bm{m}}^{\varepsilon}(\bm{r})$.

\subsection{Thermal Hall effect}
\label{sec:thermal}

Let us consider now a thermal gradient within the film of a helimagnet. The system hosts a gas of skyrmions whose collective motion is described by the semiclassical theory that we just exposed in the preceding subsections. After subtracting the divergence-free component, the energy current carried by the skyrmions reads
\begin{align}
\label{eq:JE_sky}
& \bm{J}^{\varepsilon}\left(\bm{r}\right)=\sum_n\int_{\text{BZ}}\frac{d\bm{k}}{\left(2\pi\right)^2}\,f_n\left(\bm{r},\bm{k}\right)\left(\varepsilon_{\mathbf{k}}^n-\mu\right)\dot{\bm{r}}
\\
& -\sum_n\int_{\text{BZ}}\frac{d\bm{k}}{\left(2\pi\right)^2}\frac{\Omega_n\left(\bm{k}\right)}{\hbar}\left(\bm{\hat{z}}\times\bm{\nabla}\right)
\int_{\varepsilon_{\bm{k}}^n-\mu}^{\infty}d\varepsilon\, f_{n}\left(\varepsilon\right)\varepsilon.\nonumber
\end{align}
Notice that the self-rotation contribution in Eq.~\eqref{eq:am_thermal} cancels the second term in the definition of the local current. The first term of this last equation defines the longitudinal thermal conductivity coming from deviations of the skyrmion distribution function with respect to local equilibrium. The second term gives rise to a transverse response containing a nondissipative component. This flow of energy is sustained by the circulation of skyrmions at the edges of the system; integrating the second line of Eq.~\eqref{eq:JE_sky} close to the boundary gives
\begin{align}
\bm{I}_{\textrm{edge}}^{\varepsilon}=\sum_n\int_{\text{BZ}}\frac{d\bm{k}}{\left(2\pi\right)^2}\frac{\bm{\hat{n}}\times\bm{\Omega}_n\left(\bm{k}\right)}{\hbar}
\int_{\varepsilon_{\bm{k}}^n-\mu}^{\infty}d\varepsilon\, f_{n}\left(\varepsilon\right)\varepsilon,
\end{align} 
which is the energy flow associated with the edge current in Eq.~\eqref{eq:edge_current}. This thermal current is robust against dissipative processes conserving the number of skyrmions.

\begin{figure}[t!]
\includegraphics[width=1\columnwidth]{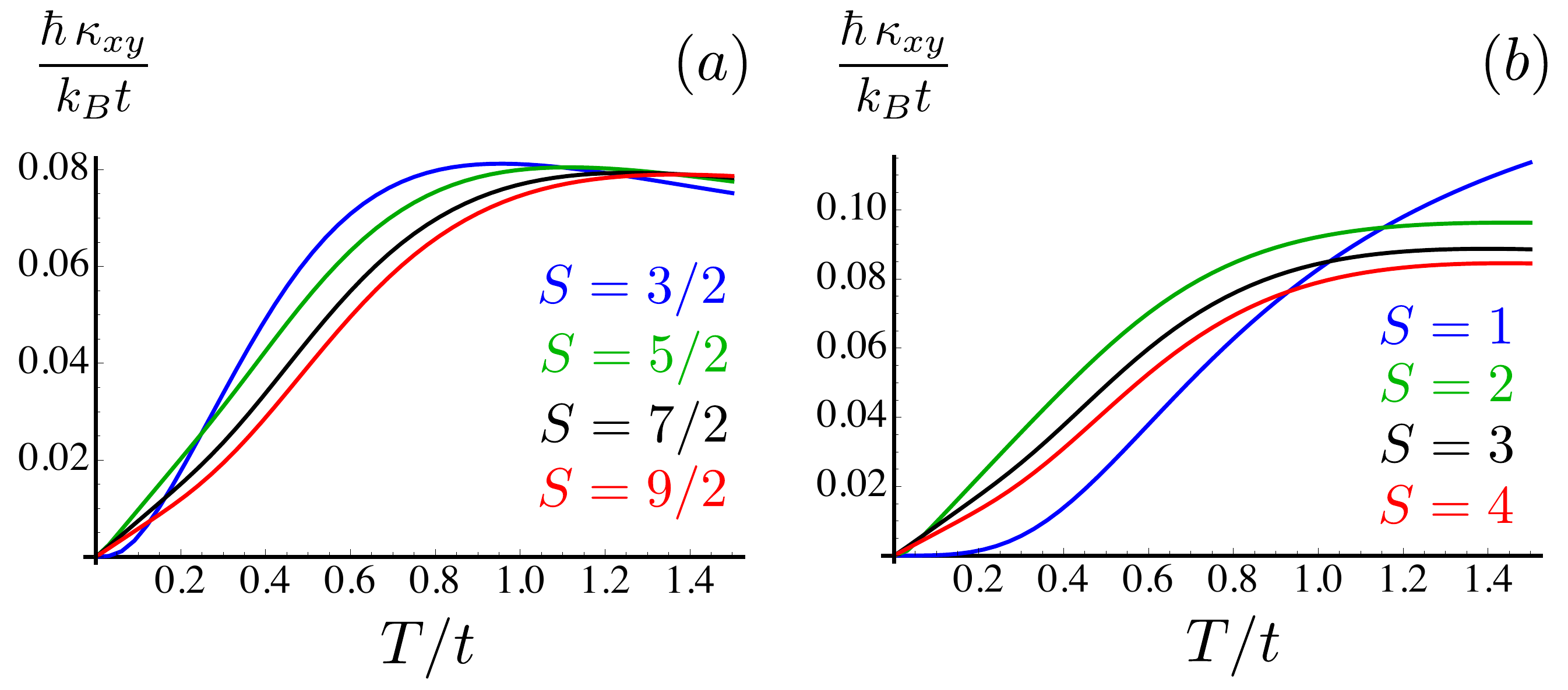}
\caption{Thermal Hall conductivity driven by skyrmions ($\mathcal{Q}=-1$) in the square lattice of half-integer ($a$) and integer ($b$) spins as a function of temperature. In all cases we include second harmonics $t'=-0.1t$ and the chemical potential is at the bottom of the lowest band.}
\label{fig:thermal1}
\end{figure}

\begin{figure*}[t!]
\begin{center}
\includegraphics[width=1\textwidth]{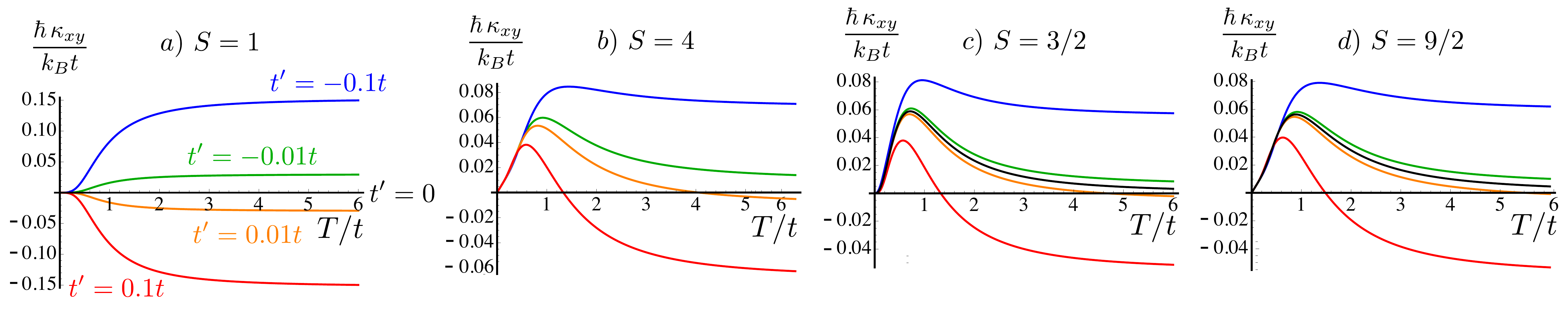}
\caption{Thermal Hall conductivity driven by skyrmions ($\mathcal{Q}=-1$) in the square lattice as a function of temperature for different spin numbers ($a-d$) and strengths of the second harmonics of the lattice, $t'$. As in Fig.~\ref{fig:thermal1}, the chemical potential is at the bottom of the lowest energy band.}
\label{fig:thermal2}
\end{center}
\end{figure*}

In thermal equilibrium, the circulation of skyrmions at opposite edges is compensated. The thermal gradient, however, generates an imbalance in the skyrmion populations at opposite edges, giving rise to a net energy current. In our linear-response calculation, we can substitute the distribution function for its value at local equilibrium, depending only on the local temperature. An expression for the thermal Hall conductivity can be inferred from the relation $J_x^{\varepsilon}=\kappa_{xy}T^2\partial_y\left(1/T\right)$, leading to\begin{align}
\kappa_{xy}=\sum_n\int_{\text{BZ}}\frac{d\bm{k}}{\left(2\pi\right)^2}\,\frac{\Omega_n\left(\bm{k}\right)}{\hbar\, T}\int_{\varepsilon_{\bm{k}}^n-\mu}^{\infty}d\varepsilon\,\frac{\partial f_n^0\left(\varepsilon\right)}{\partial\varepsilon}\,\varepsilon^2,
\end{align}
where $T$ represents the average temperature from now on. Since $f_n^0=(e^{\varepsilon/k_BT}-1)^{-1}$, we can rewrite this last expression as
\begin{align}
\label{eq:kappa}
\kappa_{xy}=-\frac{k_B^2T}{\hbar}\sum_n\int_{\text{BZ}}\frac{d\bm{k}}{\left(2\pi\right)^2}\,\, \eta\left[f_n^0\left(\varepsilon_{\bm{k}}^n-\mu\right)\right]\,\Omega_n\left(\bm{k}\right),
\end{align}
where $\eta[x]\equiv\int_0^x dy\,(\ln\frac{1+y}{y})^2$ is a function that grows monotonically with the skyrmion occupation number.

Figure~\ref{fig:thermal1} shows the thermal Hall conductivity in units of $k_B t/\hbar$ as a function of the average temperature of the system for different spin quantum numbers in the square lattice. We have included second harmonics of the lattice potential, $t'=-0.1t$. The chemical potential is at the bottom of the lowest energy band in all cases. At low temperatures, $T<t$, $\kappa_{xy}$ is extremely sensitive to details of the band structure and therefore to $S$. At higher temperatures these differences are more subtle, but still we can distinguish different behaviors for half-integer ($a$) and integer ($b$) spins. By expanding $\eta[x]\approx\pi^2/3 -1/x$ for large occupation numbers, and noticing that the sum of the Berry curvatures of the total number of bands is $0$, we obtain for the high temperature limit of Eq.~\eqref{eq:kappa}:\begin{align}
\label{eq:kappa_limit}
\lim_{T\rightarrow \infty}\kappa_{xy}=\frac{k_B}{\hbar}\sum_n\int \frac{d^2\bm{k}}{\left(2\pi\right)^2}\,\Omega_n\left(\bm{k}\right)\varepsilon_{n}\left(\bm{k}\right),
\end{align}
where the energy of the band is measured with respect to $\varepsilon_0$, the energy of the classical solution.\footnote{Note that this result is independent of $\mu$.} The Berry curvature is dominated by avoided crossings, lying approximately at opposite energies in the highest/lowest bands, so only the middle bands contribute effectively. Their effect is more prominent for integer spins; in that case, the sign of the Berry curvature is determined by the sign of $t'$ and not only the skyrmion charge.

The persistence of quantum-size effects depending on the parity of $2S$ even at temperatures of the order of the bandwidth, $T\sim t$, can be understood in terms of the counter-propagating edge modes discussed in Sec.~\ref{sec:lattice}. The behavior of $\kappa_{xy}$ at high temperatures is shown in Fig.~\ref{fig:thermal2} for different spin numbers. These curves reveal a change in the sign of the heat current depending on the sign of $t'$, as we just anticipated. The latter controls the inversion of the middle bands in the case of integer spin numbers and therefore the chirality of the associated edge modes. For example, the propagation of edge modes in the case of $S=1$ (Fig.~\ref{fig:edge}a) or the ones labeled by ($iii$) in the case of $S=2$ (Fig.~\ref{fig:edge}c) is inverted when $t'$ goes from negative to positive values. When $t'=0$, these edge modes disappear (specifically, the localization length diverges) and hence the thermal Hall conductivity is exactly $0$ for $S=1$; for larger spins, the energy carried by the remaining edge modes (for example, the modes denoted by ($i$) and ($ii$) in Fig.~\ref{fig:edge}c) flows in opposite directions and the thermal Hall conductivity goes to $0$ as $\sim 1/T$. The situation is similar to the case of half-integer spins, for which the edge modes always appear in pairs of opposite chirality. The imbalance introduced by $t'$ is only reflected in the propagation velocities, so the effect in the thermal conductivity is weaker. 

\section{Outlook}
\label{sec:outlook}

In this colloquium, we tried to extend the notions about the dynamics of skyrmion solitons beyond the micromagnetic regime, with the aim of providing a hydrodynamical description of their topological density with the account of quantum effects. We have seen that the nontrivial topology of the classical texture in real space is manifested in this semiclassical limit as the accumulation of geometrical phases in reciprocal space. The skyrmion dynamics can be monitored with electron microscopies, but the resolution of these techniques is usually limited to lengths of the order of 10 nm, where quantum fluctuations start to matter. A viable alternative, as we argued in the introduction, is thermal transport measurements. The thermal Hall effect driven by skyrmions is generically present, regardless of microscopic details like the specific lattice geometry. As we have shown, quantum interference introduces different behaviors for integer and half-integer spins, even at temperatures of the order of the bandwidth of the skyrmion spectrum. The extrapolation of these results to larger temperatures ($T\gg t$) is limited by the dissipation of the skyrmion currents due to thermal fluctuations, including the hybridization with gapped modes of the magnetization dynamics. In our treatment, these modes are separated by a large energy gap, but this truncation is questionable in the presence of sharp confining potentials. The inclusion of quantum dissipation is also beyond the semiclassical description provided here.

We have restricted the discussion to the single-particle properties of the soliton gas on top of the saturated state, in a region of the phase diagram with propensity for the formation of skyrmions in the dilute limit. Interactions between skyrmions, along with quantum fluctuations and lattice effects, might lead to the competition between different symmetry-broken phases not discussed here. This rich phenomenology can include Bose-Einstein condensation \cite{Huber2010}, localization and the Mott-superfluid transition \cite{Fisher1989}, supersolid phases \cite{Wessel2005,Melko2005}, or Wigner crystallization \cite{Wu2007}, which have been extensively explored in models of hard-core bosons in the lattice. In the case of systems with frustration, this intricate competition arises from the multiple minima in the single-particle spectrum, where a macroscopic number of bosons can condense \cite{Zhu2016,Zaletel2014,Sedrakyan2014}. These models have been realized in cold-atom systems so far \cite{Greiner2002}, but nanoscale skyrmions offer a solid-state alternative. Some works have already explored how quantum effects modify the transition to a skyrmion lattice, anticipated by a variant of Bose-Einstein condensation \cite{Takashima2016}. The associated critical behavior is strongly influenced by the position of the energy minimum in reciprocal space, and, therefore, it depends explicitly on the parity of $2S$, the number of bands. The effective spin numbers can be altered experimentally by modifying the thickness of the film. 

Although we have not considered itinerant magnets in our discussion, the inclusion of fermion degrees of freedom enriches the problem greatly. Skyrmions can trap charge and acquire nontrivial spin numbers \cite{Nomura2010,Yang2011}, just like in the quantum Hall effect \cite{Sondhi1993,YangSondhi}. The interplay of the spin, electrical, and topological charge of these excitations might lead to various thermoelectric effects along the lines discussed in this article, either subjected to thermal forces or by the application of electric fields \cite{Hurst2015}. Beyond the context of magnetism, other platforms for skyrmion physics include topological Mott insulators and heavy-fermion systems. In the former case, the condensation of $2e$ charged skyrmions drives the system into an exotic superconducting state \cite{Grover2008}. A similar paring mechanism has been discussed in the context of bilayer graphene \cite{Moon2012,Lu2012} or the heavy-fermion material URu$_2$Si$_2$ \cite{Hsu2013,Hsu2014}. Finally, the concept of \textit{Skyrme insulator} \cite{Onur2017} has been recently proposed to describe the anomalies in bulk transport and thermodynamic properties of the Kondo insulator SmB$_6$.



\begin{acknowledgments}

We thank valuable discussions with Se Kwon Kim, Oleg Tchernyshyov, and Ricardo Zarzuela over the years. Ana Asenjo-Garcia's assistance in Fig.~\ref{fig:qhe} is highly appreciated. This work has been supported by U.S. Department of Energy, Office of Basic Energy Sciences under Award No. DE-SC0012190.

\end{acknowledgments}

\appendix

\section{Symplectic reduction and spin coherent states}

\label{sec:A}

We derive here the Lagrangian in Eq.~\eqref{eq:L_main} (subjected to the constraint in Eq.~\ref{eq:a}) starting from the angular momentum algebra in Eq.~\eqref{eq:s_algebra}. For the moment, we avoid imposing constraints by applying instead a \textit{symplectic reduction} \cite{Faddeev1988}. We will recover later in Appendix~\ref{sec:B} the notion of \textit{second class} constraints to illustrate the equivalence between this approach and Dirac's method. An alternative derivation using Lagrange multipliers can be found in \onlinecite{Valenti1977}.

The difficulty in deriving the Lagrangian stems from the fact that the Landau-Lifshitz dynamics is an example of a noncanonical Hamiltonian system \cite{Morrison1998}: there are more elements in the Poisson-brackets algebra than dynamical variables. Mathematically, this observation can be traced to the existence of a Casimir invariant, $C\equiv |\bm{s}(\mathbf{r})|^2/2=s^2/2$, which is a constant of motion regardless of the symmetries of the Hamiltonian. This is a kinematic constraint built directly onto the phase space and is rooted on the underlying quantum nature of spin operators. The leaves $C=\text{constant}$ (i.e., $s=\text{constant}$) define \textit{symplectic manifolds}: spherical sections parametrized by two (field) variables for which we can define a Lagrangian dynamics, or more specifically, a \textit{symplectic 2-form} inherited from the Poisson brackets defined in the 3-dimensional phase space. Quantum mechanically, the symplectic manifolds correspond to different irreducible representations of spin rotations labelled by a (half-)integer $S$.

Let us consider then parametrizations of the spin density in terms of generic (not necessarily canonical) field variables $\bm{\zeta}=(\zeta_1,\zeta_2)$ everywhere in the sphere except at a single point $\bm{n}_0$, where the parametrization $\bm{n}(\bm{r})=\bm{n}[\bm{\zeta}(\mathbf{r})]$ fails. The Lagrangian can be written as \begin{align}
\label{eq:L}
L\left[\bm{\zeta}\right]=\int d^2\mathbf{r}\,\, \bm{A}\left[\bm{\zeta}\left(\mathbf{r}\right)\right]\cdot\dot{\bm{\zeta}}\left(\mathbf{r}\right)-H\left[\bm{\zeta}\right].
\end{align}
Here $\bm{A}\left[\bm{\zeta}\right]$ is called the \textit{canonical 1-form} (a generalized momentum), which is singular at $\bm{n}_0$. The symplectic 2-form is defined as\begin{align}
\label{eq:G}
G_{ij}\left(\mathbf{x},\mathbf{y}\right)\equiv\frac{\delta A_j\left[\bm{\zeta}\left(\mathbf{y}\right)\right]}{\delta\zeta_i\left(\mathbf{x}\right)}-\frac{\delta A_i\left[\bm{\zeta}\left(\mathbf{x}\right)\right]}{\delta\zeta_j\left(\mathbf{y}\right)},
\end{align}
in such a way that the Euler-Lagrange equations read\begin{align}
\int d\mathbf{y}\,G_{ij}\left(\mathbf{x},\mathbf{y}\right)\dot{\zeta}_j\left(\mathbf{y}\right)=\frac{\delta H}{\delta\zeta_i\left(\mathbf{x}\right)}.
\end{align}
These correspond to the Landau-Lifshitz equations as long as\begin{align}
\label{eq:demos}
\int d\mathbf{z}\,\, G_{ik}\left(\mathbf{x},\mathbf{z}\right)\left\{\zeta_k\left(\mathbf{z}\right),\zeta_j\left(\mathbf{y}\right)\right\}=\delta_{ij}\,\delta\left(\mathbf{x}-\mathbf{y}\right),
\end{align}where the Poisson brackets between generalized coordinates are given by\begin{widetext}\begin{align}
\label{eq:reduced_brackets}
\left\{\zeta_i\left(\mathbf{x}\right),\zeta_j\left(\mathbf{y}\right)\right\}=
\int d\mathbf{r}\int d\mathbf{r}'\,\left\{s_{\alpha}\left(\mathbf{r}\right),s_{\beta}\left(\mathbf{r}'\right)\right\}\,\frac{\delta\zeta_i\left(\mathbf{x}\right)}{\delta s_{\alpha}\left(\mathbf{r}\right)}\,\frac{\delta \zeta_j\left(\mathbf{y}\right)}{\delta s_{\beta}\left(\mathbf{r}'\right)}.
\end{align}
The canonical relation between $\phi$ and $s\cos\theta$ noted before follows directly from Eq.~\eqref{eq:reduced_brackets}. By combining Eqs.~\eqref{eq:demos}~and~\eqref{eq:reduced_brackets} we arrive at the following constraint for the symplectic form:
\begin{align}
\label{eq:Gbis}
G_{ij}\left(\mathbf{x},\mathbf{y}\right)=-s\int d\mathbf{z}\,\,\,\bm{n}\left(\mathbf{z}\right)\cdot\left(\frac{\delta \bm{n}\left(\mathbf{z}\right)}{\delta \zeta_i\left(\mathbf{x}\right)}\times\frac{\delta \bm{n}\left(\mathbf{z}\right)}{\delta \zeta_j\left(\mathbf{y}\right)}\right).
\end{align}
\end{widetext}
Equivalently, if we write the 1-canonical form as\begin{align}
\label{eq:A}
A_i\left[\bm{\zeta}\left(\mathbf{r}\right)\right]=s\int d\mathbf{r}'\,\bm{a}\left[\bm{n}\left(\mathbf{r}'\right)\right]\cdot\frac{\delta\bm{n}\left(\mathbf{r}'\right)}{\delta\zeta_i\left(\mathbf{r}\right)},
\end{align}
then from Eqs.~\eqref{eq:G}~and~\eqref{eq:Gbis} we arrive at Eq.~\eqref{eq:a}.

Each time the texture $\bm{n}\left(\mathbf{r}\right)$ wraps the unit sphere and therefore sweeps the singular point $\bm{n}_0$, the canonical 1-form $\bm{A}\left[\bm{\zeta}\right]$ jumps by an amount $\pm 4\pi s$. Therefore, $\bm{A}\left[\bm{\zeta}\right]$ is only a good momentum functional (i.e. generator of translations of the texture) if the number of jumps during the evolution is fixed by the boundary conditions \cite{Haldane1986}. This is in fact satisfied by skyrmion solitons on top of the uniformly ordered background: the number of jumps is given by the skyrmion charge $\mathcal{Q}$ defined in Eq.~\eqref{eq:Q} \cite{Papanicolaou1991}, leading to the central extension in the algebra of generators, Eq.~\eqref{eq:generators}.

Semiclassically, the field in Eq.~\eqref{eq:a} is just the Berry-phase connection associated with the spin-coherent representation of the macroscopic state of the magnet \cite{Kovner1989}. Specifically, the coarse-graining of the path integral gives $s\,\bm{a}[\bm{n}]\approx i\,\hbar\,\langle\Psi_{\textrm{sc}}|\bm{\nabla}_{\bm{n}} |\Psi_{\textrm{sc}}\rangle/A_c$, where the spin coherent state is defined as \cite{Klauder1979} \begin{align}
\label{eq:spin_coherent}
\left|\Psi_{\textrm{sc}}\right\rangle\equiv\bigotimes_i \left|\bm{n}_i\right\rangle,\,\text{such that}\,\,\bm{n}_i\cdot\hat{\bm{S}}_i\left|\bm{n}_i\right\rangle=S\left|\bm{n}_i\right\rangle.
\end{align}
In this equation, $\bm{n}_i\equiv \bm{n}\left(\mathbf{R}_i\right)$ is an abbreviation for the unit vector along the spin orientation at position $\mathbf{R}_i$. In this coarse-graining, the classical free energy functional corresponds to $H\approx\langle \Psi_{\textrm{sc}}| \hat{H} |\Psi_{\textrm{sc}}\rangle$, where $\hat{H}$ is a quantum Hamiltonian defined on the lattice. 
The state $|\bm{n}_i\rangle$ can be represented by an unitary rotation of $|S\rangle$, the state with maximum projection $S$ along a certain quantization axis (the $z$-axis in the most common parametrization),\begin{align}
\label{eq:coherent_skyrmion}
\left|\bm{n}_i\right\rangle=e^{-i\phi_i \hat{S}_i^z}\,e^{-i\theta_i \hat{S}_i^y}\,e^{-i\chi_i \hat{S}_i^z}\left|S\right\rangle.
\end{align}
The third Euler angle $\chi_i$ enters as a global phase. We have to impose $\chi_i=(2 n+1)\phi_i$ with $n$ an integer in order for $\left|\bm{n}_i\right\rangle$ to be single valued upon $\phi_i\rightarrow \phi_i+2\pi n$, corresponding to the same semiclassical state.\footnote{This has not to be confused with an active rotation of the quantum state, it is a gauge transformation.} The choice $\chi_i=\phi_i$ ($\chi_i=-\phi_i$) corresponds to the Dirac string lying along $\bm{n}_0=\bm{z}$ ($\bm{n}_0=-\bm{z}$). In the north-pole gauge, for example, we can rewrite Eq.~\eqref{eq:coherent_skyrmion} as \begin{align}
\label{eq:coherent_superpostion}
\left|\bm{n}_i\right\rangle=\left(\cos\frac{\theta_i}{2}\right)^{2S}\sum_{n=0}^{2S}e^{i\phi_i\left(n-2S\right)} \left(\tan\frac{\theta_i}{2}\right)^{n}\frac{\left(\hat{S}_i^{-}\right)^n}{n!}\left|S\right\rangle,
\end{align}
with $\hat{S}_i^{-}=\hat{S}_i^{x}-i\hat{S}_i^{y}$. From this last result we can interpret the skyrmion state in Eq~\eqref{eq:spin_coherent} as a coherent superposition of magnon bound states.

\section{Second-class constraints in the space of collective coordinates}

\label{sec:B}

The dynamics of the magnet constrained to the space of rigid skyrmion textures can be deduced from Eq.~\eqref{eq:Gbis} just by approximating the functional derivatives inside the integrals as derivatives with respect to the coordinates of the skyrmion center, or more formally,\begin{align}
\frac{\delta\bm{s}\left(\mathbf{r}\right)}{\delta{R_i}}\approx-\int d\mathbf{r}'\,\partial_i\bm{s}\left(\mathbf{r}'\right)\,\delta\left(\mathbf{r}-\mathbf{r}'\right).
\end{align}This last formula expresses that the dynamical variations of $\bm{s}$ are approximated by rigid translations. The symplectic form reduces to\begin{align}
\label{eq:G2}
G_{ij}=-4\pi s\mathcal{Q}\,\epsilon_{ij},
\end{align}
the \textit{gyromagnetic tensor}. The Poisson bracket is just the inverse matrix,\begin{align}
\label{eq:G3bis}
\left\{R_i,R_j\right\}=G_{ij}^{-1}=\frac{\epsilon_{ij}}{4\pi s\mathcal{Q}},
\end{align}
which corresponds to Eq.~\eqref{eq:G3}.

Alternatively, this equation can be derived from the theory of massive skyrmions following Dirac's method \cite{Dirac1964}. Dirac realized that the algebra of Poisson brackets associated with a singular Lagrangian introduces a division of constraints into two classes: the so-called \textit{first-class} and \textit{second-class} constraints. The former has zero Poisson brackets with the rest of constraints and can thus be implemented straightforwardly. Second class constraints, on the other hand, are those of the form $\phi_i=0$, $\phi_i$ being functions in phase space with $C_{ij}\equiv\{\phi_i,\phi_j\}\neq0$. The Dirac bracket (DB) between two arbitrary functions $f$ and $g$ is related to the Poisson bracket (PB) as
\begin{align}
\label{eq:Dirac_bra}
\left\{f,g\right\}_{\textrm{DB}}\equiv \left\{f,g\right\}_{\textrm{PB}}-C_{ij}^{-1}\left\{f,\phi_i\right\}_{\textrm{PB}}\left\{\phi_j,g\right\}_{\textrm{PB}}.
\end{align}
This bracket generates the dynamics in the symplectic manifold defined by the constraints.

The following discussion illustrates how the skyrmion mass acts a control parameter for the reduction of the number of degrees of freedom. Specifically, Eq.~\eqref{eq:G3} corresponds to the Dirac bracket in the theory of massive skyrmions imposing the constraint $M=0$, i.e., neglecting the effect of gapped modes. Notice first that the Hamiltonian deduced from Eq.~\eqref{eq:Leff} can be written in noncanonical variables as\begin{align}
H(\bm{R},\bm{P})=\frac{\bm{P}^2}{2M}+V(\bm{R}),
\end{align} where $\bm{P}\equiv M\dot{\bm{R}}$ is the kinetic momentum. These variables satisfy the relations\begin{subequations}
\begin{align}
& \left\{R_i,R_j\right\}_{\textrm{PB}}=0,\\
& \left\{P_i,P_j\right\}_{\textrm{PB}}=G_{ij},\\
& \left\{R_i,P_j\right\}_{\textrm{PB}}=\delta_{ij}.
\end{align}
\end{subequations}
Taking the limit of rigid textures ($M=0$) in the Hamiltonian formalism gives rise to obvious singularities. The way to recover this limit is by imposing the second class constraint $\bm{P}=0$. The Dirac bracket $\left\{R_i,R_j\right\}_{\textrm{DB}}$ equals Eq.~\eqref{eq:G3bis}, for we have $C_{ij}=G_{ij}$ in Eq.~\eqref{eq:Dirac_bra}.

\section{Wave functions in the truncated Hilbert space}

\label{sec:C}

To illustrate the ambiguity in the construction of wave functions and the associated operator-ordering issues implied by Eq.~\eqref{eq:G4}, let us consider the dynamics of skyrmions under the action of a central potential $V(|\bm{R}|)$, representing, for example, the interaction with a pinning center \cite{Lin2013}. The interaction removes the angular-momentum degeneracy, which remains a good quantum number. The problem is then diagonalized in the basis of eigenstates of the angular-momentum operator defined in Eq.~\eqref{eq:Lz}, $\hat{L}_z\left|n\right\rangle=\mp \hbar\left(n+1/2\right)\left|n\right\rangle$. In order to construct wave functions, we may consider coherent states satisfying\begin{subequations}\begin{align}
& \left\langle z\right|\hat{a}^{\dagger}=\left\langle z\right| z,\\
& \hat{a}\left|z\right\rangle=z^* \left| z\right\rangle,
\end{align}\end{subequations}
where $z\equiv(x\mp i y)/\sqrt{2}\ell_N$, $z^*\equiv(x\pm i y)/\sqrt{2}\ell_N$. 
A skyrmion in a coherent state $|z\rangle$ fluctuates around the position $\bm{R}=(x,y)$ over a characteristic length $\ell_N$. These are the states with the minimum quantum uncertainty, and therefore the closest to a single point in the constrained phase space, as depicted in Fig.~\ref{fig:qhe}(a). Due to their semiclassical properties, we can identify $|z\rangle$ as the representative in the truncated Hilbert space of the spin coherent state associated with the classical skyrmion texture centered around $\bm{R}$.\footnote{Hence, $|n\rangle$ can be interpreted as a state of $n$ bounded magnons in analogy with Eq.~\eqref{eq:coherent_superpostion}.} The set $\left\{|z\rangle\right\}$ is therefore overcomplete. The spectral decomposition of the identity in this basis is given by
\begin{align}
\label{eq:Iz}
\hat{1}=\int \frac{d z^* dz}{2\pi i}\,e^{-z^* z}\,\left|z\right\rangle\left\langle z\right|,
\end{align}
so the inner product of two states reads\begin{align}
\left\langle\Psi|\Phi\right\rangle=\int d\mu\left(z,z^*\right)\, \Psi^*\left(z\right)\Phi\left(z\right),
\end{align}
with the integration measure given by\begin{align}
\label{eq:measure}
d\mu\left(z,z^*\right)\equiv\frac{dz^*dz}{2\pi i}\,e^{-z^*z}=\frac{d^2\mathbf{r}}{2\pi\left(\ell_{N}\right)^2}\,e^{-\frac{r^2}{2\left(\ell_N\right)^2}}.
\end{align}

The wave functions in this representation are $\Psi\left(z\right)=\left\langle z|\Psi\right\rangle$. The truncated Hilbert space is then isomorphic to the space of analytic functions of $z$. In particular, for the eigenstates of $\hat{L}_z$ we have\begin{align}
\label{eq:wf}
\Psi_n\left(z\right)=\left\langle z|n\right\rangle=\frac{z^n}{\sqrt{n!}}.
\end{align} 
The operator $\hat{a}^{\dagger}$ acts on these functions by multiplication, $\hat{a}$ by differentiation:\begin{subequations}\begin{align}
\left\langle z\right|\hat{a}^{\dagger}\left|\Psi\right\rangle=z\Psi\left(z\right),\\
\left\langle z\right|\hat{a}\left|\Psi\right\rangle=\partial_z\Psi\left(z\right).
\end{align}\end{subequations}
In other words, the representation of the operator $\hat{a}$ in the space of analytic functions is $\partial_z$. Since $\left[\partial_z,z\right]=1\neq0$, we have to be careful with the ordering of operators when acting over the truncated wave functions. Specifically, the prescription is such that the operators containing both $\hat{a}$ and $\hat{a}^{\dagger}$ must be expressed in anti-normal order, i.e., $\hat{a}$'s must be moved to the left (derivatives $\partial_z$), $\hat{a}^{\dagger}$'s to the right (positions $z$) \cite{Dunne1993}. The reason for that can be understood from the spectral decomposition in Eq.~\eqref{eq:Iz}, from which we have $\left(\hat{a}\right)^l\left(\hat{a}^{\dagger}\right)^m=\int d\mu\,(z^*)^l(z)^m\,|z\rangle\langle z|$.

For comparison, let us consider for a moment the angular-momentum eigenfunctions in the effective theory for massive skyrmions, Eq.~\eqref{eq:Leff}. In the limit $M\rightarrow0$, the normalized wave functions in the lowest Landau level behave as \begin{align}
\Psi_n\left(\mathbf{r}\right)\equiv\left\langle \mathbf{r}|LL=0;\,n\right\rangle\longrightarrow\frac{2^{\frac{n+1}{2}}r^ne^{\mp in\theta}e^{-\frac{r^2}{4\ell_N^2}}}{\sqrt{\pi n!}\left(\ell_N\right)^{n+1}}.
\end{align}
These eigenfunctions are in one-to-one correspondence with the ones in Eq.~\eqref{eq:wf}, but obviously they are not the same. It cannot be otherwise, since the two spatial coordinates parametrizing the original wave functions correspond now to noncommuting operators. The normalization densities are, nonetheless, properly related:\begin{align}
d^2\mathbf{r}\,\left|\Psi_n\left(\mathbf{r}\right)\right|^2\longrightarrow d\mu\left(z\right)\,\left|\Psi_n\left(z\right)\right|^2.
\end{align}

The limiting relations between the wave functions as well as the operator-ordering prescription take different forms depending on the representation. When the microscopic lattice is taken into account, as in Sec.~\ref{sec:lattice}, it is more convenient to use a representation adapted to the symmetry of the lattice. In that case, it is also useful to introduce the concept of von Neumann lattice \cite{vonNeumann}. Notice that we can always choose a complete set of coherent states localized around positions in a lattice as long as the unit cell of this fictitious lattice is $A_c^*=2\pi(\ell_{N})^2$. More generically, a complete set can be generated, starting from an arbitrary normalized state $\left|\phi_0\right\rangle$, by applying mutually commuting displacement operators; in the representation of wave functions of Sec.~\ref{sec:translations}, we have\begin{align}
\label{eq:von_Neumann}
\left|\phi_{\alpha\beta}\right\rangle=\left(\hat{T}_1\right)^{\alpha\,N} \left(\hat{T}_2\right)^{\beta}\left|\phi_0\right\rangle,
\end{align}
with $\alpha,\beta$ integers. The elements defined through Eq.~\eqref{eq:von_Neumann} are only orthogonal if $|\phi_0\rangle$ is completely delocalized in phase space \cite{Bacry1975,Boon1978}. In other words, the elements of an orthogonal set defined in the von Neumann lattice do not posses the semiclassical properties of coherent states. This incompatibility of orthogonality and good localization properties implies that it is not possible to define Wannier functions for this problem \cite{Dana1983,Thouless1984}.

\section{Angular-momentum anomaly}
\label{sec:anomaly}

Let us consider now the angular momentum in the theory for massive skyrmions, Eq.~\eqref{eq:Leff}, where the inertial term introduces the effect of the high-energy degrees of freedom. The quantum operator associated with the generator of rotations in this theory, $L_z=|\bm{R}\wedge\bm{\Pi}|$ (where now $\bm{\Pi}\equiv\partial L_{\textrm{eff}}/\partial \dot{\bm{R}}=4\pi s \mathcal{Q}\,\bm{R}\times\mathbf{z}+M\dot{\bm{R}}$ is the canonical momentum) can be written in terms of two inter-commutating oscillator variables, $\hat{L}_z=\mp\hbar(\hat{a}^{\dagger}\hat{a}-\hat{b}^{\dagger}\hat{b})$, where the operators $\hat{a}$ and $\hat{b}$ are defined as for the Landau quantization of cyclotron orbits in an axial symmetric gauge. Here $\hat{b}$ connects different Landau levels (in particular, $\hat{b}$ annihilate states in the lowest Landau level), whereas $\hat{a}$ moves between different degenerate angular-momentum states within a given Landau level. 
In the lowest Landau level we have $\hat{L}_z\sim  \hat{a}^\dagger\hat{a}$, and the angular momentum takes integer values, in contrast to the result in Eq.~\eqref{eq:Lz}. This discrepancy reflects an important difference in the rotational symmetries for the two theories.
 
This last observation can be recast in terms of gauge invariance. 
The Lagrangian expressed in the collective coordinates does not possess an explicit gauge symmetry, but a canonical freedom remains in the choice of the kinetic term up to a total time derivative; the choice in Eq.~\eqref{eq:Leff} enforces the invariance under $z$-rotations. This global rotational symmetry can be promoted to a \textit{instantaneous} symmetry just by adding a gauge field that acts as a Lagrange multiplier for the generator of rotations, $L_z$. Specifically, we may consider an infinitesimal time-dependent rotation of the skyrmion texture with respect to the origin of collective coordinates,
\begin{align}
\delta R_i=\lambda\left(t\right)\, \epsilon_{ij}\,R_j.
\end{align}
The time dependence in the parametrization entails the covariance of time derivatives in order to preserve the symmetry of the Lagrangian,
\begin{align}
& D_t R_i\equiv\dot{R}_i-\mathcal{A}\,\epsilon_{ij}\,R_j,\,\, \text{with}\,\,\,\delta \mathcal{A}=\dot{\lambda}.
\end{align}
In this manner, we end up with a U(1) quantum mechanical ($0$ spatial dimensions) gauge theory, where a gauge transformation implements an adiabatic rotation of the skyrmion state. The gauge field defines then a map $\mathcal{A}(t):[0,T]\rightarrow \text{U(1)}\cong S_1$, classified according to the fundamental group $\pi_1\left(S_1\right)=\mathbb{Z}$, and therefore\begin{align}
\label{eq:homotopy}
\int_0^T dt \,\delta\mathcal{A}\left(t\right)=\int_0^T dt \,\dot{\lambda}\left(t\right)=2\pi n,
\end{align}
with $n$ an integer, the winding number of the transformation. Once a $d+1$-dimensional gauge theory is under discussion, with $d$ even, a Chern-Simons term is naturally present. This generalization reads \cite{Dunne1990}
\begin{align}
\label{eq:CS}
\mathcal{L}_{\textrm{CS}}=2\pi s\mathcal{Q}|D_t\bm{R}\wedge\bm{R}|+\frac{M}{2}|\dot{\bm{R}}|^2+\nu \mathcal{A},
\end{align}
supplemented by the subsidiary condition $\nu=L_z$. In the Weyl gauge ($\mathcal{A}=0$), we recover the Lagrangian in Eq.~\eqref{eq:Leff} [again, omitting $V(\mathbf{R})$].

The quantization of the Chern-Simons coupling and therefore the angular momentum follows from the gauge invariance of the generalized theory. Under a gauge transformation, the action changes by $\delta\mathcal{S}=\int_0^{T} dt\,\mathcal{L}_{\textrm{CS}}$. The angular-momentum quantum number $J$ is related to the quantum amplitude of this transformation, $e^{i\delta \mathcal{S}/\hbar}=e^{i 2\pi J}$ \cite{Wilczek1983}. Gauge invariance, on the other hand, implies that $\delta\mathcal{S}$ is quantized in units of $2\pi\hbar$, so $J$ is an integer, as we already suspected. When $M\neq 0$, only the additional topological term changes,\begin{align}
\delta\mathcal{S} = \nu \int_0^T dt \,\delta\mathcal{A}\left(t\right)=2\pi\nu n,
\end{align}
where the last result follows from Eq.~\eqref{eq:homotopy}. For arbitrary winding numbers, $\nu$ is forced to take integer multiples of $\hbar$, and so does the spectrum of $L_z$. 

Naively, one could apply the same argument for the theory with $M=0$, in clear discrepancy with Eq.~\eqref{eq:Lz}. However, one must be careful with this calculation. The functional integration is better performed in the holomorphic representation,\begin{align}
\mathcal{L}_{\textrm{CS}}\left[z,\bar{z}\right]=-\hbar\,\bar{z}\left(i\frac{d}{dt}+\mathcal{A}\right)z+\nu\mathcal{A},
\end{align}
The quantum action derived from the functional integration on these variables reads \begin{align}
\label{eq:Gamma}
\frac{i\mathcal{S}}{\hbar} & \equiv\ln\left[\int \mathcal{D}\left[z,\bar{z}\right]\,e^{\frac{i}{\hbar}\int dt\, \mathcal{L}_{\textrm{CS}}\left[z,\bar{z}\right]}\right]\\
\nonumber
& =\ln\left[\text{det}\left(i\frac{d}{dt}+\mathcal{A}\right)\right]+\frac{i \nu}{\hbar} \int dt \,\mathcal{A}\left(t\right).
\end{align}
Under a gauge transformation with winding number $n$, the determinant inside the logarithm changes by $\left(-1\right)^{n}$ \cite{anomalies_book,Dunne1990}. Thus, the total quantum action is not gauge invariant for generic winding numbers unless $\nu/\hbar$ takes half-integer values, in agreement with the angular-momentum spectrum deduced from Eq.~\eqref{eq:Lz}. Why does the additional inertial term in the Lagrangian regularize then the anomaly in the determinant of the theory? Notice that in the $M\neq0$ case we have to introduce not one but two oscillator operators to diagonalize the problem. In terms of these new canonical variables, the determinant of the theory is the product of two anomalously behaving determinants and the sign cancels out.

The behavior of the determinant in Eq.~\eqref{eq:Gamma} is the result of a quantum anomaly, i.e., the breaking of a classical conservation law by quantum fluctuations. A noteworthy example is the chiral anomaly \cite{Adler,BellJackiw}, realized in Weyl semimetals \cite{Nielsen1983}. In these materials, the conservation of the number of carriers around each of the degenerate Weyl nodes in reciprocal space (the charge associated with the emergent chiral symmetry of the low-energy field theoretical description) is broken in the presence of external electromagnetic fields. The transfer of particles from one node to another, the chiral current, results in a negative magnetoresistance signature, recently observed experimentally \cite{Li2014,Xiong2015}. The total number of particles is obviously conserved. Similarly in our case, there is a transfer of angular momentum to the internal degrees of freedom of the skyrmion \cite{Nayak1996}.\footnote{In quantum field theory, this mixed internal-rotational symmetry is usually a manifestation of the boundary conditions imposed on the fields \cite{Hasenfratz1976,Jackiw1976,Jackiw1976a}.} The symmetry under instantaneous (gauge) rotations implies that the total angular-momentum number $J=L_z\pm I$ is an integer. The fractional (half-integer) part of the angular momentum $L_z$ is then compensated by the internal isospin $I$ of the wave function in the truncated Hilbert space. 
The high-energy modes entering as an inertial term provides the $\textit{completion}$ of the skyrmion Lagrangian, regularizing the low-energy theory for skyrmion translations.

\section{Skyrmion self-rotation and angular-momentum currents}
\label{sec:currents}

The inclusion of inhomogeneous forces in the skyrmion dynamics arising from the exchange of linear and angular momenta with other degrees of freedom makes the global (i.e., time-independent) translations to be ill-defined, provided that now the generators vary in time: part of the skyrmion momentum is transferred to the background of magnonic excitations, and only the total linear momentum of both subsystems is conserved during the evolution. This is akin to the dynamics of condensates \cite{Volovik1986,Balatsky1987}, where the superfluid transfers linear and angular momenta to the normal component. 
In order to reflect these additional degrees of freedom, we can proceed as in Appendix~\ref{sec:anomaly} and promote the translations to local (i.e., instantaneous) operators by introducing a gauge field that acts as a Lagrange multiplier for the generators of the translation group. Specifically, the phase-space translations are promoted to gauge transformations of the form
\begin{align}
\hat{T}_{n.m}\longrightarrow \exp\left[\frac{i}{\hbar}\int_n^{m} d\mathbf{r}\cdot\bm{\mathcal{A}}\left(\mathbf{r},t\right)\right]\times \hat{T}_{n,m},
\end{align}
subjected to the condition $\partial_t\bm{\mathcal{A}}(\mathbf{r},t)=\bm{F}(\mathbf{r},t)$, so the crystal momentum in the lattice Hamiltonian is replaced by a gauge-invariant momentum operator according to the Peierls substitution. The gauge field $\bm{\mathcal{A}}(\mathbf{r},t)$ incorporates the contribution from reactive torques to the lowest order in gradients of the order parameter exerted by the magnetization dynamics. The wave packet in this gauge reads now
\begin{align}
\left| \Psi_n'\right\rangle=e^{-\frac{i}{\hbar}\bm{\mathcal{A}}\left(\bm{r},t\right)\cdot\hat{\bm{R}}}\,\left| \Psi_n\right\rangle,
\end{align}
where $\bm{\mathcal{A}}\left(\bm{r},t\right)$ is approximated by its value at the center of the wave packet in the adiabatic limit.

The correction to the kinetic term in Eq.~\eqref{eq:L_0} amounts to the substitution $\hbar\bm{k}\rightarrow\hbar\bm{k}-\bm{\mathcal{A}}$. The correction to the semiclassical Hamiltonian comes from
\begin{align}
\left\langle \Psi_n'\left| \hat{H} \right|\Psi_n'\right\rangle=\left\langle \Psi_n\left| \hat{H}' \right|\Psi_n\right\rangle,
\end{align}
where $\hat{H}'$ corresponds to the original Hamiltonian in minimal coupling. The energy of the wave packet to the lowest order in a gradient expansion is just $E_n\left(\bm{k}\right)=\varepsilon_{\bm{k}}^n+\bm{L}_n\left(\bm{k}\right)\cdot\bm{B}$, where $\bm{B}\left(\bm{r},t\right)=\bm{\nabla}_{\bm{r}}\times\bm{\mathcal{A}}\left(\bm{r},t\right)(\equiv B\bm{\hat{z}})$ is the total angular-momentum density transferred from the incoherent background, and $\bm{L}_n(\bm{k})=L_n(\bm{k})\bm{\hat{z}}$ is the orbital moment of the skyrmion band \cite{Chang1995b,Sundaram1999},
\begin{align}
\bm{L}_n\left(\mathbf{k}\right)=\frac{i}{\hbar}\left\langle\bm{\nabla}_{\mathbf{k}}\Psi_{\mathbf{k},n}\left|\times\left(\mathcal{H}_{\mathbf{k}}-\varepsilon^n_{\mathbf{k}}\right)\right| \bm{\nabla}_{\mathbf{k}}\Psi_{\mathbf{k},n} \right\rangle.
\end{align}

The inclusion of these additional degrees of freedom has an impact on the symplectic structure of the semiclassical phase space. This is better captured in the Hamiltonian formulation of the semiclassical dynamics. The equations of motion can be regarded as the Hamiltonian dynamics $\Theta_{\alpha\beta}\,\dot{\zeta}_{\beta}=\partial \mathcal{H}_n/\partial\zeta_{\alpha}$ expressed in non-canonical variables $\bm{\zeta}=\left(\bm{r},\hbar\bm{k}\right)$, where the Hamiltonian is $\mathcal{H}_n(\bm{r},\bm{k})=E_n(\bm{k})-\bm{F}\cdot\bm{r}$, and the symplectic 2-form is given by
\begin{align}
\Theta=\left(\begin{array}{cc}
-\epsilon_{ij}\, B & -\delta_{ij}\\
\delta_{ij} & \epsilon_{ij}\,\Omega_n/\hbar
\end{array}\right).
\end{align}
The volume form in the semiclassical phase space reads then
\begin{align}
dV=\sqrt{\text{det}\,\Theta}\,\frac{d\bm{\zeta}}{\hbar^2}=\left(1+\frac{\bm{B}\cdot\bm{\Omega}}{\hbar}\right)d\bm{r}d\bm{k}.
\end{align}
This is the element of volume conserved during the Hamiltonian evolution according to Liouville's theorem \cite{Xiao2005,Duval2006}.

Let us consider now the distribution function of skyrmions in band $n$, which is in general a function of the phase-space coordinates and time, $f_n\left(\bm{r},\bm{k},t\right)$. This function remains constant during the Hamiltonian evolution by definition, $\dot{f}_n=\partial_t f_n+\{f_n,\mathcal{H}_n\}=0$ (assuming no dissipation for the moment, i.e., $\tau\rightarrow\infty$ in Eq.~\ref{eq:Boltzman}). The (conserved) probability of finding a skyrmion in a wave packet centered at $(\bm{r}$, $\bm{k})$ in a band $n$ is, therefore, $f_n\,dV=f_n(1+\bm{B}\cdot\bm{\Omega}/\hbar)d\bm{r}d\bm{k}$. We can define the probability density as
\begin{align}
\label{eq:varrho}
\varrho_n\left(\bm{r},\bm{k},t\right)\equiv f_n\left(\bm{r},\bm{k},t\right)\left(1+\frac{\bm{B}\cdot\bm{\Omega}}{\hbar}\right),
\end{align}
which actually satisfies a continuity equation, $\partial_{t}\varrho_n+\partial_{\alpha}(\dot{\zeta}_{\alpha}\varrho_n)=0$, as inferred directly from the equations of motion. Thus, the probability density $\varrho_n\left(\bm{r},\bm{k},t\right)$ (and not the distribution function) provides then a measure of the number of quantum states in an element of volume $d\bm{r}d\bm{k}$.

This correction to the density of quantum states in the semiclassical phase space affects the evaluation of ensemble-averaged observables. For example, $\bm{m}(\bm{r})$, which is just the response of the system to the transfer of angular momentum $\bm{B}(\bm{r})$, can be related to the variation of the free energy of the skyrmion ensemble as
\begin{align}
\label{eq:L_def}
\bm{m}\left(\bm{r}\right)=\frac{1}{A}\frac{\partial F}{\partial \bm{B}\left(\bm{r}\right)}.
\end{align}
Here the derivative is taken at constant temperature and chemical potential. In order to compute $F$ in the present semiclassical approximation, notice first that the skyrmion density is defined as
\begin{align}
-\frac{1}{A}\frac{\partial F}{\partial\mu\left(\bm{r}\right)}\equiv\varrho\left(\bm{r}\right)=\sum_n\int_{\text{BZ}}\frac{d\bm{k}}{\left(2\pi\right)^2}\,\varrho_{n}\left(\bm{r},\bm{k}\right).
\end{align}
In the second equality we have coarse-grained the skyrmion density from the probability density $\varrho_{n}\left(\bm{r},\bm{k}\right)$. When the chemical potential is uniform, this last equation can be formally integrated as
\begin{align}
F/A=-\int_{-\infty}^{\mu}d\mu'\,\sum_n\int_{\text{BZ}}\frac{d\bm{k}}{\left(2\pi\right)^2}\,\varrho_{n}\left(\bm{r},\bm{k}\right).
\end{align}
Then, using Eq.~\eqref{eq:varrho} and the fact that $f_n$ in thermal equilibrium is a function of $E_n(\bm{r},\bm{k})-\mu$ , we can change the integration variable, $\mu'\longrightarrow \varepsilon=E_n(\bm{r},\bm{k})-\mu'$, and express the free-energy density as
\begin{align}
 F/A=& -\sum_n\int_{\text{BZ}}\frac{d\bm{k}}{\left(2\pi\right)^2}\left(1+\frac{\bm{B}\left(\bm{r}\right)\cdot\bm{\Omega}_n\left(\bm{k}\right)}{\hbar}\right)
\nonumber\\
& \times \int_{E_n\left(\bm{r},\bm{k}\right)-\mu}^{\infty}d\varepsilon\, f_{n}\left(\varepsilon\right).
\end{align}
Then, plugging this result into Eq.~\eqref{eq:L_def} gives the final result in Eq.~\eqref{eq:am_part}. Note that the term associated with the skyrmion edge currents comes precisely from the correction to the probability density. 

\bibliography{biblio/miscellaneous,biblio/QHE,biblio/quantum_sky,biblio/skyrmions,biblio/transport,biblio/classical_mag,biblio/quantum_mag,biblio/quantum_mechanics,biblio/lattice,biblio/outlook}

\begin{thebibliography}{174}%
\makeatletter
\providecommand \@ifxundefined [1]{%
 \@ifx{#1\undefined}
}%
\providecommand \@ifnum [1]{%
 \ifnum #1\expandafter \@firstoftwo
 \else \expandafter \@secondoftwo
 \fi
}%
\providecommand \@ifx [1]{%
 \ifx #1\expandafter \@firstoftwo
 \else \expandafter \@secondoftwo
 \fi
}%
\providecommand \natexlab [1]{#1}%
\providecommand \enquote  [1]{``#1''}%
\providecommand \bibnamefont  [1]{#1}%
\providecommand \bibfnamefont [1]{#1}%
\providecommand \citenamefont [1]{#1}%
\providecommand \href@noop [0]{\@secondoftwo}%
\providecommand \href [0]{\begingroup \@sanitize@url \@href}%
\providecommand \@href[1]{\@@startlink{#1}\@@href}%
\providecommand \@@href[1]{\endgroup#1\@@endlink}%
\providecommand \@sanitize@url [0]{\catcode `\\12\catcode `\$12\catcode
  `\&12\catcode `\#12\catcode `\^12\catcode `\_12\catcode `\%12\relax}%
\providecommand \@@startlink[1]{}%
\providecommand \@@endlink[0]{}%
\providecommand \url  [0]{\begingroup\@sanitize@url \@url }%
\providecommand \@url [1]{\endgroup\@href {#1}{\urlprefix }}%
\providecommand \urlprefix  [0]{URL }%
\providecommand \Eprint [0]{\href }%
\providecommand \doibase [0]{http://dx.doi.org/}%
\providecommand \selectlanguage [0]{\@gobble}%
\providecommand \bibinfo  [0]{\@secondoftwo}%
\providecommand \bibfield  [0]{\@secondoftwo}%
\providecommand \translation [1]{[#1]}%
\providecommand \BibitemOpen [0]{}%
\providecommand \bibitemStop [0]{}%
\providecommand \bibitemNoStop [0]{.\EOS\space}%
\providecommand \EOS [0]{\spacefactor3000\relax}%
\providecommand \BibitemShut  [1]{\csname bibitem#1\endcsname}%
\let\auto@bib@innerbib\@empty
\bibitem [{\citenamefont {Abrikosov}(1957)}]{Abrikosov1957}%
  \BibitemOpen
  \bibfield  {author} {\bibinfo {author} {\bibnamefont {Abrikosov},
  \bibfnamefont {A.~A.}}} (\bibinfo {year} {1957}),\ \href@noop {} {\bibfield
  {journal} {\bibinfo  {journal} {Journal of Physics and Chemistry of Solids}\
  }\textbf {\bibinfo {volume} {2}}~(\bibinfo {number} {3}),\ \bibinfo {pages}
  {199}}\BibitemShut {NoStop}%
\bibitem [{\citenamefont {Adams}\ \emph {et~al.}(2012)\citenamefont {Adams},
  \citenamefont {Chacon}, \citenamefont {Wagner}, \citenamefont {Bauer},
  \citenamefont {Brandl}, \citenamefont {Pedersen}, \citenamefont {Berger},
  \citenamefont {Lemmens},\ and\ \citenamefont {Pfleiderer}}]{Adams2012}%
  \BibitemOpen
  \bibfield  {author} {\bibinfo {author} {\bibnamefont {Adams}, \bibfnamefont
  {T.}}, \bibinfo {author} {\bibfnamefont {A.}~\bibnamefont {Chacon}}, \bibinfo
  {author} {\bibfnamefont {M.}~\bibnamefont {Wagner}}, \bibinfo {author}
  {\bibfnamefont {A.}~\bibnamefont {Bauer}}, \bibinfo {author} {\bibfnamefont
  {G.}~\bibnamefont {Brandl}}, \bibinfo {author} {\bibfnamefont
  {B.}~\bibnamefont {Pedersen}}, \bibinfo {author} {\bibfnamefont
  {H.}~\bibnamefont {Berger}}, \bibinfo {author} {\bibfnamefont
  {P.}~\bibnamefont {Lemmens}}, \ and\ \bibinfo {author} {\bibfnamefont
  {C.}~\bibnamefont {Pfleiderer}}} (\bibinfo {year} {2012}),\ \href {\doibase
  10.1103/PhysRevLett.108.237204} {\bibfield  {journal} {\bibinfo  {journal}
  {Physical Review Letters}\ }\textbf {\bibinfo {volume} {108}}~(\bibinfo
  {number} {23}),\ \bibinfo {pages} {237204}}\BibitemShut {NoStop}%
\bibitem [{\citenamefont {Adler}(1969)}]{Adler}%
  \BibitemOpen
  \bibfield  {author} {\bibinfo {author} {\bibnamefont {Adler}, \bibfnamefont
  {S.~L.}}} (\bibinfo {year} {1969}),\ \href@noop {} {\bibfield  {journal}
  {\bibinfo  {journal} {Physical Review}\ }\textbf {\bibinfo {volume} {177}},\
  \bibinfo {pages} {2426}}\BibitemShut {NoStop}%
\bibitem [{\citenamefont {Al~Khawaja}\ and\ \citenamefont
  {Stoof}(2001)}]{Usama2001}%
  \BibitemOpen
  \bibfield  {author} {\bibinfo {author} {\bibnamefont {Al~Khawaja},
  \bibfnamefont {U.}}, \ and\ \bibinfo {author} {\bibfnamefont
  {H.}~\bibnamefont {Stoof}}} (\bibinfo {year} {2001}),\ \href@noop {}
  {\bibfield  {journal} {\bibinfo  {journal} {Nature}\ }\textbf {\bibinfo
  {volume} {411}},\ \bibinfo {pages} {918}}\BibitemShut {NoStop}%
\bibitem [{\citenamefont {Anderson}\ and\ \citenamefont
  {Toulouse}(1977)}]{Anderson1977}%
  \BibitemOpen
  \bibfield  {author} {\bibinfo {author} {\bibnamefont {Anderson},
  \bibfnamefont {P.~W.}}, \ and\ \bibinfo {author} {\bibfnamefont
  {G.}~\bibnamefont {Toulouse}}} (\bibinfo {year} {1977}),\ \href@noop {}
  {\bibfield  {journal} {\bibinfo  {journal} {Physical Review Letters}\
  }\textbf {\bibinfo {volume} {38}}~(\bibinfo {number} {9}),\ \bibinfo {pages}
  {508}}\BibitemShut {NoStop}%
\bibitem [{\citenamefont {Avron}\ \emph {et~al.}(1983)\citenamefont {Avron},
  \citenamefont {Seiler},\ and\ \citenamefont {Simon}}]{Avron1983}%
  \BibitemOpen
  \bibfield  {author} {\bibinfo {author} {\bibnamefont {Avron}, \bibfnamefont
  {J.~E.}}, \bibinfo {author} {\bibfnamefont {R.}~\bibnamefont {Seiler}}, \
  and\ \bibinfo {author} {\bibfnamefont {B.}~\bibnamefont {Simon}}} (\bibinfo
  {year} {1983}),\ \href@noop {} {\bibfield  {journal} {\bibinfo  {journal}
  {Physical Review Letters}\ }\textbf {\bibinfo {volume} {51}},\ \bibinfo
  {pages} {51}}\BibitemShut {NoStop}%
\bibitem [{\citenamefont {Bacry}\ \emph {et~al.}(1975)\citenamefont {Bacry},
  \citenamefont {Grossmann},\ and\ \citenamefont {Zak}}]{Bacry1975}%
  \BibitemOpen
  \bibfield  {author} {\bibinfo {author} {\bibnamefont {Bacry}, \bibfnamefont
  {H.}}, \bibinfo {author} {\bibfnamefont {A.}~\bibnamefont {Grossmann}}, \
  and\ \bibinfo {author} {\bibfnamefont {J.}~\bibnamefont {Zak}}} (\bibinfo
  {year} {1975}),\ \href {\doibase 10.1103/PhysRevB.12.1118} {\bibfield
  {journal} {\bibinfo  {journal} {Physical Review B}\ }\textbf {\bibinfo
  {volume} {12}}~(\bibinfo {number} {4}),\ \bibinfo {pages} {1118}}\BibitemShut
  {NoStop}%
\bibitem [{\citenamefont {Balatsky}(1987)}]{Balatsky1987}%
  \BibitemOpen
  \bibfield  {author} {\bibinfo {author} {\bibnamefont {Balatsky},
  \bibfnamefont {A.~V.}}} (\bibinfo {year} {1987}),\ \href@noop {} {\bibfield
  {journal} {\bibinfo  {journal} {Physics Letters A}\ }\textbf {\bibinfo
  {volume} {123}}~(\bibinfo {number} {1}),\ \bibinfo {pages} {27}}\BibitemShut
  {NoStop}%
\bibitem [{\citenamefont {Bayot}\ \emph {et~al.}(1996)\citenamefont {Bayot},
  \citenamefont {Grivei}, \citenamefont {Melinte}, \citenamefont {Santos},\
  and\ \citenamefont {Shayegan}}]{Bayot1996}%
  \BibitemOpen
  \bibfield  {author} {\bibinfo {author} {\bibnamefont {Bayot}, \bibfnamefont
  {V.}}, \bibinfo {author} {\bibfnamefont {E.}~\bibnamefont {Grivei}}, \bibinfo
  {author} {\bibfnamefont {S.}~\bibnamefont {Melinte}}, \bibinfo {author}
  {\bibfnamefont {M.~B.}\ \bibnamefont {Santos}}, \ and\ \bibinfo {author}
  {\bibfnamefont {M.}~\bibnamefont {Shayegan}}} (\bibinfo {year} {1996}),\
  \href@noop {} {\bibfield  {journal} {\bibinfo  {journal} {Physical Review
  Letters}\ }\textbf {\bibinfo {volume} {76}}~(\bibinfo {number} {24}),\
  \bibinfo {pages} {4584}}\BibitemShut {NoStop}%
\bibitem [{\citenamefont {Belavin}\ and\ \citenamefont
  {Polyakov}(1975)}]{A.Belavin1975}%
  \BibitemOpen
  \bibfield  {author} {\bibinfo {author} {\bibnamefont {Belavin}, \bibfnamefont
  {A.~A.}}, \ and\ \bibinfo {author} {\bibfnamefont {A.~M.}\ \bibnamefont
  {Polyakov}}} (\bibinfo {year} {1975}),\ \href@noop {} {\bibfield  {journal}
  {\bibinfo  {journal} {JEPT Lett.}\ }\textbf {\bibinfo {volume}
  {22}}~(\bibinfo {number} {10}),\ \bibinfo {pages} {245}}\BibitemShut
  {NoStop}%
\bibitem [{\citenamefont {Bell}\ and\ \citenamefont
  {Jackiw}(1969)}]{BellJackiw}%
  \BibitemOpen
  \bibfield  {author} {\bibinfo {author} {\bibnamefont {Bell}, \bibfnamefont
  {J.~S.}}, \ and\ \bibinfo {author} {\bibfnamefont {R.}~\bibnamefont
  {Jackiw}}} (\bibinfo {year} {1969}),\ \href@noop {} {\bibfield  {journal}
  {\bibinfo  {journal} {Il Nuovo Cimento A}\ }\textbf {\bibinfo {volume}
  {60}},\ \bibinfo {pages} {47}}\BibitemShut {NoStop}%
\bibitem [{\citenamefont {Berg}\ and\ \citenamefont
  {L{\"{u}}scher}(1981)}]{Berg1981}%
  \BibitemOpen
  \bibfield  {author} {\bibinfo {author} {\bibnamefont {Berg}, \bibfnamefont
  {B.}}, \ and\ \bibinfo {author} {\bibfnamefont {M.}~\bibnamefont
  {L{\"{u}}scher}}} (\bibinfo {year} {1981}),\ \href {\doibase
  10.1016/0550-3213(81)90568-X} {\bibfield  {journal} {\bibinfo  {journal}
  {Nuclear Physics B}\ }\textbf {\bibinfo {volume} {190}}~(\bibinfo {number}
  {2}),\ \bibinfo {pages} {412}}\BibitemShut {NoStop}%
\bibitem [{\citenamefont {Berry}(1984)}]{Berry1984}%
  \BibitemOpen
  \bibfield  {author} {\bibinfo {author} {\bibnamefont {Berry}, \bibfnamefont
  {M.~V.}}} (\bibinfo {year} {1984}),\ \href@noop {} {\bibfield  {journal}
  {\bibinfo  {journal} {Proc. R. Soc. Lond. A}\ }\textbf {\bibinfo {volume}
  {392}},\ \bibinfo {pages} {45}}\BibitemShut {NoStop}%
\bibitem [{\citenamefont {Binz}\ \emph {et~al.}(2006)\citenamefont {Binz},
  \citenamefont {Vishwanath},\ and\ \citenamefont {Aji}}]{Binz2006a}%
  \BibitemOpen
  \bibfield  {author} {\bibinfo {author} {\bibnamefont {Binz}, \bibfnamefont
  {B.}}, \bibinfo {author} {\bibfnamefont {A.}~\bibnamefont {Vishwanath}}, \
  and\ \bibinfo {author} {\bibfnamefont {V.}~\bibnamefont {Aji}}} (\bibinfo
  {year} {2006}),\ \href {\doibase 10.1103/PhysRevLett.96.207202} {\bibfield
  {journal} {\bibinfo  {journal} {Physical Review Letters}\ }\textbf {\bibinfo
  {volume} {96}}~(\bibinfo {number} {20}),\ \bibinfo {pages}
  {207202}}\BibitemShut {NoStop}%
\bibitem [{\citenamefont {Bogdanov}\ and\ \citenamefont
  {Hubert}(1994)}]{Bogdanov1994}%
  \BibitemOpen
  \bibfield  {author} {\bibinfo {author} {\bibnamefont {Bogdanov},
  \bibfnamefont {A.}}, \ and\ \bibinfo {author} {\bibfnamefont
  {A.}~\bibnamefont {Hubert}}} (\bibinfo {year} {1994}),\ \href {\doibase
  10.1016/0304-8853(94)90046-9} {\bibfield  {journal} {\bibinfo  {journal}
  {Journal of Magnetism and Magnetic Materials}\ }\textbf {\bibinfo {volume}
  {138}}~(\bibinfo {number} {3}),\ \bibinfo {pages} {255}}\BibitemShut
  {NoStop}%
\bibitem [{\citenamefont {Bogdanov}\ and\ \citenamefont
  {Yablonskii}(1989)}]{Bogdanov1989}%
  \BibitemOpen
  \bibfield  {author} {\bibinfo {author} {\bibnamefont {Bogdanov},
  \bibfnamefont {A.}}, \ and\ \bibinfo {author} {\bibfnamefont
  {D.}~\bibnamefont {Yablonskii}}} (\bibinfo {year} {1989}),\ \href {\doibase
  10.1016/0304-8853(94)90046-9} {\bibfield  {journal} {\bibinfo  {journal} {Zh.
  Eksp. Teor. Fiz}\ }\textbf {\bibinfo {volume} {95}}~(\bibinfo {number} {1}),\
  \bibinfo {pages} {178}}\BibitemShut {NoStop}%
\bibitem [{\citenamefont {Boon}\ and\ \citenamefont {Zak}(1978)}]{Boon1978}%
  \BibitemOpen
  \bibfield  {author} {\bibinfo {author} {\bibnamefont {Boon}, \bibfnamefont
  {M.}}, \ and\ \bibinfo {author} {\bibfnamefont {J.}~\bibnamefont {Zak}}}
  (\bibinfo {year} {1978}),\ \href {\doibase 10.1103/PhysRevB.18.6744}
  {\bibfield  {journal} {\bibinfo  {journal} {Physical Review B}\ }\textbf
  {\bibinfo {volume} {18}}~(\bibinfo {number} {12}),\ \bibinfo {pages}
  {6744}},\ \Eprint {http://arxiv.org/abs/arXiv:1011.1669v3}
  {arXiv:arXiv:1011.1669v3} \BibitemShut {NoStop}%
\bibitem [{\citenamefont {Braun}\ and\ \citenamefont {Loss}(1996)}]{Braun1996}%
  \BibitemOpen
  \bibfield  {author} {\bibinfo {author} {\bibnamefont {Braun}, \bibfnamefont
  {H.-B.}}, \ and\ \bibinfo {author} {\bibfnamefont {D.}~\bibnamefont {Loss}}}
  (\bibinfo {year} {1996}),\ \href {\doibase 10.1103/PhysRevB.53.3237}
  {\bibfield  {journal} {\bibinfo  {journal} {Physical Review B}\ }\textbf
  {\bibinfo {volume} {53}}~(\bibinfo {number} {6}),\ \bibinfo {pages}
  {3237}}\BibitemShut {NoStop}%
\bibitem [{\citenamefont {Brazovskii}(1975)}]{Brazovskii1975}%
  \BibitemOpen
  \bibfield  {author} {\bibinfo {author} {\bibnamefont {Brazovskii},
  \bibfnamefont {S.~A.}}} (\bibinfo {year} {1975}),\ \href {\doibase
  10.1002/mc.20806} {\bibfield  {journal} {\bibinfo  {journal} {Sov. Phys.
  JEPT}\ }\textbf {\bibinfo {volume} {41}},\ \bibinfo {pages} {85}}\BibitemShut
  {NoStop}%
\bibitem [{\citenamefont {Brazovskii}\ \emph {et~al.}(1987)\citenamefont
  {Brazovskii}, \citenamefont {Dzyaloshinskii},\ and\ \citenamefont
  {Muratov}}]{Brazovskii1987}%
  \BibitemOpen
  \bibfield  {author} {\bibinfo {author} {\bibnamefont {Brazovskii},
  \bibfnamefont {S.~A.}}, \bibinfo {author} {\bibfnamefont {I.~E.}\
  \bibnamefont {Dzyaloshinskii}}, \ and\ \bibinfo {author} {\bibfnamefont
  {A.~R.}\ \bibnamefont {Muratov}}} (\bibinfo {year} {1987}),\ \href@noop {}
  {\bibfield  {journal} {\bibinfo  {journal} {Phys. JEPT}\ }\textbf {\bibinfo
  {volume} {66}},\ \bibinfo {pages} {625}}\BibitemShut {NoStop}%
\bibitem [{\citenamefont {Brey}\ \emph {et~al.}(1995)\citenamefont {Brey},
  \citenamefont {Fertig}, \citenamefont {C{\^{o}}t{\'{e}}},\ and\ \citenamefont
  {MacDonald}}]{Brey1995}%
  \BibitemOpen
  \bibfield  {author} {\bibinfo {author} {\bibnamefont {Brey}, \bibfnamefont
  {L.}}, \bibinfo {author} {\bibfnamefont {H.~A.}\ \bibnamefont {Fertig}},
  \bibinfo {author} {\bibfnamefont {R.}~\bibnamefont {C{\^{o}}t{\'{e}}}}, \
  and\ \bibinfo {author} {\bibfnamefont {A.~H.}\ \bibnamefont {MacDonald}}}
  (\bibinfo {year} {1995}),\ \href {\doibase 10.1103/PhysRevLett.75.2562}
  {\bibfield  {journal} {\bibinfo  {journal} {Physical Review Letters}\
  }\textbf {\bibinfo {volume} {75}}~(\bibinfo {number} {13}),\ \bibinfo {pages}
  {2562}}\BibitemShut {NoStop}%
\bibitem [{\citenamefont {Brown}(1963)}]{Brown1963}%
  \BibitemOpen
  \bibfield  {author} {\bibinfo {author} {\bibnamefont {Brown}, \bibfnamefont
  {W.~F.}}} (\bibinfo {year} {1963}),\ \href@noop {} {\emph {\bibinfo {title}
  {Micromagnetics}}}\ (\bibinfo  {publisher} {Wiley},\ \bibinfo {address} {New
  York})\BibitemShut {NoStop}%
\bibitem [{\citenamefont {Buhrandt}\ and\ \citenamefont
  {Fritz}(2013)}]{Buhrandt2013}%
  \BibitemOpen
  \bibfield  {author} {\bibinfo {author} {\bibnamefont {Buhrandt},
  \bibfnamefont {S.}}, \ and\ \bibinfo {author} {\bibfnamefont
  {L.}~\bibnamefont {Fritz}}} (\bibinfo {year} {2013}),\ \href {\doibase
  10.1103/PhysRevB.88.195137} {\bibfield  {journal} {\bibinfo  {journal}
  {Physical Review B}\ }\textbf {\bibinfo {volume} {88}}~(\bibinfo {number}
  {19}),\ \bibinfo {pages} {195137}}\BibitemShut {NoStop}%
\bibitem [{\citenamefont {B{\"{u}}ttiker}(1988)}]{Buttiker1988}%
  \BibitemOpen
  \bibfield  {author} {\bibinfo {author} {\bibnamefont {B{\"{u}}ttiker},
  \bibfnamefont {M.}}} (\bibinfo {year} {1988}),\ \href {\doibase
  10.1103/PhysRevB.38.9375} {\bibfield  {journal} {\bibinfo  {journal}
  {Physical Review B}\ }\textbf {\bibinfo {volume} {38}}~(\bibinfo {number}
  {14}),\ \bibinfo {pages} {9375}}\BibitemShut {NoStop}%
\bibitem [{\citenamefont {B{\"{u}}ttner}\ \emph {et~al.}(2015)\citenamefont
  {B{\"{u}}ttner}, \citenamefont {Moutafis}, \citenamefont {Schneider},
  \citenamefont {Kr{\"{u}}ger}, \citenamefont {G{\"{u}}nther}, \citenamefont
  {Geilhufe}, \citenamefont {Kor~Schmising}, \citenamefont {Mohanty},
  \citenamefont {Pfau}, \citenamefont {Schaffert}, \citenamefont {Bisig},
  \citenamefont {Foerster}, \citenamefont {Schulz}, \citenamefont {Vaz},
  \citenamefont {Franken}, \citenamefont {Swagten}, \citenamefont
  {Kl{\"{a}}ui},\ and\ \citenamefont {Eisebitt}}]{Buttner2015}%
  \BibitemOpen
  \bibfield  {author} {\bibinfo {author} {\bibnamefont {B{\"{u}}ttner},
  \bibfnamefont {F.}}, \bibinfo {author} {\bibfnamefont {C.}~\bibnamefont
  {Moutafis}}, \bibinfo {author} {\bibfnamefont {M.}~\bibnamefont {Schneider}},
  \bibinfo {author} {\bibfnamefont {B.}~\bibnamefont {Kr{\"{u}}ger}}, \bibinfo
  {author} {\bibfnamefont {C.~M.}\ \bibnamefont {G{\"{u}}nther}}, \bibinfo
  {author} {\bibfnamefont {J.}~\bibnamefont {Geilhufe}}, \bibinfo {author}
  {\bibfnamefont {C.~v.}\ \bibnamefont {Kor~Schmising}}, \bibinfo {author}
  {\bibfnamefont {J.}~\bibnamefont {Mohanty}}, \bibinfo {author} {\bibfnamefont
  {B.}~\bibnamefont {Pfau}}, \bibinfo {author} {\bibfnamefont {S.}~\bibnamefont
  {Schaffert}}, \bibinfo {author} {\bibfnamefont {A.}~\bibnamefont {Bisig}},
  \bibinfo {author} {\bibfnamefont {M.}~\bibnamefont {Foerster}}, \bibinfo
  {author} {\bibfnamefont {T.}~\bibnamefont {Schulz}}, \bibinfo {author}
  {\bibfnamefont {C.~A.~F.}\ \bibnamefont {Vaz}}, \bibinfo {author}
  {\bibfnamefont {J.~H.}\ \bibnamefont {Franken}}, \bibinfo {author}
  {\bibfnamefont {H.~J.~M.}\ \bibnamefont {Swagten}}, \bibinfo {author}
  {\bibfnamefont {M.}~\bibnamefont {Kl{\"{a}}ui}}, \ and\ \bibinfo {author}
  {\bibfnamefont {S.}~\bibnamefont {Eisebitt}}} (\bibinfo {year} {2015}),\
  \href@noop {} {\bibfield  {journal} {\bibinfo  {journal} {Nature Physics}\
  }\textbf {\bibinfo {volume} {11}},\ \bibinfo {pages} {225}}\BibitemShut
  {NoStop}%
\bibitem [{\citenamefont {Cai}\ \emph {et~al.}(2012)\citenamefont {Cai},
  \citenamefont {Chudnovsky},\ and\ \citenamefont {Garanin}}]{Cai2012}%
  \BibitemOpen
  \bibfield  {author} {\bibinfo {author} {\bibnamefont {Cai}, \bibfnamefont
  {L.}}, \bibinfo {author} {\bibfnamefont {E.~M.}\ \bibnamefont {Chudnovsky}},
  \ and\ \bibinfo {author} {\bibfnamefont {D.~A.}\ \bibnamefont {Garanin}}}
  (\bibinfo {year} {2012}),\ \href@noop {} {\bibfield  {journal} {\bibinfo
  {journal} {Physical Review B}\ }\textbf {\bibinfo {volume} {86}},\ \bibinfo
  {pages} {024429}}\BibitemShut {NoStop}%
\bibitem [{\citenamefont {Chang}\ and\ \citenamefont {Niu}(1995)}]{Chang1995a}%
  \BibitemOpen
  \bibfield  {author} {\bibinfo {author} {\bibnamefont {Chang}, \bibfnamefont
  {M.~C.}}, \ and\ \bibinfo {author} {\bibfnamefont {Q.}~\bibnamefont {Niu}}}
  (\bibinfo {year} {1995}),\ \href {\doibase 10.1103/PhysRevLett.75.1348}
  {\bibfield  {journal} {\bibinfo  {journal} {Physical Review Letters}\
  }\textbf {\bibinfo {volume} {75}}~(\bibinfo {number} {7}),\ \bibinfo {pages}
  {1348}}\BibitemShut {NoStop}%
\bibitem [{\citenamefont {Chang}\ and\ \citenamefont {Niu}(1996)}]{Chang1995b}%
  \BibitemOpen
  \bibfield  {author} {\bibinfo {author} {\bibnamefont {Chang}, \bibfnamefont
  {M.~C.}}, \ and\ \bibinfo {author} {\bibfnamefont {Q.}~\bibnamefont {Niu}}}
  (\bibinfo {year} {1996}),\ \href@noop {} {\bibfield  {journal} {\bibinfo
  {journal} {Physical Review B}\ }\textbf {\bibinfo {volume} {53}}~(\bibinfo
  {number} {11}),\ \bibinfo {pages} {7010}}\BibitemShut {NoStop}%
\bibitem [{\citenamefont {Cooper}\ \emph {et~al.}(1997)\citenamefont {Cooper},
  \citenamefont {Halperin},\ and\ \citenamefont {Ruzin}}]{Cooper1997}%
  \BibitemOpen
  \bibfield  {author} {\bibinfo {author} {\bibnamefont {Cooper}, \bibfnamefont
  {N.~R.}}, \bibinfo {author} {\bibfnamefont {B.~I.}\ \bibnamefont {Halperin}},
  \ and\ \bibinfo {author} {\bibfnamefont {I.~M.}\ \bibnamefont {Ruzin}}}
  (\bibinfo {year} {1997}),\ \href {\doibase 10.1103/PhysRevB.55.2344}
  {\bibfield  {journal} {\bibinfo  {journal} {Physical Review B}\ }\textbf
  {\bibinfo {volume} {55}}~(\bibinfo {number} {4}),\ \bibinfo {pages}
  {2344}}\BibitemShut {NoStop}%
\bibitem [{\citenamefont {Dana}\ and\ \citenamefont {Zak}(1983)}]{Dana1983}%
  \BibitemOpen
  \bibfield  {author} {\bibinfo {author} {\bibnamefont {Dana}, \bibfnamefont
  {I.}}, \ and\ \bibinfo {author} {\bibfnamefont {J.}~\bibnamefont {Zak}}}
  (\bibinfo {year} {1983}),\ \href@noop {} {\bibfield  {journal} {\bibinfo
  {journal} {Physical Review B}\ }\textbf {\bibinfo {volume} {28}}~(\bibinfo
  {number} {2}),\ \bibinfo {pages} {811}}\BibitemShut {NoStop}%
\bibitem [{\citenamefont {Das~Sarma}\ and\ \citenamefont
  {Pinczuk}(1997)}]{DasSarmaPinczuk}%
  \BibitemOpen
  \bibinfo {editor} {\bibnamefont {Das~Sarma}, \bibfnamefont {S.}}, \ and\
  \bibinfo {editor} {\bibfnamefont {A.}~\bibnamefont {Pinczuk}},\ Eds.
  (\bibinfo {year} {1997}),\ \href@noop {} {\emph {\bibinfo {title}
  {Perspectives in Quantum Hall Effects: Novel Quantum Liquids in
  Low‐Dimensional Semiconductor Structures}}}\ (\bibinfo  {publisher} {Wiley
  and Sons},\ \bibinfo {address} {New York})\BibitemShut {NoStop}%
\bibitem [{\citenamefont {Derras-Chouk}\ \emph {et~al.}(2018)\citenamefont
  {Derras-Chouk}, \citenamefont {Chudnovsky},\ and\ \citenamefont
  {Garanin}}]{Derras-Chouk2018}%
  \BibitemOpen
  \bibfield  {author} {\bibinfo {author} {\bibnamefont {Derras-Chouk},
  \bibfnamefont {A.}}, \bibinfo {author} {\bibfnamefont {E.~M.}\ \bibnamefont
  {Chudnovsky}}, \ and\ \bibinfo {author} {\bibfnamefont {D.~A.}\ \bibnamefont
  {Garanin}}} (\bibinfo {year} {2018}),\ \href@noop {} {\bibfield  {journal}
  {\bibinfo  {journal} {Physical Review B}\ }\textbf {\bibinfo {volume} {98}},\
  \bibinfo {pages} {024423}}\BibitemShut {NoStop}%
\bibitem [{\citenamefont {Diaz}\ and\ \citenamefont {Arovas}(2016)}]{Diaz2016}%
  \BibitemOpen
  \bibfield  {author} {\bibinfo {author} {\bibnamefont {Diaz}, \bibfnamefont
  {S.~A.}}, \ and\ \bibinfo {author} {\bibfnamefont {D.~P.}\ \bibnamefont
  {Arovas}}} (\bibinfo {year} {2016}),\ \href {http://arxiv.org/abs/1604.04010}
  {\bibinfo  {journal} {arXiv:1604.04010}\ }\BibitemShut {NoStop}%
\bibitem [{\citenamefont {Dirac}(1964)}]{Dirac1964}%
  \BibitemOpen
\bibfield  {journal} {  }\bibfield  {author} {\bibinfo {author} {\bibnamefont
  {Dirac}, \bibfnamefont {P.}}} (\bibinfo {year} {1964}),\ \href@noop {} {\emph
  {\bibinfo {title} {Lecture on Quantum Mechanics}}}\ (\bibinfo  {publisher}
  {Yeshiva Press},\ \bibinfo {address} {New York})\BibitemShut {NoStop}%
\bibitem [{\citenamefont {Dresselhaus}\ \emph {et~al.}(2008)\citenamefont
  {Dresselhaus}, \citenamefont {Dresselhaus},\ and\ \citenamefont
  {Jorio}}]{Dresselhaus}%
  \BibitemOpen
  \bibfield  {author} {\bibinfo {author} {\bibnamefont {Dresselhaus},
  \bibfnamefont {M.~S.}}, \bibinfo {author} {\bibfnamefont {G.}~\bibnamefont
  {Dresselhaus}}, \ and\ \bibinfo {author} {\bibfnamefont {A.}~\bibnamefont
  {Jorio}}} (\bibinfo {year} {2008}),\ \href@noop {} {\emph {\bibinfo {title}
  {Group Theory: Application to the Physics of Condensed Matter}}}\ (\bibinfo
  {publisher} {Springer},\ \bibinfo {address} {New York})\BibitemShut {NoStop}%
\bibitem [{\citenamefont {Dunne}\ and\ \citenamefont
  {Jackiw}(1993)}]{Dunne1993}%
  \BibitemOpen
  \bibfield  {author} {\bibinfo {author} {\bibnamefont {Dunne}, \bibfnamefont
  {G.}}, \ and\ \bibinfo {author} {\bibfnamefont {R.}~\bibnamefont {Jackiw}}}
  (\bibinfo {year} {1993}),\ \href@noop {} {\bibfield  {journal} {\bibinfo
  {journal} {Nuclear Physics B (Proceedings Supplements)}\ }\textbf {\bibinfo
  {volume} {33}},\ \bibinfo {pages} {114}}\BibitemShut {NoStop}%
\bibitem [{\citenamefont {Dunne}\ \emph {et~al.}(1990)\citenamefont {Dunne},
  \citenamefont {Jackiw},\ and\ \citenamefont {Trugenberger}}]{Dunne1990}%
  \BibitemOpen
  \bibfield  {author} {\bibinfo {author} {\bibnamefont {Dunne}, \bibfnamefont
  {G.~V.}}, \bibinfo {author} {\bibfnamefont {R.}~\bibnamefont {Jackiw}}, \
  and\ \bibinfo {author} {\bibfnamefont {C.~A.}\ \bibnamefont {Trugenberger}}}
  (\bibinfo {year} {1990}),\ \href {\doibase 10.1103/PhysRevD.41.661}
  {\bibfield  {journal} {\bibinfo  {journal} {Physical Review D}\ }\textbf
  {\bibinfo {volume} {41}},\ \bibinfo {pages} {661}}\BibitemShut {NoStop}%
\bibitem [{\citenamefont {Duval}\ \emph {et~al.}(2006)\citenamefont {Duval},
  \citenamefont {Horva\'ath}, \citenamefont {Horv\'athy}, \citenamefont
  {Martina},\ and\ \citenamefont {Stichel}}]{Duval2006}%
  \BibitemOpen
  \bibfield  {author} {\bibinfo {author} {\bibnamefont {Duval}, \bibfnamefont
  {C.}}, \bibinfo {author} {\bibfnamefont {Z.}~\bibnamefont {Horva\'ath}},
  \bibinfo {author} {\bibfnamefont {P.~A.}\ \bibnamefont {Horv\'athy}},
  \bibinfo {author} {\bibfnamefont {L.}~\bibnamefont {Martina}}, \ and\
  \bibinfo {author} {\bibfnamefont {P.~C.}\ \bibnamefont {Stichel}}} (\bibinfo
  {year} {2006}),\ \href@noop {} {\bibfield  {journal} {\bibinfo  {journal}
  {Modern Physics Letters B}\ }\textbf {\bibinfo {volume} {20}}~(\bibinfo
  {number} {7}),\ \bibinfo {pages} {373}}\BibitemShut {NoStop}%
\bibitem [{\citenamefont {Dzyaloshinskii}(1957)}]{Dzyaloshinskii1957}%
  \BibitemOpen
  \bibfield  {author} {\bibinfo {author} {\bibnamefont {Dzyaloshinskii},
  \bibfnamefont {I.~E.}}} (\bibinfo {year} {1957}),\ \href@noop {} {\bibfield
  {journal} {\bibinfo  {journal} {Sov. Phys. JETP}\ }\textbf {\bibinfo {volume}
  {5}}~(\bibinfo {number} {6}),\ \bibinfo {pages} {1259}}\BibitemShut {NoStop}%
\bibitem [{\citenamefont {Dzyaloshinskii}\ and\ \citenamefont
  {Volovick}(1980)}]{Dzyaloshinskii1980}%
  \BibitemOpen
  \bibfield  {author} {\bibinfo {author} {\bibnamefont {Dzyaloshinskii},
  \bibfnamefont {I.~E.}}, \ and\ \bibinfo {author} {\bibfnamefont {G.~E.}\
  \bibnamefont {Volovick}}} (\bibinfo {year} {1980}),\ \href {\doibase
  10.1016/0003-4916(80)90119-0} {\bibfield  {journal} {\bibinfo  {journal}
  {Annals of Physics}\ }\textbf {\bibinfo {volume} {125}},\ \bibinfo {pages}
  {67}}\BibitemShut {NoStop}%
\bibitem [{\citenamefont {Dzyaloshinsky}(1958)}]{Dzyaloshinsky1958}%
  \BibitemOpen
  \bibfield  {author} {\bibinfo {author} {\bibnamefont {Dzyaloshinsky},
  \bibfnamefont {I.}}} (\bibinfo {year} {1958}),\ \href {\doibase
  10.1016/0022-3697(58)90076-3} {\bibfield  {journal} {\bibinfo  {journal}
  {Journal of Physics and Chemistry of Solids}\ }\textbf {\bibinfo {volume}
  {4}}~(\bibinfo {number} {4}),\ \bibinfo {pages} {241}}\BibitemShut {NoStop}%
\bibitem [{\citenamefont {Erten}\ \emph {et~al.}(2017)\citenamefont {Erten},
  \citenamefont {Chang}, \citenamefont {Coleman},\ and\ \citenamefont
  {Tsvelik}}]{Onur2017}%
  \BibitemOpen
  \bibfield  {author} {\bibinfo {author} {\bibnamefont {Erten}, \bibfnamefont
  {O.}}, \bibinfo {author} {\bibfnamefont {P.-Y.}\ \bibnamefont {Chang}},
  \bibinfo {author} {\bibfnamefont {P.}~\bibnamefont {Coleman}}, \ and\
  \bibinfo {author} {\bibfnamefont {A.~M.}\ \bibnamefont {Tsvelik}}} (\bibinfo
  {year} {2017}),\ \href@noop {} {\bibfield  {journal} {\bibinfo  {journal}
  {Physical Review Letters}\ }\textbf {\bibinfo {volume} {119}},\ \bibinfo
  {pages} {057603}}\BibitemShut {NoStop}%
\bibitem [{\citenamefont {Ezawa}\ \emph {et~al.}(2003)\citenamefont {Ezawa},
  \citenamefont {Tsisishvili},\ and\ \citenamefont {Hasebe}}]{Ezawa2003}%
  \BibitemOpen
  \bibfield  {author} {\bibinfo {author} {\bibnamefont {Ezawa}, \bibfnamefont
  {Z.~F.}}, \bibinfo {author} {\bibfnamefont {G.}~\bibnamefont {Tsisishvili}},
  \ and\ \bibinfo {author} {\bibfnamefont {K.}~\bibnamefont {Hasebe}}}
  (\bibinfo {year} {2003}),\ \href@noop {} {\bibfield  {journal} {\bibinfo
  {journal} {Physical Review B}\ }\textbf {\bibinfo {volume} {67}},\ \bibinfo
  {pages} {125314}}\BibitemShut {NoStop}%
\bibitem [{\citenamefont {Faddeev}\ and\ \citenamefont
  {Jackiw}(1988)}]{Faddeev1988}%
  \BibitemOpen
  \bibfield  {author} {\bibinfo {author} {\bibnamefont {Faddeev}, \bibfnamefont
  {L.}}, \ and\ \bibinfo {author} {\bibfnamefont {R.}~\bibnamefont {Jackiw}}}
  (\bibinfo {year} {1988}),\ \href {\doibase 10.1103/PhysRevLett.60.1692}
  {\bibfield  {journal} {\bibinfo  {journal} {Physical Review Letters}\
  }\textbf {\bibinfo {volume} {60}}~(\bibinfo {number} {17}),\ \bibinfo {pages}
  {1692}}\BibitemShut {NoStop}%
\bibitem [{\citenamefont {Fisher}\ \emph {et~al.}(1989)\citenamefont {Fisher},
  \citenamefont {Weichman}, \citenamefont {Watson},\ and\ \citenamefont
  {Fisher}}]{Fisher1989}%
  \BibitemOpen
  \bibfield  {author} {\bibinfo {author} {\bibnamefont {Fisher}, \bibfnamefont
  {M.~P.~A.}}, \bibinfo {author} {\bibfnamefont {P.~B.}\ \bibnamefont
  {Weichman}}, \bibinfo {author} {\bibfnamefont {J.}~\bibnamefont {Watson}}, \
  and\ \bibinfo {author} {\bibfnamefont {D.~S.}\ \bibnamefont {Fisher}}}
  (\bibinfo {year} {1989}),\ \href@noop {} {\bibfield  {journal} {\bibinfo
  {journal} {Physical Review B}\ }\textbf {\bibinfo {volume} {40}},\ \bibinfo
  {pages} {546}}\BibitemShut {NoStop}%
\bibitem [{\citenamefont {Fujita}\ and\ \citenamefont
  {Sato}(2017)}]{Fujita2017}%
  \BibitemOpen
  \bibfield  {author} {\bibinfo {author} {\bibnamefont {Fujita}, \bibfnamefont
  {H.}}, \ and\ \bibinfo {author} {\bibfnamefont {M.}~\bibnamefont {Sato}}}
  (\bibinfo {year} {2017}),\ \href@noop {} {\bibfield  {journal} {\bibinfo
  {journal} {Physical Review B}\ }\textbf {\bibinfo {volume} {95}},\ \bibinfo
  {pages} {054421}}\BibitemShut {NoStop}%
\bibitem [{\citenamefont {Galkin}\ and\ \citenamefont
  {Ivanov}(2007)}]{Galkin2007}%
  \BibitemOpen
  \bibfield  {author} {\bibinfo {author} {\bibnamefont {Galkin}, \bibfnamefont
  {A.~Y.}}, \ and\ \bibinfo {author} {\bibfnamefont {B.~A.}\ \bibnamefont
  {Ivanov}}} (\bibinfo {year} {2007}),\ \href {\doibase
  10.1134/S1063776107050123} {\bibfield  {journal} {\bibinfo  {journal}
  {Journal of Experimental and Theoretical Physics (JETP)}\ }\textbf {\bibinfo
  {volume} {104}}~(\bibinfo {number} {5}),\ \bibinfo {pages} {775}}\BibitemShut
  {NoStop}%
\bibitem [{\citenamefont {Gallagher}\ \emph {et~al.}(2017)\citenamefont
  {Gallagher}, \citenamefont {Meng}, \citenamefont {Brangham}, \citenamefont
  {Wang}, \citenamefont {Esser}, \citenamefont {McComb},\ and\ \citenamefont
  {Yang}}]{Gallagher2017}%
  \BibitemOpen
  \bibfield  {author} {\bibinfo {author} {\bibnamefont {Gallagher},
  \bibfnamefont {J.~C.}}, \bibinfo {author} {\bibfnamefont {K.~Y.}\
  \bibnamefont {Meng}}, \bibinfo {author} {\bibfnamefont {J.~T.}\ \bibnamefont
  {Brangham}}, \bibinfo {author} {\bibfnamefont {H.~L.}\ \bibnamefont {Wang}},
  \bibinfo {author} {\bibfnamefont {B.~D.}\ \bibnamefont {Esser}}, \bibinfo
  {author} {\bibfnamefont {D.~W.}\ \bibnamefont {McComb}}, \ and\ \bibinfo
  {author} {\bibfnamefont {F.~Y.}\ \bibnamefont {Yang}}} (\bibinfo {year}
  {2017}),\ \href@noop {} {\bibfield  {journal} {\bibinfo  {journal} {Physical
  Review Letters}\ }\textbf {\bibinfo {volume} {118}},\ \bibinfo {pages}
  {027201}}\BibitemShut {NoStop}%
\bibitem [{\citenamefont {de~Gennes}\ and\ \citenamefont
  {Prost}(1995)}]{deGennes_book}%
  \BibitemOpen
  \bibfield  {author} {\bibinfo {author} {\bibnamefont {de~Gennes},
  \bibfnamefont {P.~G.}}, \ and\ \bibinfo {author} {\bibfnamefont
  {J.}~\bibnamefont {Prost}}} (\bibinfo {year} {1995}),\ \href@noop {} {\emph
  {\bibinfo {title} {The Physics of Liquid Crystals}}}\ (\bibinfo  {publisher}
  {Claredon Press},\ \bibinfo {address} {Oxford})\BibitemShut {NoStop}%
\bibitem [{\citenamefont {Gilbert}(2004)}]{Gilbert2004}%
  \BibitemOpen
  \bibfield  {author} {\bibinfo {author} {\bibnamefont {Gilbert}, \bibfnamefont
  {T.}}} (\bibinfo {year} {2004}),\ \href {\doibase 10.1109/TMAG.2004.836740}
  {\bibfield  {journal} {\bibinfo  {journal} {IEEE Transactions on Magnetics}\
  }\textbf {\bibinfo {volume} {40}}~(\bibinfo {number} {6}),\ \bibinfo {pages}
  {3443}}\BibitemShut {NoStop}%
\bibitem [{\citenamefont {Girvin}\ and\ \citenamefont
  {Jach}(1984)}]{Girvin1984}%
  \BibitemOpen
  \bibfield  {author} {\bibinfo {author} {\bibnamefont {Girvin}, \bibfnamefont
  {S.}}, \ and\ \bibinfo {author} {\bibfnamefont {T.}~\bibnamefont {Jach}}}
  (\bibinfo {year} {1984}),\ \href {\doibase 10.1103/PhysRevB.29.5617}
  {\bibfield  {journal} {\bibinfo  {journal} {Physical Review B}\ }\textbf
  {\bibinfo {volume} {29}}~(\bibinfo {number} {10}),\ \bibinfo {pages}
  {5617}}\BibitemShut {NoStop}%
\bibitem [{\citenamefont {Greiner}\ \emph {et~al.}(2002)\citenamefont
  {Greiner}, \citenamefont {Mandel}, \citenamefont {Esslinger}, \citenamefont
  {H\"ansch},\ and\ \citenamefont {Bloch}}]{Greiner2002}%
  \BibitemOpen
  \bibfield  {author} {\bibinfo {author} {\bibnamefont {Greiner}, \bibfnamefont
  {M.}}, \bibinfo {author} {\bibfnamefont {O.}~\bibnamefont {Mandel}}, \bibinfo
  {author} {\bibfnamefont {T.}~\bibnamefont {Esslinger}}, \bibinfo {author}
  {\bibfnamefont {T.~W.}\ \bibnamefont {H\"ansch}}, \ and\ \bibinfo {author}
  {\bibfnamefont {I.}~\bibnamefont {Bloch}}} (\bibinfo {year} {2002}),\
  \href@noop {} {\bibfield  {journal} {\bibinfo  {journal} {Nature}\ }\textbf
  {\bibinfo {volume} {415}},\ \bibinfo {pages} {39}}\BibitemShut {NoStop}%
\bibitem [{\citenamefont {Grover}\ and\ \citenamefont
  {Senthil}(2008)}]{Grover2008}%
  \BibitemOpen
  \bibfield  {author} {\bibinfo {author} {\bibnamefont {Grover}, \bibfnamefont
  {T.}}, \ and\ \bibinfo {author} {\bibfnamefont {T.}~\bibnamefont {Senthil}}}
  (\bibinfo {year} {2008}),\ \href {\doibase 10.1103/PhysRevLett.100.156804}
  {\bibfield  {journal} {\bibinfo  {journal} {Physical Review Letters}\
  }\textbf {\bibinfo {volume} {100}}~(\bibinfo {number} {15}),\ \bibinfo
  {pages} {156804}}\BibitemShut {NoStop}%
\bibitem [{\citenamefont {Haldane}(1986)}]{Haldane1986}%
  \BibitemOpen
  \bibfield  {author} {\bibinfo {author} {\bibnamefont {Haldane}, \bibfnamefont
  {F.~D.~M.}}} (\bibinfo {year} {1986}),\ \href {\doibase
  10.1103/PhysRevLett.57.1488} {\bibfield  {journal} {\bibinfo  {journal}
  {Physical Review Letters}\ }\textbf {\bibinfo {volume} {57}}~(\bibinfo
  {number} {12}),\ \bibinfo {pages} {1488}}\BibitemShut {NoStop}%
\bibitem [{\citenamefont {Halperin}(1982)}]{Halperin1982}%
  \BibitemOpen
  \bibfield  {author} {\bibinfo {author} {\bibnamefont {Halperin},
  \bibfnamefont {B.~I.}}} (\bibinfo {year} {1982}),\ \href@noop {} {\bibfield
  {journal} {\bibinfo  {journal} {Physical Review B}\ }\textbf {\bibinfo
  {volume} {25}},\ \bibinfo {pages} {2185}}\BibitemShut {NoStop}%
\bibitem [{\citenamefont {Halperin}\ and\ \citenamefont
  {Hohenberg}(1969)}]{Halperin1969}%
  \BibitemOpen
  \bibfield  {author} {\bibinfo {author} {\bibnamefont {Halperin},
  \bibfnamefont {B.~I.}}, \ and\ \bibinfo {author} {\bibfnamefont {P.~C.}\
  \bibnamefont {Hohenberg}}} (\bibinfo {year} {1969}),\ \href {\doibase
  10.1103/PhysRev.188.898} {\bibfield  {journal} {\bibinfo  {journal} {Physical
  Review}\ }\textbf {\bibinfo {volume} {188}}~(\bibinfo {number} {2}),\
  \bibinfo {pages} {898}}\BibitemShut {NoStop}%
\bibitem [{\citenamefont {Harper}(1955)}]{Harper1955}%
  \BibitemOpen
  \bibfield  {author} {\bibinfo {author} {\bibnamefont {Harper}, \bibfnamefont
  {P.~G.}}} (\bibinfo {year} {1955}),\ \href {\doibase
  10.1088/0370-1298/68/10/304} {\bibfield  {journal} {\bibinfo  {journal}
  {Proceedings of the Physical Society: Section A}\ }\textbf {\bibinfo {volume}
  {68}}~(\bibinfo {number} {10}),\ \bibinfo {pages} {874}}\BibitemShut
  {NoStop}%
\bibitem [{\citenamefont {Hasenfratz}\ and\ \citenamefont {{'t
  Hooft}}(1976)}]{Hasenfratz1976}%
  \BibitemOpen
  \bibfield  {author} {\bibinfo {author} {\bibnamefont {Hasenfratz},
  \bibfnamefont {P.}}, \ and\ \bibinfo {author} {\bibfnamefont
  {G.}~\bibnamefont {{'t Hooft}}}} (\bibinfo {year} {1976}),\ \href {\doibase
  10.1103/PhysRevLett.36.1119} {\bibfield  {journal} {\bibinfo  {journal}
  {Physical Review Letters}\ }\textbf {\bibinfo {volume} {36}}~(\bibinfo
  {number} {19}),\ \bibinfo {pages} {1119}}\BibitemShut {NoStop}%
\bibitem [{\citenamefont {Hatsugai}(1993)}]{Hatsugai1993}%
  \BibitemOpen
  \bibfield  {author} {\bibinfo {author} {\bibnamefont {Hatsugai},
  \bibfnamefont {Y.}}} (\bibinfo {year} {1993}),\ \href {\doibase
  10.1103/PhysRevLett.71.3697} {\bibfield  {journal} {\bibinfo  {journal}
  {Physical Review Letters}\ }\textbf {\bibinfo {volume} {71}}~(\bibinfo
  {number} {22}),\ \bibinfo {pages} {3697}}\BibitemShut {NoStop}%
\bibitem [{\citenamefont {Heinze}\ \emph {et~al.}(2011)\citenamefont {Heinze},
  \citenamefont {von Bergmann}, \citenamefont {Menzel}, \citenamefont {Brede},
  \citenamefont {Kubetzka}, \citenamefont {Wiesendanger}, \citenamefont
  {Bihlmayer},\ and\ \citenamefont {Bl\"ugel}}]{Heinze2011}%
  \BibitemOpen
  \bibfield  {author} {\bibinfo {author} {\bibnamefont {Heinze}, \bibfnamefont
  {S.}}, \bibinfo {author} {\bibfnamefont {K.}~\bibnamefont {von Bergmann}},
  \bibinfo {author} {\bibfnamefont {M.}~\bibnamefont {Menzel}}, \bibinfo
  {author} {\bibfnamefont {J.}~\bibnamefont {Brede}}, \bibinfo {author}
  {\bibfnamefont {A.}~\bibnamefont {Kubetzka}}, \bibinfo {author}
  {\bibfnamefont {R.}~\bibnamefont {Wiesendanger}}, \bibinfo {author}
  {\bibfnamefont {G.}~\bibnamefont {Bihlmayer}}, \ and\ \bibinfo {author}
  {\bibfnamefont {S.}~\bibnamefont {Bl\"ugel}}} (\bibinfo {year} {2011}),\
  \href {\doibase 10.1038/nphys2045} {\bibfield  {journal} {\bibinfo  {journal}
  {Nature Physics}\ }\textbf {\bibinfo {volume} {7}}~(\bibinfo {number} {9}),\
  \bibinfo {pages} {713}}\BibitemShut {NoStop}%
\bibitem [{\citenamefont {Hirschberger}\ \emph
  {et~al.}(2015{\natexlab{a}})\citenamefont {Hirschberger}, \citenamefont
  {Chisnell}, \citenamefont {Lee},\ and\ \citenamefont
  {Ong}}]{Hirschberger2015}%
  \BibitemOpen
  \bibfield  {author} {\bibinfo {author} {\bibnamefont {Hirschberger},
  \bibfnamefont {M.}}, \bibinfo {author} {\bibfnamefont {R.}~\bibnamefont
  {Chisnell}}, \bibinfo {author} {\bibfnamefont {Y.~S.}\ \bibnamefont {Lee}}, \
  and\ \bibinfo {author} {\bibfnamefont {N.~P.}\ \bibnamefont {Ong}}} (\bibinfo
  {year} {2015}{\natexlab{a}}),\ \href {\doibase
  10.1103/PhysRevLett.115.106603} {\bibfield  {journal} {\bibinfo  {journal}
  {Physical Review Letters}\ }\textbf {\bibinfo {volume} {115}}~(\bibinfo
  {number} {10}),\ \bibinfo {pages} {106603}}\BibitemShut {NoStop}%
\bibitem [{\citenamefont {Hirschberger}\ \emph
  {et~al.}(2015{\natexlab{b}})\citenamefont {Hirschberger}, \citenamefont
  {Krizan}, \citenamefont {Cava},\ and\ \citenamefont
  {Ong}}]{Hirschberger2015a}%
  \BibitemOpen
  \bibfield  {author} {\bibinfo {author} {\bibnamefont {Hirschberger},
  \bibfnamefont {M.}}, \bibinfo {author} {\bibfnamefont {J.~W.}\ \bibnamefont
  {Krizan}}, \bibinfo {author} {\bibfnamefont {R.~J.}\ \bibnamefont {Cava}}, \
  and\ \bibinfo {author} {\bibfnamefont {N.~P.}\ \bibnamefont {Ong}}} (\bibinfo
  {year} {2015}{\natexlab{b}}),\ \href {\doibase 10.1126/science.1257340}
  {\bibfield  {journal} {\bibinfo  {journal} {Science}\ }\textbf {\bibinfo
  {volume} {348}}~(\bibinfo {number} {6230}),\ \bibinfo {pages}
  {106}}\BibitemShut {NoStop}%
\bibitem [{\citenamefont {Ho}(1998)}]{Ho1998}%
  \BibitemOpen
  \bibfield  {author} {\bibinfo {author} {\bibnamefont {Ho}, \bibfnamefont
  {T.-L.}}} (\bibinfo {year} {1998}),\ \href {\doibase
  10.1103/PhysRevLett.81.742} {\bibfield  {journal} {\bibinfo  {journal}
  {Physical Review Letters}\ }\textbf {\bibinfo {volume} {81}}~(\bibinfo
  {number} {4}),\ \bibinfo {pages} {742}}\BibitemShut {NoStop}%
\bibitem [{\citenamefont {Hofstadter}(1976)}]{Hofstadter1976}%
  \BibitemOpen
  \bibfield  {author} {\bibinfo {author} {\bibnamefont {Hofstadter},
  \bibfnamefont {D.~R.}}} (\bibinfo {year} {1976}),\ \href@noop {} {\bibfield
  {journal} {\bibinfo  {journal} {Physical Review B}\ }\textbf {\bibinfo
  {volume} {14}}~(\bibinfo {number} {6}),\ \bibinfo {pages} {2239}}\BibitemShut
  {NoStop}%
\bibitem [{\citenamefont {Hsu}\ and\ \citenamefont
  {Chakravarty}(2013)}]{Hsu2013}%
  \BibitemOpen
  \bibfield  {author} {\bibinfo {author} {\bibnamefont {Hsu}, \bibfnamefont
  {C.~H.}}, \ and\ \bibinfo {author} {\bibfnamefont {S.}~\bibnamefont
  {Chakravarty}}} (\bibinfo {year} {2013}),\ \href {\doibase
  10.1103/PhysRevB.87.085114} {\bibfield  {journal} {\bibinfo  {journal}
  {Physical Review B}\ }\textbf {\bibinfo {volume} {87}}~(\bibinfo {number}
  {8}),\ \bibinfo {pages} {1}}\BibitemShut {NoStop}%
\bibitem [{\citenamefont {Hsu}\ and\ \citenamefont
  {Chakravarty}(2014)}]{Hsu2014}%
  \BibitemOpen
  \bibfield  {author} {\bibinfo {author} {\bibnamefont {Hsu}, \bibfnamefont
  {C.~H.}}, \ and\ \bibinfo {author} {\bibfnamefont {S.}~\bibnamefont
  {Chakravarty}}} (\bibinfo {year} {2014}),\ \href {\doibase
  10.1103/PhysRevB.90.134507} {\bibfield  {journal} {\bibinfo  {journal}
  {Physical Review B}\ }\textbf {\bibinfo {volume} {90}}~(\bibinfo {number}
  {13}),\ 10.1103/PhysRevB.90.134507}\BibitemShut {NoStop}%
\bibitem [{\citenamefont {Huber}\ and\ \citenamefont
  {Altman}(2010)}]{Huber2010}%
  \BibitemOpen
  \bibfield  {author} {\bibinfo {author} {\bibnamefont {Huber}, \bibfnamefont
  {S.~D.}}, \ and\ \bibinfo {author} {\bibfnamefont {E.}~\bibnamefont
  {Altman}}} (\bibinfo {year} {2010}),\ \href@noop {} {\bibfield  {journal}
  {\bibinfo  {journal} {Physical Review B}\ }\textbf {\bibinfo {volume} {82}},\
  \bibinfo {pages} {184502}}\BibitemShut {NoStop}%
\bibitem [{\citenamefont {Hurst}\ \emph {et~al.}(2015)\citenamefont {Hurst},
  \citenamefont {Efimkin}, \citenamefont {Zang},\ and\ \citenamefont
  {Galitski}}]{Hurst2015}%
  \BibitemOpen
  \bibfield  {author} {\bibinfo {author} {\bibnamefont {Hurst}, \bibfnamefont
  {H.~M.}}, \bibinfo {author} {\bibfnamefont {D.~K.}\ \bibnamefont {Efimkin}},
  \bibinfo {author} {\bibfnamefont {J.}~\bibnamefont {Zang}}, \ and\ \bibinfo
  {author} {\bibfnamefont {V.}~\bibnamefont {Galitski}}} (\bibinfo {year}
  {2015}),\ \href {\doibase 10.1103/PhysRevB.91.060401} {\bibfield  {journal}
  {\bibinfo  {journal} {Physical Review B}\ }\textbf {\bibinfo {volume} {91}},\
  \bibinfo {pages} {060401}}\BibitemShut {NoStop}%
\bibitem [{\citenamefont {Ivanov}\ \emph {et~al.}(2007)\citenamefont {Ivanov},
  \citenamefont {Sheka}, \citenamefont {Kryvonos},\ and\ \citenamefont
  {Mertens}}]{Ivanov2007}%
  \BibitemOpen
  \bibfield  {author} {\bibinfo {author} {\bibnamefont {Ivanov}, \bibfnamefont
  {B.~A.}}, \bibinfo {author} {\bibfnamefont {D.~D.}\ \bibnamefont {Sheka}},
  \bibinfo {author} {\bibfnamefont {V.~V.}\ \bibnamefont {Kryvonos}}, \ and\
  \bibinfo {author} {\bibfnamefont {F.~G.}\ \bibnamefont {Mertens}}} (\bibinfo
  {year} {2007}),\ \href@noop {} {\bibfield  {journal} {\bibinfo  {journal}
  {Physical Review B}\ }\textbf {\bibinfo {volume} {75}},\ \bibinfo {pages}
  {132401}}\BibitemShut {NoStop}%
\bibitem [{\citenamefont {Jackiw}\ and\ \citenamefont
  {Rebbi}(1976{\natexlab{a}})}]{Jackiw1976a}%
  \BibitemOpen
  \bibfield  {author} {\bibinfo {author} {\bibnamefont {Jackiw}, \bibfnamefont
  {R.}}, \ and\ \bibinfo {author} {\bibfnamefont {C.}~\bibnamefont {Rebbi}}}
  (\bibinfo {year} {1976}{\natexlab{a}}),\ \href {\doibase
  10.1103/PhysRevD.13.3398} {\bibfield  {journal} {\bibinfo  {journal}
  {Physical Review D}\ }\textbf {\bibinfo {volume} {13}}~(\bibinfo {number}
  {12}),\ \bibinfo {pages} {3398}}\BibitemShut {NoStop}%
\bibitem [{\citenamefont {Jackiw}\ and\ \citenamefont
  {Rebbi}(1976{\natexlab{b}})}]{Jackiw1976}%
  \BibitemOpen
  \bibfield  {author} {\bibinfo {author} {\bibnamefont {Jackiw}, \bibfnamefont
  {R.}}, \ and\ \bibinfo {author} {\bibfnamefont {C.}~\bibnamefont {Rebbi}}}
  (\bibinfo {year} {1976}{\natexlab{b}}),\ \href@noop {} {\bibfield  {journal}
  {\bibinfo  {journal} {Physical Review Letters}\ }\textbf {\bibinfo {volume}
  {36}}~(\bibinfo {number} {19}),\ \bibinfo {pages} {1116}}\BibitemShut
  {NoStop}%
\bibitem [{\citenamefont {Janoschek}\ \emph {et~al.}(2013)\citenamefont
  {Janoschek}, \citenamefont {Garst}, \citenamefont {Bauer}, \citenamefont
  {Krautscheid}, \citenamefont {Georgii}, \citenamefont {B{\"{o}}ni},\ and\
  \citenamefont {Pfleiderer}}]{Janoschek2013}%
  \BibitemOpen
  \bibfield  {author} {\bibinfo {author} {\bibnamefont {Janoschek},
  \bibfnamefont {M.}}, \bibinfo {author} {\bibfnamefont {M.}~\bibnamefont
  {Garst}}, \bibinfo {author} {\bibfnamefont {A.}~\bibnamefont {Bauer}},
  \bibinfo {author} {\bibfnamefont {P.}~\bibnamefont {Krautscheid}}, \bibinfo
  {author} {\bibfnamefont {R.}~\bibnamefont {Georgii}}, \bibinfo {author}
  {\bibfnamefont {P.}~\bibnamefont {B{\"{o}}ni}}, \ and\ \bibinfo {author}
  {\bibfnamefont {C.}~\bibnamefont {Pfleiderer}}} (\bibinfo {year} {2013}),\
  \href@noop {} {\bibfield  {journal} {\bibinfo  {journal} {Physical Review B}\
  }\textbf {\bibinfo {volume} {87}},\ \bibinfo {pages} {134407}}\BibitemShut
  {NoStop}%
\bibitem [{\citenamefont {Jiang}\ \emph {et~al.}(2015)\citenamefont {Jiang},
  \citenamefont {Upadhyaya}, \citenamefont {Zhang}, \citenamefont {Yu},
  \citenamefont {Jungfleisch}, \citenamefont {Fradin}, \citenamefont {Pearson},
  \citenamefont {Tserkovnyak}, \citenamefont {Wang}, \citenamefont {Heinonen},
  \citenamefont {te~Velthuis},\ and\ \citenamefont {Hoffmann}}]{Jiang2015}%
  \BibitemOpen
  \bibfield  {author} {\bibinfo {author} {\bibnamefont {Jiang}, \bibfnamefont
  {W.}}, \bibinfo {author} {\bibfnamefont {P.}~\bibnamefont {Upadhyaya}},
  \bibinfo {author} {\bibfnamefont {W.}~\bibnamefont {Zhang}}, \bibinfo
  {author} {\bibfnamefont {G.}~\bibnamefont {Yu}}, \bibinfo {author}
  {\bibfnamefont {M.~B.}\ \bibnamefont {Jungfleisch}}, \bibinfo {author}
  {\bibfnamefont {F.~Y.}\ \bibnamefont {Fradin}}, \bibinfo {author}
  {\bibfnamefont {J.~E.}\ \bibnamefont {Pearson}}, \bibinfo {author}
  {\bibfnamefont {Y.}~\bibnamefont {Tserkovnyak}}, \bibinfo {author}
  {\bibfnamefont {K.~L.}\ \bibnamefont {Wang}}, \bibinfo {author}
  {\bibfnamefont {O.}~\bibnamefont {Heinonen}}, \bibinfo {author}
  {\bibfnamefont {S.~G.~E.}\ \bibnamefont {te~Velthuis}}, \ and\ \bibinfo
  {author} {\bibfnamefont {A.}~\bibnamefont {Hoffmann}}} (\bibinfo {year}
  {2015}),\ \href {\doibase 10.1126/science.aaa1442} {\bibfield  {journal}
  {\bibinfo  {journal} {Science}\ }\textbf {\bibinfo {volume} {349}}~(\bibinfo
  {number} {6245}),\ \bibinfo {pages} {283}}\BibitemShut {NoStop}%
\bibitem [{\citenamefont {Jiang}\ \emph {et~al.}(2016)\citenamefont {Jiang},
  \citenamefont {Zhang}, \citenamefont {Yu}, \citenamefont {Zhang},
  \citenamefont {Wang}, \citenamefont {Benjamin~Jungfleisch}, \citenamefont
  {Pearson}, \citenamefont {Cheng}, \citenamefont {Heinonen}, \citenamefont
  {Wang}, \citenamefont {Zhou}, \citenamefont {Hoffmann},\ and\ \citenamefont
  {te~Velthuis}}]{Jiang2016}%
  \BibitemOpen
  \bibfield  {author} {\bibinfo {author} {\bibnamefont {Jiang}, \bibfnamefont
  {W.}}, \bibinfo {author} {\bibfnamefont {X.}~\bibnamefont {Zhang}}, \bibinfo
  {author} {\bibfnamefont {G.}~\bibnamefont {Yu}}, \bibinfo {author}
  {\bibfnamefont {W.}~\bibnamefont {Zhang}}, \bibinfo {author} {\bibfnamefont
  {X.}~\bibnamefont {Wang}}, \bibinfo {author} {\bibfnamefont {M.}~\bibnamefont
  {Benjamin~Jungfleisch}}, \bibinfo {author} {\bibfnamefont {J.~E.}\
  \bibnamefont {Pearson}}, \bibinfo {author} {\bibfnamefont {X.}~\bibnamefont
  {Cheng}}, \bibinfo {author} {\bibfnamefont {O.}~\bibnamefont {Heinonen}},
  \bibinfo {author} {\bibfnamefont {K.~L.}\ \bibnamefont {Wang}}, \bibinfo
  {author} {\bibfnamefont {Y.}~\bibnamefont {Zhou}}, \bibinfo {author}
  {\bibfnamefont {A.}~\bibnamefont {Hoffmann}}, \ and\ \bibinfo {author}
  {\bibfnamefont {S.~G.~E.}\ \bibnamefont {te~Velthuis}}} (\bibinfo {year}
  {2016}),\ \href {\doibase 10.1038/nphys3883} {\bibfield  {journal} {\bibinfo
  {journal} {Nature Physics}\ }\textbf {\bibinfo {volume} {13}}~(\bibinfo
  {number} {2}),\ \bibinfo {pages} {162}}\BibitemShut {NoStop}%
\bibitem [{\citenamefont {Jonietz}\ \emph {et~al.}(2010)\citenamefont
  {Jonietz}, \citenamefont {M\"uhlbauer}, \citenamefont {Pfleiderer},
  \citenamefont {Neubauer}, \citenamefont {M\"unzer}, \citenamefont {Bauer},
  \citenamefont {Adams}, \citenamefont {Georgii}, \citenamefont {B\"oni},
  \citenamefont {Duine}, \citenamefont {Everschor}, \citenamefont {Garst},\
  and\ \citenamefont {Rosch}}]{Jonietz2010}%
  \BibitemOpen
  \bibfield  {author} {\bibinfo {author} {\bibnamefont {Jonietz}, \bibfnamefont
  {F.}}, \bibinfo {author} {\bibfnamefont {S.}~\bibnamefont {M\"uhlbauer}},
  \bibinfo {author} {\bibfnamefont {C.}~\bibnamefont {Pfleiderer}}, \bibinfo
  {author} {\bibfnamefont {A.}~\bibnamefont {Neubauer}}, \bibinfo {author}
  {\bibfnamefont {W.}~\bibnamefont {M\"unzer}}, \bibinfo {author}
  {\bibfnamefont {A.}~\bibnamefont {Bauer}}, \bibinfo {author} {\bibfnamefont
  {T.}~\bibnamefont {Adams}}, \bibinfo {author} {\bibfnamefont
  {R.}~\bibnamefont {Georgii}}, \bibinfo {author} {\bibfnamefont
  {P.}~\bibnamefont {B\"oni}}, \bibinfo {author} {\bibfnamefont {R.~A.}\
  \bibnamefont {Duine}}, \bibinfo {author} {\bibfnamefont {K.}~\bibnamefont
  {Everschor}}, \bibinfo {author} {\bibfnamefont {M.}~\bibnamefont {Garst}}, \
  and\ \bibinfo {author} {\bibfnamefont {A.}~\bibnamefont {Rosch}}} (\bibinfo
  {year} {2010}),\ \href@noop {} {\bibfield  {journal} {\bibinfo  {journal}
  {Science}\ }\textbf {\bibinfo {volume} {330}}~(\bibinfo {number}
  {December}),\ \bibinfo {pages} {1648}}\BibitemShut {NoStop}%
\bibitem [{\citenamefont {Kanazawa}\ \emph {et~al.}(2011)\citenamefont
  {Kanazawa}, \citenamefont {Onose}, \citenamefont {Arima}, \citenamefont
  {Okuyama}, \citenamefont {Ohoyama}, \citenamefont {Wakimoto}, \citenamefont
  {Kakurai}, \citenamefont {Ishiwata},\ and\ \citenamefont
  {Tokura}}]{Kanazawa2011}%
  \BibitemOpen
  \bibfield  {author} {\bibinfo {author} {\bibnamefont {Kanazawa},
  \bibfnamefont {N.}}, \bibinfo {author} {\bibfnamefont {Y.}~\bibnamefont
  {Onose}}, \bibinfo {author} {\bibfnamefont {T.}~\bibnamefont {Arima}},
  \bibinfo {author} {\bibfnamefont {D.}~\bibnamefont {Okuyama}}, \bibinfo
  {author} {\bibfnamefont {K.}~\bibnamefont {Ohoyama}}, \bibinfo {author}
  {\bibfnamefont {S.}~\bibnamefont {Wakimoto}}, \bibinfo {author}
  {\bibfnamefont {K.}~\bibnamefont {Kakurai}}, \bibinfo {author} {\bibfnamefont
  {S.}~\bibnamefont {Ishiwata}}, \ and\ \bibinfo {author} {\bibfnamefont
  {Y.}~\bibnamefont {Tokura}}} (\bibinfo {year} {2011}),\ \href@noop {}
  {\bibfield  {journal} {\bibinfo  {journal} {Physical Review Letters}\
  }\textbf {\bibinfo {volume} {106}},\ \bibinfo {pages} {156603}}\BibitemShut
  {NoStop}%
\bibitem [{\citenamefont {Kasahara}\ \emph {et~al.}(2018)\citenamefont
  {Kasahara}, \citenamefont {Sugii}, \citenamefont {Ohnishi}, \citenamefont
  {Shimozawa}, \citenamefont {Yamashita}, \citenamefont {Kurita}, \citenamefont
  {Tanaka}, \citenamefont {Nasu}, \citenamefont {Motome}, \citenamefont
  {Shibauchi},\ and\ \citenamefont {Matsuda}}]{Kasahara}%
  \BibitemOpen
  \bibfield  {author} {\bibinfo {author} {\bibnamefont {Kasahara},
  \bibfnamefont {Y.}}, \bibinfo {author} {\bibfnamefont {K.}~\bibnamefont
  {Sugii}}, \bibinfo {author} {\bibfnamefont {T.}~\bibnamefont {Ohnishi}},
  \bibinfo {author} {\bibfnamefont {M.}~\bibnamefont {Shimozawa}}, \bibinfo
  {author} {\bibfnamefont {M.}~\bibnamefont {Yamashita}}, \bibinfo {author}
  {\bibfnamefont {N.}~\bibnamefont {Kurita}}, \bibinfo {author} {\bibfnamefont
  {H.}~\bibnamefont {Tanaka}}, \bibinfo {author} {\bibfnamefont
  {J.}~\bibnamefont {Nasu}}, \bibinfo {author} {\bibfnamefont {Y.}~\bibnamefont
  {Motome}}, \bibinfo {author} {\bibfnamefont {T.}~\bibnamefont {Shibauchi}}, \
  and\ \bibinfo {author} {\bibfnamefont {Y.}~\bibnamefont {Matsuda}}} (\bibinfo
  {year} {2018}),\ \href@noop {} {\bibfield  {journal} {\bibinfo  {journal}
  {Physical Review Letters}\ }\textbf {\bibinfo {volume} {120}},\ \bibinfo
  {pages} {217205}}\BibitemShut {NoStop}%
\bibitem [{\citenamefont {Kim}\ and\ \citenamefont
  {Shapere}(2016)}]{KimSoo2016}%
  \BibitemOpen
  \bibfield  {author} {\bibinfo {author} {\bibnamefont {Kim}, \bibfnamefont
  {B.~S.}}, \ and\ \bibinfo {author} {\bibfnamefont {A.~D.}\ \bibnamefont
  {Shapere}}} (\bibinfo {year} {2016}),\ \href {\doibase
  10.1103/PhysRevLett.117.116805} {\bibfield  {journal} {\bibinfo  {journal}
  {Physical Review Letters}\ }\textbf {\bibinfo {volume} {117}},\ \bibinfo
  {pages} {116805}}\BibitemShut {NoStop}%
\bibitem [{\citenamefont {Klauder}(1979)}]{Klauder1979}%
  \BibitemOpen
  \bibfield  {author} {\bibinfo {author} {\bibnamefont {Klauder}, \bibfnamefont
  {J.~R.}}} (\bibinfo {year} {1979}),\ \href@noop {} {\bibfield  {journal}
  {\bibinfo  {journal} {Physical Review D}\ }\textbf {\bibinfo {volume}
  {19}}~(\bibinfo {number} {8}),\ \bibinfo {pages} {2349}}\BibitemShut
  {NoStop}%
\bibitem [{\citenamefont {Knigavko}\ \emph {et~al.}(1999)\citenamefont
  {Knigavko}, \citenamefont {Rosenstein},\ and\ \citenamefont
  {Chen}}]{Knigavko1999}%
  \BibitemOpen
  \bibfield  {author} {\bibinfo {author} {\bibnamefont {Knigavko},
  \bibfnamefont {A.}}, \bibinfo {author} {\bibfnamefont {B.}~\bibnamefont
  {Rosenstein}}, \ and\ \bibinfo {author} {\bibfnamefont {Y.~F.}\ \bibnamefont
  {Chen}}} (\bibinfo {year} {1999}),\ \href@noop {} {\bibfield  {journal}
  {\bibinfo  {journal} {Physical Review B}\ }\textbf {\bibinfo {volume} {60}},\
  \bibinfo {pages} {550}}\BibitemShut {NoStop}%
\bibitem [{\citenamefont {Kohn}(1959)}]{Kohn1959}%
  \BibitemOpen
  \bibfield  {author} {\bibinfo {author} {\bibnamefont {Kohn}, \bibfnamefont
  {W.}}} (\bibinfo {year} {1959}),\ \href {\doibase 10.1103/PhysRev.115.809}
  {\bibfield  {journal} {\bibinfo  {journal} {Physical Review}\ }\textbf
  {\bibinfo {volume} {115}}~(\bibinfo {number} {4}),\ \bibinfo {pages}
  {809}}\BibitemShut {NoStop}%
\bibitem [{\citenamefont {Kong}\ and\ \citenamefont {Zang}(2013)}]{Kong2013}%
  \BibitemOpen
  \bibfield  {author} {\bibinfo {author} {\bibnamefont {Kong}, \bibfnamefont
  {L.}}, \ and\ \bibinfo {author} {\bibfnamefont {J.}~\bibnamefont {Zang}}}
  (\bibinfo {year} {2013}),\ \href {\doibase 10.1103/PhysRevLett.111.067203}
  {\bibfield  {journal} {\bibinfo  {journal} {Physical Review Letters}\
  }\textbf {\bibinfo {volume} {111}}~(\bibinfo {number} {6}),\ \bibinfo {pages}
  {067203}},\ \Eprint {http://arxiv.org/abs/1308.2343} {1308.2343} \BibitemShut
  {NoStop}%
\bibitem [{\citenamefont {Koshibae}\ and\ \citenamefont
  {Nagaosa}(2014)}]{Koshibae2014}%
  \BibitemOpen
  \bibfield  {author} {\bibinfo {author} {\bibnamefont {Koshibae},
  \bibfnamefont {W.}}, \ and\ \bibinfo {author} {\bibfnamefont
  {N.}~\bibnamefont {Nagaosa}}} (\bibinfo {year} {2014}),\ \href {\doibase
  10.1038/ncomms6148} {\bibfield  {journal} {\bibinfo  {journal} {Nature
  Communications}\ }\textbf {\bibinfo {volume} {5}},\ \bibinfo {pages}
  {5148}}\BibitemShut {NoStop}%
\bibitem [{\citenamefont {Kovner}(1989)}]{Kovner1989}%
  \BibitemOpen
  \bibfield  {author} {\bibinfo {author} {\bibnamefont {Kovner}, \bibfnamefont
  {A.}}} (\bibinfo {year} {1989}),\ \href {\doibase 10.1103/PhysRevA.40.545}
  {\bibfield  {journal} {\bibinfo  {journal} {Physical Review A}\ }\textbf
  {\bibinfo {volume} {40}}~(\bibinfo {number} {2}),\ \bibinfo {pages}
  {545}}\BibitemShut {NoStop}%
\bibitem [{\citenamefont {Kravchuk}\ \emph {et~al.}(2018)\citenamefont
  {Kravchuk}, \citenamefont {Sheka}, \citenamefont {R\"o\ss~ler}, \citenamefont
  {van~den Brink},\ and\ \citenamefont {Gaididei}}]{Kravchuk2018}%
  \BibitemOpen
  \bibfield  {author} {\bibinfo {author} {\bibnamefont {Kravchuk},
  \bibfnamefont {V.~P.}}, \bibinfo {author} {\bibfnamefont {D.~D.}\
  \bibnamefont {Sheka}}, \bibinfo {author} {\bibfnamefont {U.~K.}\ \bibnamefont
  {R\"o\ss~ler}}, \bibinfo {author} {\bibfnamefont {J.}~\bibnamefont {van~den
  Brink}}, \ and\ \bibinfo {author} {\bibfnamefont {Y.}~\bibnamefont
  {Gaididei}}} (\bibinfo {year} {2018}),\ \href@noop {} {\bibfield  {journal}
  {\bibinfo  {journal} {Physical Review B}\ }\textbf {\bibinfo {volume} {97}},\
  \bibinfo {pages} {064403}}\BibitemShut {NoStop}%
\bibitem [{\citenamefont {Landau}\ and\ \citenamefont
  {Lifshitz}(1935)}]{Landau1}%
  \BibitemOpen
  \bibfield  {author} {\bibinfo {author} {\bibnamefont {Landau}, \bibfnamefont
  {L.~D.}}, \ and\ \bibinfo {author} {\bibfnamefont {E.~M.}\ \bibnamefont
  {Lifshitz}}} (\bibinfo {year} {1935}),\ \href@noop {} {\bibfield  {journal}
  {\bibinfo  {journal} {Phys. Zeitsch. der Sow}\ }\textbf {\bibinfo {volume}
  {8}},\ \bibinfo {pages} {153}}\BibitemShut {NoStop}%
\bibitem [{\citenamefont {Landau}\ \emph {et~al.}(1980)\citenamefont {Landau},
  \citenamefont {Lifshitz},\ and\ \citenamefont {Pitaevskii}}]{Landau2}%
  \BibitemOpen
  \bibfield  {author} {\bibinfo {author} {\bibnamefont {Landau}, \bibfnamefont
  {L.~D.}}, \bibinfo {author} {\bibfnamefont {E.~M.}\ \bibnamefont {Lifshitz}},
  \ and\ \bibinfo {author} {\bibfnamefont {L.~P.}\ \bibnamefont {Pitaevskii}}}
  (\bibinfo {year} {1980}),\ \href@noop {} {\emph {\bibinfo {title}
  {Statistical Physics, Part 2}}}\ (\bibinfo  {publisher} {Pergamon},\ \bibinfo
  {address} {Oxford})\BibitemShut {NoStop}%
\bibitem [{\citenamefont {Lee}\ \emph {et~al.}(2015)\citenamefont {Lee},
  \citenamefont {Han},\ and\ \citenamefont {Lee}}]{Lee2015}%
  \BibitemOpen
  \bibfield  {author} {\bibinfo {author} {\bibnamefont {Lee}, \bibfnamefont
  {H.}}, \bibinfo {author} {\bibfnamefont {J.~H.}\ \bibnamefont {Han}}, \ and\
  \bibinfo {author} {\bibfnamefont {P.~A.}\ \bibnamefont {Lee}}} (\bibinfo
  {year} {2015}),\ \href {\doibase 10.1103/PhysRevB.91.125413} {\bibfield
  {journal} {\bibinfo  {journal} {Physical Review B}\ }\textbf {\bibinfo
  {volume} {91}}~(\bibinfo {number} {12}),\ \bibinfo {pages}
  {125413}}\BibitemShut {NoStop}%
\bibitem [{\citenamefont {Leonov}\ and\ \citenamefont
  {Mostovoy}(2015)}]{Leonov2015}%
  \BibitemOpen
  \bibfield  {author} {\bibinfo {author} {\bibnamefont {Leonov}, \bibfnamefont
  {A.~O.}}, \ and\ \bibinfo {author} {\bibfnamefont {M.}~\bibnamefont
  {Mostovoy}}} (\bibinfo {year} {2015}),\ \href {\doibase 10.1038/ncomms9275}
  {\bibfield  {journal} {\bibinfo  {journal} {Nature Communications}\ }\textbf
  {\bibinfo {volume} {6}},\ \bibinfo {pages} {8275}}\BibitemShut {NoStop}%
\bibitem [{\citenamefont {Li}\ \emph {et~al.}(2014)\citenamefont {Li},
  \citenamefont {Kharzeev}, \citenamefont {Zhang}, \citenamefont {Huang},
  \citenamefont {Pletikosic}, \citenamefont {Fedorov}, \citenamefont {Zhong},
  \citenamefont {Schneeloch}, \citenamefont {Gu},\ and\ \citenamefont
  {Valla}}]{Li2014}%
  \BibitemOpen
  \bibfield  {author} {\bibinfo {author} {\bibnamefont {Li}, \bibfnamefont
  {Q.}}, \bibinfo {author} {\bibfnamefont {D.~E.}\ \bibnamefont {Kharzeev}},
  \bibinfo {author} {\bibfnamefont {C.}~\bibnamefont {Zhang}}, \bibinfo
  {author} {\bibfnamefont {Y.}~\bibnamefont {Huang}}, \bibinfo {author}
  {\bibfnamefont {I.}~\bibnamefont {Pletikosic}}, \bibinfo {author}
  {\bibfnamefont {A.~V.}\ \bibnamefont {Fedorov}}, \bibinfo {author}
  {\bibfnamefont {R.~D.}\ \bibnamefont {Zhong}}, \bibinfo {author}
  {\bibfnamefont {J.~A.}\ \bibnamefont {Schneeloch}}, \bibinfo {author}
  {\bibfnamefont {G.~D.}\ \bibnamefont {Gu}}, \ and\ \bibinfo {author}
  {\bibfnamefont {T.}~\bibnamefont {Valla}}} (\bibinfo {year} {2014}),\ \href
  {\doibase 10.1038/nphys3648} {\bibfield  {journal} {\bibinfo  {journal}
  {Nature Physics}\ }\textbf {\bibinfo {volume} {12}}~(\bibinfo {number}
  {June}),\ \bibinfo {pages} {550}}\BibitemShut {NoStop}%
\bibitem [{\citenamefont {Li}\ \emph {et~al.}(2009)\citenamefont {Li},
  \citenamefont {Toner},\ and\ \citenamefont {Belitz}}]{Li2009}%
  \BibitemOpen
  \bibfield  {author} {\bibinfo {author} {\bibnamefont {Li}, \bibfnamefont
  {Q.}}, \bibinfo {author} {\bibfnamefont {J.}~\bibnamefont {Toner}}, \ and\
  \bibinfo {author} {\bibfnamefont {D.}~\bibnamefont {Belitz}}} (\bibinfo
  {year} {2009}),\ \href {\doibase 10.1103/PhysRevB.79.014517} {\bibfield
  {journal} {\bibinfo  {journal} {Physical Review B}\ }\textbf {\bibinfo
  {volume} {79}}~(\bibinfo {number} {1}),\ \bibinfo {pages}
  {014517}}\BibitemShut {NoStop}%
\bibitem [{\citenamefont {Li}\ \emph {et~al.}(2013)\citenamefont {Li},
  \citenamefont {Kanazawa}, \citenamefont {Yu}, \citenamefont {Tsukazaki},
  \citenamefont {Kawasaki}, \citenamefont {Ichikawa}, \citenamefont {Jin},
  \citenamefont {Kagawa},\ and\ \citenamefont {Tokura}}]{Li2013}%
  \BibitemOpen
  \bibfield  {author} {\bibinfo {author} {\bibnamefont {Li}, \bibfnamefont
  {Y.}}, \bibinfo {author} {\bibfnamefont {N.}~\bibnamefont {Kanazawa}},
  \bibinfo {author} {\bibfnamefont {X.~Z.}\ \bibnamefont {Yu}}, \bibinfo
  {author} {\bibfnamefont {A.}~\bibnamefont {Tsukazaki}}, \bibinfo {author}
  {\bibfnamefont {M.}~\bibnamefont {Kawasaki}}, \bibinfo {author}
  {\bibfnamefont {M.}~\bibnamefont {Ichikawa}}, \bibinfo {author}
  {\bibfnamefont {X.~F.}\ \bibnamefont {Jin}}, \bibinfo {author} {\bibfnamefont
  {F.}~\bibnamefont {Kagawa}}, \ and\ \bibinfo {author} {\bibfnamefont
  {Y.}~\bibnamefont {Tokura}}} (\bibinfo {year} {2013}),\ \href {\doibase
  10.1103/PhysRevLett.110.117202} {\bibfield  {journal} {\bibinfo  {journal}
  {Physical Review Letters}\ }\textbf {\bibinfo {volume} {110}}~(\bibinfo
  {number} {11}),\ \bibinfo {pages} {117202}}\BibitemShut {NoStop}%
\bibitem [{\citenamefont {Lin}(2017)}]{Lin2017}%
  \BibitemOpen
  \bibfield  {author} {\bibinfo {author} {\bibnamefont {Lin}, \bibfnamefont
  {S.~Z.}}} (\bibinfo {year} {2017}),\ \href@noop {} {\bibfield  {journal}
  {\bibinfo  {journal} {Physical Review B}\ }\textbf {\bibinfo {volume} {96}},\
  \bibinfo {pages} {014407}}\BibitemShut {NoStop}%
\bibitem [{\citenamefont {Lin}\ \emph {et~al.}(2014)\citenamefont {Lin},
  \citenamefont {Batista}, \citenamefont {Reichhardt},\ and\ \citenamefont
  {Saxena}}]{Lin2014}%
  \BibitemOpen
  \bibfield  {author} {\bibinfo {author} {\bibnamefont {Lin}, \bibfnamefont
  {S.~Z.}}, \bibinfo {author} {\bibfnamefont {C.~D.}\ \bibnamefont {Batista}},
  \bibinfo {author} {\bibfnamefont {C.}~\bibnamefont {Reichhardt}}, \ and\
  \bibinfo {author} {\bibfnamefont {A.}~\bibnamefont {Saxena}}} (\bibinfo
  {year} {2014}),\ \href {\doibase 10.1103/PhysRevLett.112.187203} {\bibfield
  {journal} {\bibinfo  {journal} {Physical Review Letters}\ }\textbf {\bibinfo
  {volume} {112}}~(\bibinfo {number} {18}),\ \bibinfo {pages}
  {187203}}\BibitemShut {NoStop}%
\bibitem [{\citenamefont {Lin}\ and\ \citenamefont
  {Bulaevskii}(2013)}]{Lin2013}%
  \BibitemOpen
  \bibfield  {author} {\bibinfo {author} {\bibnamefont {Lin}, \bibfnamefont
  {S.~Z.}}, \ and\ \bibinfo {author} {\bibfnamefont {L.~N.}\ \bibnamefont
  {Bulaevskii}}} (\bibinfo {year} {2013}),\ \href {\doibase
  10.1103/PhysRevB.88.060404} {\bibfield  {journal} {\bibinfo  {journal}
  {Physical Review B}\ }\textbf {\bibinfo {volume} {88}}~(\bibinfo {number}
  {6}),\ \bibinfo {pages} {060404}}\BibitemShut {NoStop}%
\bibitem [{\citenamefont {Lin}\ and\ \citenamefont {Hayami}(2016)}]{Lin2016}%
  \BibitemOpen
  \bibfield  {author} {\bibinfo {author} {\bibnamefont {Lin}, \bibfnamefont
  {S.~Z.}}, \ and\ \bibinfo {author} {\bibfnamefont {S.}~\bibnamefont
  {Hayami}}} (\bibinfo {year} {2016}),\ \href {\doibase
  10.1103/PhysRevB.93.064430} {\bibfield  {journal} {\bibinfo  {journal}
  {Physical Review B}\ }\textbf {\bibinfo {volume} {93}}~(\bibinfo {number}
  {6}),\ \bibinfo {pages} {064430}}\BibitemShut {NoStop}%
\bibitem [{\citenamefont {Litzius}\ \emph {et~al.}(2016)\citenamefont
  {Litzius}, \citenamefont {Lemesh}, \citenamefont {Kr{\"{u}}ger},
  \citenamefont {Bassirian}, \citenamefont {Caretta}, \citenamefont {Richter},
  \citenamefont {B{\"{u}}ttner}, \citenamefont {Sato}, \citenamefont
  {Tretiakov}, \citenamefont {F{\"{o}}rster}, \citenamefont {Reeve},
  \citenamefont {Weigand}, \citenamefont {Bykova}, \citenamefont {Stoll},
  \citenamefont {Sch{\"{u}}tz}, \citenamefont {Beach},\ and\ \citenamefont
  {Kl{\"{a}}ui}}]{Litzius2016}%
  \BibitemOpen
  \bibfield  {author} {\bibinfo {author} {\bibnamefont {Litzius}, \bibfnamefont
  {K.}}, \bibinfo {author} {\bibfnamefont {I.}~\bibnamefont {Lemesh}}, \bibinfo
  {author} {\bibfnamefont {B.}~\bibnamefont {Kr{\"{u}}ger}}, \bibinfo {author}
  {\bibfnamefont {P.}~\bibnamefont {Bassirian}}, \bibinfo {author}
  {\bibfnamefont {L.}~\bibnamefont {Caretta}}, \bibinfo {author} {\bibfnamefont
  {K.}~\bibnamefont {Richter}}, \bibinfo {author} {\bibfnamefont
  {F.}~\bibnamefont {B{\"{u}}ttner}}, \bibinfo {author} {\bibfnamefont
  {K.}~\bibnamefont {Sato}}, \bibinfo {author} {\bibfnamefont {O.~A.}\
  \bibnamefont {Tretiakov}}, \bibinfo {author} {\bibfnamefont {J.}~\bibnamefont
  {F{\"{o}}rster}}, \bibinfo {author} {\bibfnamefont {R.~M.}\ \bibnamefont
  {Reeve}}, \bibinfo {author} {\bibfnamefont {M.}~\bibnamefont {Weigand}},
  \bibinfo {author} {\bibfnamefont {I.}~\bibnamefont {Bykova}}, \bibinfo
  {author} {\bibfnamefont {H.}~\bibnamefont {Stoll}}, \bibinfo {author}
  {\bibfnamefont {G.}~\bibnamefont {Sch{\"{u}}tz}}, \bibinfo {author}
  {\bibfnamefont {G.~S.~D.}\ \bibnamefont {Beach}}, \ and\ \bibinfo {author}
  {\bibfnamefont {M.}~\bibnamefont {Kl{\"{a}}ui}}} (\bibinfo {year} {2016}),\
  \href {\doibase 10.1038/nphys4000} {\bibfield  {journal} {\bibinfo  {journal}
  {Nature Physics}\ }\textbf {\bibinfo {volume} {13}}~(\bibinfo {number} {2}),\
  \bibinfo {pages} {170}}\BibitemShut {NoStop}%
\bibitem [{\citenamefont {Lu}\ and\ \citenamefont {Herbut}(2012)}]{Lu2012}%
  \BibitemOpen
  \bibfield  {author} {\bibinfo {author} {\bibnamefont {Lu}, \bibfnamefont
  {C.~K.}}, \ and\ \bibinfo {author} {\bibfnamefont {I.~F.}\ \bibnamefont
  {Herbut}}} (\bibinfo {year} {2012}),\ \href {\doibase
  10.1103/PhysRevLett.108.266402} {\bibfield  {journal} {\bibinfo  {journal}
  {Physical Review Letters}\ }\textbf {\bibinfo {volume} {108}}~(\bibinfo
  {number} {26}),\ \bibinfo {pages} {266402}},\ \Eprint
  {http://arxiv.org/abs/1202.6047} {1202.6047} \BibitemShut {NoStop}%
\bibitem [{\citenamefont {Makhfudz}\ \emph {et~al.}(2012)\citenamefont
  {Makhfudz}, \citenamefont {Kruger},\ and\ \citenamefont
  {Tchernyshyov}}]{Makhfudz2012}%
  \BibitemOpen
  \bibfield  {author} {\bibinfo {author} {\bibnamefont {Makhfudz},
  \bibfnamefont {I.}}, \bibinfo {author} {\bibfnamefont {B.}~\bibnamefont
  {Kruger}}, \ and\ \bibinfo {author} {\bibfnamefont {O.}~\bibnamefont
  {Tchernyshyov}}} (\bibinfo {year} {2012}),\ \href {\doibase
  10.1103/PhysRevLett.109.217201} {\bibfield  {journal} {\bibinfo  {journal}
  {Physical Review Letters}\ }\textbf {\bibinfo {volume} {109}}~(\bibinfo
  {number} {21}),\ \bibinfo {pages} {217201}}\BibitemShut {NoStop}%
\bibitem [{\citenamefont {Matsumoto}\ and\ \citenamefont
  {Murakami}(2011{\natexlab{a}})}]{Matsumoto2011}%
  \BibitemOpen
  \bibfield  {author} {\bibinfo {author} {\bibnamefont {Matsumoto},
  \bibfnamefont {R.}}, \ and\ \bibinfo {author} {\bibfnamefont
  {S.}~\bibnamefont {Murakami}}} (\bibinfo {year} {2011}{\natexlab{a}}),\ \href
  {\doibase 10.1103/PhysRevB.84.184406} {\bibfield  {journal} {\bibinfo
  {journal} {Physical Review B}\ }\textbf {\bibinfo {volume} {84}}~(\bibinfo
  {number} {18}),\ \bibinfo {pages} {184406}}\BibitemShut {NoStop}%
\bibitem [{\citenamefont {Matsumoto}\ and\ \citenamefont
  {Murakami}(2011{\natexlab{b}})}]{Matsumoto2011a}%
  \BibitemOpen
  \bibfield  {author} {\bibinfo {author} {\bibnamefont {Matsumoto},
  \bibfnamefont {R.}}, \ and\ \bibinfo {author} {\bibfnamefont
  {S.}~\bibnamefont {Murakami}}} (\bibinfo {year} {2011}{\natexlab{b}}),\ \href
  {\doibase 10.1103/PhysRevLett.106.197202} {\bibfield  {journal} {\bibinfo
  {journal} {Physical Review Letters}\ }\textbf {\bibinfo {volume}
  {106}}~(\bibinfo {number} {19}),\ \bibinfo {pages} {197202}}\BibitemShut
  {NoStop}%
\bibitem [{\citenamefont {Melko}\ \emph {et~al.}(2005)\citenamefont {Melko},
  \citenamefont {Paramekanti}, \citenamefont {Burkov}, \citenamefont
  {Vishwanath}, \citenamefont {Sheng},\ and\ \citenamefont
  {Balents}}]{Melko2005}%
  \BibitemOpen
  \bibfield  {author} {\bibinfo {author} {\bibnamefont {Melko}, \bibfnamefont
  {R.~G.}}, \bibinfo {author} {\bibfnamefont {A.}~\bibnamefont {Paramekanti}},
  \bibinfo {author} {\bibfnamefont {A.~A.}\ \bibnamefont {Burkov}}, \bibinfo
  {author} {\bibfnamefont {A.}~\bibnamefont {Vishwanath}}, \bibinfo {author}
  {\bibfnamefont {D.~N.}\ \bibnamefont {Sheng}}, \ and\ \bibinfo {author}
  {\bibfnamefont {L.}~\bibnamefont {Balents}}} (\bibinfo {year} {2005}),\
  \href@noop {} {\bibfield  {journal} {\bibinfo  {journal} {Physical Review
  Letters}\ }\textbf {\bibinfo {volume} {95}},\ \bibinfo {pages}
  {127207}}\BibitemShut {NoStop}%
\bibitem [{\citenamefont {Mermin}\ and\ \citenamefont {Ho}(1976)}]{Mermin-Ho}%
  \BibitemOpen
  \bibfield  {author} {\bibinfo {author} {\bibnamefont {Mermin}, \bibfnamefont
  {N.~D.}}, \ and\ \bibinfo {author} {\bibfnamefont {T.-L.}\ \bibnamefont
  {Ho}}} (\bibinfo {year} {1976}),\ \href@noop {} {\bibfield  {journal}
  {\bibinfo  {journal} {Physical Review Letters}\ }\textbf {\bibinfo {volume}
  {36}},\ \bibinfo {pages} {594}}\BibitemShut {NoStop}%
\bibitem [{\citenamefont {Mizushima}\ \emph {et~al.}(2002)\citenamefont
  {Mizushima}, \citenamefont {Machida},\ and\ \citenamefont
  {Kita}}]{Mizushima2002}%
  \BibitemOpen
  \bibfield  {author} {\bibinfo {author} {\bibnamefont {Mizushima},
  \bibfnamefont {T.}}, \bibinfo {author} {\bibfnamefont {K.}~\bibnamefont
  {Machida}}, \ and\ \bibinfo {author} {\bibfnamefont {T.}~\bibnamefont
  {Kita}}} (\bibinfo {year} {2002}),\ \href@noop {} {\bibfield  {journal}
  {\bibinfo  {journal} {Physical Review Letters}\ }\textbf {\bibinfo {volume}
  {89}},\ \bibinfo {pages} {030401}}\BibitemShut {NoStop}%
\bibitem [{\citenamefont {Mochizuki}\ \emph {et~al.}(2015)\citenamefont
  {Mochizuki}, \citenamefont {Yu}, \citenamefont {Seki}, \citenamefont
  {Kanazawa}, \citenamefont {Koshibae}, \citenamefont {Zang}, \citenamefont
  {Mostovoy}, \citenamefont {Tokura},\ and\ \citenamefont
  {Nagaosa}}]{Mochizuki2015}%
  \BibitemOpen
  \bibfield  {author} {\bibinfo {author} {\bibnamefont {Mochizuki},
  \bibfnamefont {M.}}, \bibinfo {author} {\bibfnamefont {X.~Z.}\ \bibnamefont
  {Yu}}, \bibinfo {author} {\bibfnamefont {S.}~\bibnamefont {Seki}}, \bibinfo
  {author} {\bibfnamefont {N.}~\bibnamefont {Kanazawa}}, \bibinfo {author}
  {\bibfnamefont {W.}~\bibnamefont {Koshibae}}, \bibinfo {author}
  {\bibfnamefont {J.}~\bibnamefont {Zang}}, \bibinfo {author} {\bibfnamefont
  {M.}~\bibnamefont {Mostovoy}}, \bibinfo {author} {\bibfnamefont
  {Y.}~\bibnamefont {Tokura}}, \ and\ \bibinfo {author} {\bibfnamefont
  {N.}~\bibnamefont {Nagaosa}}} (\bibinfo {year} {2015}),\ \href {\doibase
  10.1038/nmat3862} {\bibfield  {journal} {\bibinfo  {journal} {Nature
  Materials}\ }\textbf {\bibinfo {volume} {13}}~(\bibinfo {number} {March}),\
  \bibinfo {pages} {241}}\BibitemShut {NoStop}%
\bibitem [{\citenamefont {Moon}(2012)}]{Moon2012}%
  \BibitemOpen
  \bibfield  {author} {\bibinfo {author} {\bibnamefont {Moon}, \bibfnamefont
  {E.~G.}}} (\bibinfo {year} {2012}),\ \href {\doibase
  10.1103/PhysRevB.85.245123} {\bibfield  {journal} {\bibinfo  {journal}
  {Physical Review B}\ }\textbf {\bibinfo {volume} {85}}~(\bibinfo {number}
  {24}),\ \bibinfo {pages} {245123}},\ \Eprint
  {http://arxiv.org/abs/arXiv:1202.5389v1} {arXiv:1202.5389v1} \BibitemShut
  {NoStop}%
\bibitem [{\citenamefont {Moriya}(1956)}]{Moriya1956}%
  \BibitemOpen
  \bibfield  {author} {\bibinfo {author} {\bibnamefont {Moriya}, \bibfnamefont
  {T.}}} (\bibinfo {year} {1956}),\ \href {\doibase
  https://doi.org/10.1103/PhysRev.120.91} {\bibfield  {journal} {\bibinfo
  {journal} {Physical Review}\ }\textbf {\bibinfo {volume} {249}}~(\bibinfo
  {number} {1949}),\ \bibinfo {pages} {91}}\BibitemShut {NoStop}%
\bibitem [{\citenamefont {Morrison}(1998)}]{Morrison1998}%
  \BibitemOpen
  \bibfield  {author} {\bibinfo {author} {\bibnamefont {Morrison},
  \bibfnamefont {P.~J.}}} (\bibinfo {year} {1998}),\ \href {\doibase
  10.1103/RevModPhys.70.467} {\bibfield  {journal} {\bibinfo  {journal}
  {Reviews of Modern Physics}\ }\textbf {\bibinfo {volume} {70}}~(\bibinfo
  {number} {2}),\ \bibinfo {pages} {467}}\BibitemShut {NoStop}%
\bibitem [{\citenamefont {Moutafis}\ \emph {et~al.}(2009)\citenamefont
  {Moutafis}, \citenamefont {Komineas},\ and\ \citenamefont
  {Bland}}]{Moutafis2009}%
  \BibitemOpen
  \bibfield  {author} {\bibinfo {author} {\bibnamefont {Moutafis},
  \bibfnamefont {C.}}, \bibinfo {author} {\bibfnamefont {S.}~\bibnamefont
  {Komineas}}, \ and\ \bibinfo {author} {\bibfnamefont {J.~A.~C.}\ \bibnamefont
  {Bland}}} (\bibinfo {year} {2009}),\ \href {\doibase
  10.1103/PhysRevB.79.224429} {\bibfield  {journal} {\bibinfo  {journal}
  {Physical Review B}\ }\textbf {\bibinfo {volume} {79}}~(\bibinfo {number}
  {22}),\ \bibinfo {pages} {224429}}\BibitemShut {NoStop}%
\bibitem [{\citenamefont {M{\"{u}}hlbauer}\ \emph {et~al.}(2009)\citenamefont
  {M{\"{u}}hlbauer}, \citenamefont {Binz}, \citenamefont {Jonietz},
  \citenamefont {Pfleiderer}, \citenamefont {Rosch}, \citenamefont {Neubauer},
  \citenamefont {Georgii},\ and\ \citenamefont {B{\"{o}}ni}}]{Muhlbauer2009}%
  \BibitemOpen
  \bibfield  {author} {\bibinfo {author} {\bibnamefont {M{\"{u}}hlbauer},
  \bibfnamefont {S.}}, \bibinfo {author} {\bibfnamefont {B.}~\bibnamefont
  {Binz}}, \bibinfo {author} {\bibfnamefont {F.}~\bibnamefont {Jonietz}},
  \bibinfo {author} {\bibfnamefont {C.}~\bibnamefont {Pfleiderer}}, \bibinfo
  {author} {\bibfnamefont {A.}~\bibnamefont {Rosch}}, \bibinfo {author}
  {\bibfnamefont {A.}~\bibnamefont {Neubauer}}, \bibinfo {author}
  {\bibfnamefont {R.}~\bibnamefont {Georgii}}, \ and\ \bibinfo {author}
  {\bibfnamefont {P.}~\bibnamefont {B{\"{o}}ni}}} (\bibinfo {year} {2009}),\
  \href {\doibase 10.1126/science.333.6048.1381-b} {\bibfield  {journal}
  {\bibinfo  {journal} {Science}\ }\textbf {\bibinfo {volume} {323}}~(\bibinfo
  {number} {September}),\ \bibinfo {pages} {915}}\BibitemShut {NoStop}%
\bibitem [{\citenamefont {Nayak}\ and\ \citenamefont
  {Wilczek}(1996)}]{Nayak1996}%
  \BibitemOpen
  \bibfield  {author} {\bibinfo {author} {\bibnamefont {Nayak}, \bibfnamefont
  {C.}}, \ and\ \bibinfo {author} {\bibfnamefont {F.}~\bibnamefont {Wilczek}}}
  (\bibinfo {year} {1996}),\ \href
  {http://www.ncbi.nlm.nih.gov/pubmed/10062533} {\bibfield  {journal} {\bibinfo
   {journal} {Physical Review Letters}\ }\textbf {\bibinfo {volume}
  {77}}~(\bibinfo {number} {21}),\ \bibinfo {pages} {4418}}\BibitemShut
  {NoStop}%
\bibitem [{\citenamefont {Nenciu}(1991)}]{Nenciu}%
  \BibitemOpen
  \bibfield  {author} {\bibinfo {author} {\bibnamefont {Nenciu}, \bibfnamefont
  {G.}}} (\bibinfo {year} {1991}),\ \href@noop {} {\bibfield  {journal}
  {\bibinfo  {journal} {Reviews of Modern Physics}\ }\textbf {\bibinfo {volume}
  {63}},\ \bibinfo {pages} {91}}\BibitemShut {NoStop}%
\bibitem [{\citenamefont {Neubauer}\ \emph {et~al.}(2009)\citenamefont
  {Neubauer}, \citenamefont {Pfleiderer}, \citenamefont {Binz}, \citenamefont
  {Rosch}, \citenamefont {Ritz}, \citenamefont {Niklowitz},\ and\ \citenamefont
  {B{\"{o}}ni}}]{Neubauer2009}%
  \BibitemOpen
  \bibfield  {author} {\bibinfo {author} {\bibnamefont {Neubauer},
  \bibfnamefont {A.}}, \bibinfo {author} {\bibfnamefont {C.}~\bibnamefont
  {Pfleiderer}}, \bibinfo {author} {\bibfnamefont {B.}~\bibnamefont {Binz}},
  \bibinfo {author} {\bibfnamefont {A.}~\bibnamefont {Rosch}}, \bibinfo
  {author} {\bibfnamefont {R.}~\bibnamefont {Ritz}}, \bibinfo {author}
  {\bibfnamefont {P.~G.}\ \bibnamefont {Niklowitz}}, \ and\ \bibinfo {author}
  {\bibfnamefont {P.}~\bibnamefont {B{\"{o}}ni}}} (\bibinfo {year} {2009}),\
  \href@noop {} {\bibfield  {journal} {\bibinfo  {journal} {Physical Review
  Letters}\ }\textbf {\bibinfo {volume} {102}},\ \bibinfo {pages}
  {186602}}\BibitemShut {NoStop}%
\bibitem [{\citenamefont {von Neumann}(1955)}]{vonNeumann}%
  \BibitemOpen
  \bibfield  {author} {\bibinfo {author} {\bibnamefont {von Neumann},
  \bibfnamefont {J.}}} (\bibinfo {year} {1955}),\ \href@noop {} {\emph
  {\bibinfo {title} {Mathematical Foundation of Quantum Mechanics}}}\ (\bibinfo
   {publisher} {Princeton University Press},\ \bibinfo {address}
  {Princeton})\BibitemShut {NoStop}%
\bibitem [{\citenamefont {Nielsen}\ and\ \citenamefont
  {Ninomiya}(1981)}]{Nielsen1981}%
  \BibitemOpen
  \bibfield  {author} {\bibinfo {author} {\bibnamefont {Nielsen}, \bibfnamefont
  {H.~B.}}, \ and\ \bibinfo {author} {\bibfnamefont {M.}~\bibnamefont
  {Ninomiya}}} (\bibinfo {year} {1981}),\ \href {\doibase
  10.1016/0370-2693(81)91026-1} {\bibfield  {journal} {\bibinfo  {journal}
  {Physics Letters B}\ }\textbf {\bibinfo {volume} {105}}~(\bibinfo {number}
  {2-3}),\ \bibinfo {pages} {219}}\BibitemShut {NoStop}%
\bibitem [{\citenamefont {Nielsen}\ and\ \citenamefont
  {Ninomiya}(1983)}]{Nielsen1983}%
  \BibitemOpen
  \bibfield  {author} {\bibinfo {author} {\bibnamefont {Nielsen}, \bibfnamefont
  {H.~B.}}, \ and\ \bibinfo {author} {\bibfnamefont {M.}~\bibnamefont
  {Ninomiya}}} (\bibinfo {year} {1983}),\ \href {\doibase
  10.1016/0370-2693(83)91529-0} {\bibfield  {journal} {\bibinfo  {journal}
  {Physics Letters B}\ }\textbf {\bibinfo {volume} {130}}~(\bibinfo {number}
  {6}),\ \bibinfo {pages} {389}}\BibitemShut {NoStop}%
\bibitem [{\citenamefont {Nikiforov}\ and\ \citenamefont
  {Sonin}(1983)}]{Nikiforov1983}%
  \BibitemOpen
  \bibfield  {author} {\bibinfo {author} {\bibnamefont {Nikiforov},
  \bibfnamefont {A.~V.}}, \ and\ \bibinfo {author} {\bibfnamefont {E.~B.}\
  \bibnamefont {Sonin}}} (\bibinfo {year} {1983}),\ \href
  {http://www.jetp.ac.ru/cgi-bin/r/index/r/85/2/p642?a=list{\%}5Cnhttp://www.jetp.ac.ru/cgi-bin/r/index/e/58/2/p373?a=list}
  {\bibfield  {journal} {\bibinfo  {journal} {Sov. Phys. JETP}\ }\textbf
  {\bibinfo {volume} {58}}~(\bibinfo {number} {2}),\ \bibinfo {pages}
  {373}}\BibitemShut {NoStop}%
\bibitem [{\citenamefont {Nomura}\ and\ \citenamefont
  {Nagaosa}(2010)}]{Nomura2010}%
  \BibitemOpen
  \bibfield  {author} {\bibinfo {author} {\bibnamefont {Nomura}, \bibfnamefont
  {K.}}, \ and\ \bibinfo {author} {\bibfnamefont {N.}~\bibnamefont {Nagaosa}}}
  (\bibinfo {year} {2010}),\ \href {\doibase 10.1103/PhysRevB.82.161401}
  {\bibfield  {journal} {\bibinfo  {journal} {Physical Review B}\ }\textbf
  {\bibinfo {volume} {82}},\ \bibinfo {pages} {161401(R)}}\BibitemShut
  {NoStop}%
\bibitem [{\citenamefont {Okubo}\ \emph {et~al.}(2012)\citenamefont {Okubo},
  \citenamefont {Chung},\ and\ \citenamefont {Kawamura}}]{Okubo2012}%
  \BibitemOpen
  \bibfield  {author} {\bibinfo {author} {\bibnamefont {Okubo}, \bibfnamefont
  {T.}}, \bibinfo {author} {\bibfnamefont {S.}~\bibnamefont {Chung}}, \ and\
  \bibinfo {author} {\bibfnamefont {H.}~\bibnamefont {Kawamura}}} (\bibinfo
  {year} {2012}),\ \href {\doibase 10.1103/PhysRevLett.108.017206} {\bibfield
  {journal} {\bibinfo  {journal} {Physical Review Letters}\ }\textbf {\bibinfo
  {volume} {108}}~(\bibinfo {number} {1}),\ \bibinfo {pages}
  {017206}}\BibitemShut {NoStop}%
\bibitem [{\citenamefont {Papanicolaou}\ and\ \citenamefont
  {Tomaras}(1991)}]{Papanicolaou1991}%
  \BibitemOpen
  \bibfield  {author} {\bibinfo {author} {\bibnamefont {Papanicolaou},
  \bibfnamefont {N.}}, \ and\ \bibinfo {author} {\bibfnamefont {T.~N.}\
  \bibnamefont {Tomaras}}} (\bibinfo {year} {1991}),\ \href@noop {} {\bibfield
  {journal} {\bibinfo  {journal} {Nuclear Physics B}\ }\textbf {\bibinfo
  {volume} {360}},\ \bibinfo {pages} {425}}\BibitemShut {NoStop}%
\bibitem [{\citenamefont {Psaroudaki}\ \emph {et~al.}(2017)\citenamefont
  {Psaroudaki}, \citenamefont {Hoffman}, \citenamefont {Klinovaja},\ and\
  \citenamefont {Loss}}]{Psaroudaki2016}%
  \BibitemOpen
  \bibfield  {author} {\bibinfo {author} {\bibnamefont {Psaroudaki},
  \bibfnamefont {C.}}, \bibinfo {author} {\bibfnamefont {S.}~\bibnamefont
  {Hoffman}}, \bibinfo {author} {\bibfnamefont {J.}~\bibnamefont {Klinovaja}},
  \ and\ \bibinfo {author} {\bibfnamefont {D.}~\bibnamefont {Loss}}} (\bibinfo
  {year} {2017}),\ \href {http://arxiv.org/abs/1612.01885} {\bibfield
  {journal} {\bibinfo  {journal} {Physical Review X}\ }\textbf {\bibinfo
  {volume} {7}},\ \bibinfo {pages} {041045}}\BibitemShut {NoStop}%
\bibitem [{\citenamefont {Roessler}\ \emph {et~al.}(2006)\citenamefont
  {Roessler}, \citenamefont {Bogdanov},\ and\ \citenamefont
  {Pfleiderer}}]{Roessler2006}%
  \BibitemOpen
  \bibfield  {author} {\bibinfo {author} {\bibnamefont {Roessler},
  \bibfnamefont {U.~K.}}, \bibinfo {author} {\bibfnamefont {A.~N.}\
  \bibnamefont {Bogdanov}}, \ and\ \bibinfo {author} {\bibfnamefont
  {C.}~\bibnamefont {Pfleiderer}}} (\bibinfo {year} {2006}),\ \href {\doibase
  10.1038/nature05056} {\bibfield  {journal} {\bibinfo  {journal} {Nature}\
  }\textbf {\bibinfo {volume} {442}}~(\bibinfo {number} {August}),\ \bibinfo
  {pages} {797}}\BibitemShut {NoStop}%
\bibitem [{\citenamefont {Rold\'an-Molina}\ \emph {et~al.}(2015)\citenamefont
  {Rold\'an-Molina}, \citenamefont {Santander}, \citenamefont {Nunez},\ and\
  \citenamefont {Fern\'andez-Rossier}}]{Roldan2015}%
  \BibitemOpen
  \bibfield  {author} {\bibinfo {author} {\bibnamefont {Rold\'an-Molina},
  \bibfnamefont {A.}}, \bibinfo {author} {\bibfnamefont {M.~J.}\ \bibnamefont
  {Santander}}, \bibinfo {author} {\bibfnamefont {A.~S.}\ \bibnamefont
  {Nunez}}, \ and\ \bibinfo {author} {\bibfnamefont {J.}~\bibnamefont
  {Fern\'andez-Rossier}}} (\bibinfo {year} {2015}),\ \href@noop {} {\bibfield
  {journal} {\bibinfo  {journal} {Physical Review B}\ }\textbf {\bibinfo
  {volume} {92}},\ \bibinfo {pages} {245436}}\BibitemShut {NoStop}%
\bibitem [{\citenamefont {Romming}\ \emph {et~al.}(2013)\citenamefont
  {Romming}, \citenamefont {Hanneken}, \citenamefont {Menzel}, \citenamefont
  {Bickel}, \citenamefont {Wolter}, \citenamefont {von Bergmann}, \citenamefont
  {Kubetzka},\ and\ \citenamefont {Wiesendanger}}]{Romming2013}%
  \BibitemOpen
  \bibfield  {author} {\bibinfo {author} {\bibnamefont {Romming}, \bibfnamefont
  {N.}}, \bibinfo {author} {\bibfnamefont {C.}~\bibnamefont {Hanneken}},
  \bibinfo {author} {\bibfnamefont {M.}~\bibnamefont {Menzel}}, \bibinfo
  {author} {\bibfnamefont {J.~E.}\ \bibnamefont {Bickel}}, \bibinfo {author}
  {\bibfnamefont {B.}~\bibnamefont {Wolter}}, \bibinfo {author} {\bibfnamefont
  {K.}~\bibnamefont {von Bergmann}}, \bibinfo {author} {\bibfnamefont
  {A.}~\bibnamefont {Kubetzka}}, \ and\ \bibinfo {author} {\bibfnamefont
  {R.}~\bibnamefont {Wiesendanger}}} (\bibinfo {year} {2013}),\ \href {\doibase
  10.1038/nphys2045} {\bibfield  {journal} {\bibinfo  {journal} {Science}\
  }\textbf {\bibinfo {volume} {341}}~(\bibinfo {number} {August}),\ \bibinfo
  {pages} {636}}\BibitemShut {NoStop}%
\bibitem [{\citenamefont {Salomaa}\ and\ \citenamefont
  {Volovik}(1987)}]{Salomaa1987}%
  \BibitemOpen
  \bibfield  {author} {\bibinfo {author} {\bibnamefont {Salomaa}, \bibfnamefont
  {M.~M.}}, \ and\ \bibinfo {author} {\bibfnamefont {G.~E.}\ \bibnamefont
  {Volovik}}} (\bibinfo {year} {1987}),\ \href {\doibase
  10.1103/RevModPhys.59.533} {\bibfield  {journal} {\bibinfo  {journal}
  {Reviews of Modern Physics}\ }\textbf {\bibinfo {volume} {59}}~(\bibinfo
  {number} {3}),\ \bibinfo {pages} {533}}\BibitemShut {NoStop}%
\bibitem [{\citenamefont {Schmeller}\ \emph {et~al.}(1995)\citenamefont
  {Schmeller}, \citenamefont {Eisenstein}, \citenamefont {Pfeiffer},\ and\
  \citenamefont {West}}]{Schmeller1995}%
  \BibitemOpen
  \bibfield  {author} {\bibinfo {author} {\bibnamefont {Schmeller},
  \bibfnamefont {A.}}, \bibinfo {author} {\bibfnamefont {J.~P.}\ \bibnamefont
  {Eisenstein}}, \bibinfo {author} {\bibfnamefont {L.~N.}\ \bibnamefont
  {Pfeiffer}}, \ and\ \bibinfo {author} {\bibfnamefont {K.~W.}\ \bibnamefont
  {West}}} (\bibinfo {year} {1995}),\ \href@noop {} {\bibfield  {journal}
  {\bibinfo  {journal} {Physical Review Letters}\ }\textbf {\bibinfo {volume}
  {75}}~(\bibinfo {number} {23}),\ \bibinfo {pages} {4290}}\BibitemShut
  {NoStop}%
\bibitem [{\citenamefont {Sch{\"{u}}tte}\ and\ \citenamefont
  {Garst}(2014)}]{Schutte2014}%
  \BibitemOpen
  \bibfield  {author} {\bibinfo {author} {\bibnamefont {Sch{\"{u}}tte},
  \bibfnamefont {C.}}, \ and\ \bibinfo {author} {\bibfnamefont
  {M.}~\bibnamefont {Garst}}} (\bibinfo {year} {2014}),\ \href {\doibase
  10.1103/PhysRevB.90.094423} {\bibfield  {journal} {\bibinfo  {journal}
  {Physical Review B}\ }\textbf {\bibinfo {volume} {90}}~(\bibinfo {number}
  {9}),\ \bibinfo {pages} {094423}}\BibitemShut {NoStop}%
\bibitem [{\citenamefont {Sedrakyan}\ \emph {et~al.}(2014)\citenamefont
  {Sedrakyan}, \citenamefont {Glazman},\ and\ \citenamefont
  {Kamenev}}]{Sedrakyan2014}%
  \BibitemOpen
  \bibfield  {author} {\bibinfo {author} {\bibnamefont {Sedrakyan},
  \bibfnamefont {T.~A.}}, \bibinfo {author} {\bibfnamefont {L.~I.}\
  \bibnamefont {Glazman}}, \ and\ \bibinfo {author} {\bibfnamefont
  {A.}~\bibnamefont {Kamenev}}} (\bibinfo {year} {2014}),\ \href@noop {}
  {\bibfield  {journal} {\bibinfo  {journal} {Physical Review B}\ }\textbf
  {\bibinfo {volume} {89}},\ \bibinfo {pages} {201112(R)}}\BibitemShut
  {NoStop}%
\bibitem [{\citenamefont {Seki}\ \emph
  {et~al.}(2012{\natexlab{a}})\citenamefont {Seki}, \citenamefont {Kim},
  \citenamefont {Inosov}, \citenamefont {Georgii}, \citenamefont {Keimer},
  \citenamefont {Ishiwata},\ and\ \citenamefont {Tokura}}]{Seki2012b}%
  \BibitemOpen
  \bibfield  {author} {\bibinfo {author} {\bibnamefont {Seki}, \bibfnamefont
  {S.}}, \bibinfo {author} {\bibfnamefont {J.-H.}\ \bibnamefont {Kim}},
  \bibinfo {author} {\bibfnamefont {D.~S.}\ \bibnamefont {Inosov}}, \bibinfo
  {author} {\bibfnamefont {R.}~\bibnamefont {Georgii}}, \bibinfo {author}
  {\bibfnamefont {B.}~\bibnamefont {Keimer}}, \bibinfo {author} {\bibfnamefont
  {S.}~\bibnamefont {Ishiwata}}, \ and\ \bibinfo {author} {\bibfnamefont
  {Y.}~\bibnamefont {Tokura}}} (\bibinfo {year} {2012}{\natexlab{a}}),\
  \href@noop {} {\bibfield  {journal} {\bibinfo  {journal} {Physical Review B}\
  }\textbf {\bibinfo {volume} {85}},\ \bibinfo {pages} {220406(R)}}\BibitemShut
  {NoStop}%
\bibitem [{\citenamefont {Seki}\ \emph
  {et~al.}(2012{\natexlab{b}})\citenamefont {Seki}, \citenamefont {Yu},
  \citenamefont {Ishiwata},\ and\ \citenamefont {Tokura}}]{Seki2012}%
  \BibitemOpen
  \bibfield  {author} {\bibinfo {author} {\bibnamefont {Seki}, \bibfnamefont
  {S.}}, \bibinfo {author} {\bibfnamefont {X.~Z.}\ \bibnamefont {Yu}}, \bibinfo
  {author} {\bibfnamefont {S.}~\bibnamefont {Ishiwata}}, \ and\ \bibinfo
  {author} {\bibfnamefont {Y.}~\bibnamefont {Tokura}}} (\bibinfo {year}
  {2012}{\natexlab{b}}),\ \href {\doibase 10.1126/science.1214143} {\bibfield
  {journal} {\bibinfo  {journal} {Science}\ }\textbf {\bibinfo {volume}
  {336}}~(\bibinfo {number} {April}),\ \bibinfo {pages} {198}}\BibitemShut
  {NoStop}%
\bibitem [{\citenamefont {Shankar}(1977)}]{Shankar1977}%
  \BibitemOpen
  \bibfield  {author} {\bibinfo {author} {\bibnamefont {Shankar}, \bibfnamefont
  {R.}}} (\bibinfo {year} {1977}),\ \href@noop {} {\bibfield  {journal}
  {\bibinfo  {journal} {Journal de Physique}\ }\textbf {\bibinfo {volume}
  {38}},\ \bibinfo {pages} {1405}}\BibitemShut {NoStop}%
\bibitem [{\citenamefont {Skyrme}(1962)}]{skyrme1962}%
  \BibitemOpen
  \bibfield  {author} {\bibinfo {author} {\bibnamefont {Skyrme}, \bibfnamefont
  {T.~H.~R.}}} (\bibinfo {year} {1962}),\ \href@noop {} {\bibfield  {journal}
  {\bibinfo  {journal} {Nuclear Physics}\ }\textbf {\bibinfo {volume} {31}},\
  \bibinfo {pages} {556}}\BibitemShut {NoStop}%
\bibitem [{\citenamefont {Sokoloff}(1985)}]{Sokoloff}%
  \BibitemOpen
  \bibfield  {author} {\bibinfo {author} {\bibnamefont {Sokoloff},
  \bibfnamefont {J.~B.}}} (\bibinfo {year} {1985}),\ \href@noop {} {\bibfield
  {journal} {\bibinfo  {journal} {Physics Reports}\ }\textbf {\bibinfo {volume}
  {126}},\ \bibinfo {pages} {189}}\BibitemShut {NoStop}%
\bibitem [{\citenamefont {Sondhi}\ \emph {et~al.}(1993)\citenamefont {Sondhi},
  \citenamefont {Karlhede}, \citenamefont {Kivelson},\ and\ \citenamefont
  {Rezayi}}]{Sondhi1993}%
  \BibitemOpen
  \bibfield  {author} {\bibinfo {author} {\bibnamefont {Sondhi}, \bibfnamefont
  {S.~L.}}, \bibinfo {author} {\bibfnamefont {A.}~\bibnamefont {Karlhede}},
  \bibinfo {author} {\bibfnamefont {S.~A.}\ \bibnamefont {Kivelson}}, \ and\
  \bibinfo {author} {\bibfnamefont {E.~H.}\ \bibnamefont {Rezayi}}} (\bibinfo
  {year} {1993}),\ \href {\doibase 10.1103/PhysRevB.47.16419} {\bibfield
  {journal} {\bibinfo  {journal} {Physical Review B}\ }\textbf {\bibinfo
  {volume} {47}}~(\bibinfo {number} {24}),\ \bibinfo {pages}
  {16419}}\BibitemShut {NoStop}%
\bibitem [{\citenamefont {Sundaram}\ and\ \citenamefont
  {Niu}(1999)}]{Sundaram1999}%
  \BibitemOpen
  \bibfield  {author} {\bibinfo {author} {\bibnamefont {Sundaram},
  \bibfnamefont {G.}}, \ and\ \bibinfo {author} {\bibfnamefont
  {Q.}~\bibnamefont {Niu}}} (\bibinfo {year} {1999}),\ \href {\doibase
  10.1103/PhysRevB.59.14915} {\bibfield  {journal} {\bibinfo  {journal}
  {Physical Review B}\ }\textbf {\bibinfo {volume} {59}}~(\bibinfo {number}
  {23}),\ \bibinfo {pages} {14915}}\BibitemShut {NoStop}%
\bibitem [{\citenamefont {Takashima}\ \emph {et~al.}(2016)\citenamefont
  {Takashima}, \citenamefont {Ishizuka},\ and\ \citenamefont
  {Balents}}]{Takashima2016}%
  \BibitemOpen
  \bibfield  {author} {\bibinfo {author} {\bibnamefont {Takashima},
  \bibfnamefont {R.}}, \bibinfo {author} {\bibfnamefont {H.}~\bibnamefont
  {Ishizuka}}, \ and\ \bibinfo {author} {\bibfnamefont {L.}~\bibnamefont
  {Balents}}} (\bibinfo {year} {2016}),\ \href {\doibase
  10.1103/PhysRevB.94.134415} {\bibfield  {journal} {\bibinfo  {journal}
  {Physical Review B}\ }\textbf {\bibinfo {volume} {94}}~(\bibinfo {number}
  {13}),\ \bibinfo {pages} {134415}}\BibitemShut {NoStop}%
\bibitem [{\citenamefont {Tchernyshyov}(2015)}]{Tchernyshyov2015}%
  \BibitemOpen
  \bibfield  {author} {\bibinfo {author} {\bibnamefont {Tchernyshyov},
  \bibfnamefont {O.}}} (\bibinfo {year} {2015}),\ \href {\doibase
  10.1016/j.aop.2015.09.004} {\bibfield  {journal} {\bibinfo  {journal} {Annals
  of Physics}\ }\textbf {\bibinfo {volume} {363}},\ \bibinfo {pages}
  {98}}\BibitemShut {NoStop}%
\bibitem [{\citenamefont {Tewari}\ \emph {et~al.}(2006)\citenamefont {Tewari},
  \citenamefont {Belitz},\ and\ \citenamefont {Kirkpatrick}}]{Tewari2006}%
  \BibitemOpen
  \bibfield  {author} {\bibinfo {author} {\bibnamefont {Tewari}, \bibfnamefont
  {S.}}, \bibinfo {author} {\bibfnamefont {D.}~\bibnamefont {Belitz}}, \ and\
  \bibinfo {author} {\bibfnamefont {T.~R.}\ \bibnamefont {Kirkpatrick}}}
  (\bibinfo {year} {2006}),\ \href {\doibase 10.1103/PhysRevLett.96.047207}
  {\bibfield  {journal} {\bibinfo  {journal} {Physical Review Letters}\
  }\textbf {\bibinfo {volume} {96}}~(\bibinfo {number} {4}),\ \bibinfo {pages}
  {047207}}\BibitemShut {NoStop}%
\bibitem [{\citenamefont {Thiele}(1973)}]{Thiele1973}%
  \BibitemOpen
  \bibfield  {author} {\bibinfo {author} {\bibnamefont {Thiele}, \bibfnamefont
  {A.~A.}}} (\bibinfo {year} {1973}),\ \href {\doibase
  10.1103/PhysRevLett.30.230} {\bibfield  {journal} {\bibinfo  {journal}
  {Physical Review Letters}\ }\textbf {\bibinfo {volume} {30}}~(\bibinfo
  {number} {6}),\ \bibinfo {pages} {230}}\BibitemShut {NoStop}%
\bibitem [{\citenamefont {Thouless}(1983)}]{Thouless1983}%
  \BibitemOpen
  \bibfield  {author} {\bibinfo {author} {\bibnamefont {Thouless},
  \bibfnamefont {D.~J.}}} (\bibinfo {year} {1983}),\ \href@noop {} {\bibfield
  {journal} {\bibinfo  {journal} {Physical Review B}\ }\textbf {\bibinfo
  {volume} {27}}~(\bibinfo {number} {10}),\ \bibinfo {pages}
  {6083}}\BibitemShut {NoStop}%
\bibitem [{\citenamefont {Thouless}(1984)}]{Thouless1984}%
  \BibitemOpen
  \bibfield  {author} {\bibinfo {author} {\bibnamefont {Thouless},
  \bibfnamefont {D.~J.}}} (\bibinfo {year} {1984}),\ \href {\doibase
  10.1088/0022-3719/17/12/003} {\bibfield  {journal} {\bibinfo  {journal}
  {Journal of Physics C: Solid State Physics}\ }\textbf {\bibinfo {volume}
  {17}}~(\bibinfo {number} {12}),\ \bibinfo {pages} {L325}}\BibitemShut
  {NoStop}%
\bibitem [{\citenamefont {Thouless}\ \emph {et~al.}(1996)\citenamefont
  {Thouless}, \citenamefont {Ao},\ and\ \citenamefont {Niu}}]{Thouless1996}%
  \BibitemOpen
  \bibfield  {author} {\bibinfo {author} {\bibnamefont {Thouless},
  \bibfnamefont {D.~J.}}, \bibinfo {author} {\bibfnamefont {P.}~\bibnamefont
  {Ao}}, \ and\ \bibinfo {author} {\bibfnamefont {Q.}~\bibnamefont {Niu}}}
  (\bibinfo {year} {1996}),\ \href@noop {} {\bibfield  {journal} {\bibinfo
  {journal} {Physical Review Letters}\ }\textbf {\bibinfo {volume} {76}},\
  \bibinfo {pages} {3758}}\BibitemShut {NoStop}%
\bibitem [{\citenamefont {Thouless}\ \emph {et~al.}(1982)\citenamefont
  {Thouless}, \citenamefont {Kohmoto}, \citenamefont {Nightingale},\ and\
  \citenamefont {den Nijs}}]{Thouless1982}%
  \BibitemOpen
  \bibfield  {author} {\bibinfo {author} {\bibnamefont {Thouless},
  \bibfnamefont {D.~J.}}, \bibinfo {author} {\bibfnamefont {M.}~\bibnamefont
  {Kohmoto}}, \bibinfo {author} {\bibfnamefont {M.~P.}\ \bibnamefont
  {Nightingale}}, \ and\ \bibinfo {author} {\bibfnamefont {M.}~\bibnamefont
  {den Nijs}}} (\bibinfo {year} {1982}),\ \href@noop {} {\bibfield  {journal}
  {\bibinfo  {journal} {Physical Review Letters}\ }\textbf {\bibinfo {volume}
  {49}},\ \bibinfo {pages} {405}}\BibitemShut {NoStop}%
\bibitem [{\citenamefont {Treima}\ \emph {et~al.}(1985)\citenamefont {Treima},
  \citenamefont {Jackiw}, \citenamefont {Zumino},\ and\ \citenamefont
  {Witten}}]{anomalies_book}%
  \BibitemOpen
  \bibfield  {author} {\bibinfo {author} {\bibnamefont {Treima}, \bibfnamefont
  {S.}}, \bibinfo {author} {\bibfnamefont {R.}~\bibnamefont {Jackiw}}, \bibinfo
  {author} {\bibfnamefont {B.}~\bibnamefont {Zumino}}, \ and\ \bibinfo {author}
  {\bibfnamefont {E.}~\bibnamefont {Witten}}} (\bibinfo {year} {1985}),\
  \href@noop {} {\emph {\bibinfo {title} {Current Algebras and Anomalies}}}\
  (\bibinfo  {publisher} {Princeton University Press, NJ/World Scientific},\
  \bibinfo {address} {Singapore})\BibitemShut {NoStop}%
\bibitem [{\citenamefont {Tretiakov}\ \emph {et~al.}(2008)\citenamefont
  {Tretiakov}, \citenamefont {Clarke}, \citenamefont {Chern}, \citenamefont
  {Bazaliy},\ and\ \citenamefont {Tchernyshyov}}]{Tretiakov2008}%
  \BibitemOpen
  \bibfield  {author} {\bibinfo {author} {\bibnamefont {Tretiakov},
  \bibfnamefont {O.~A.}}, \bibinfo {author} {\bibfnamefont {D.}~\bibnamefont
  {Clarke}}, \bibinfo {author} {\bibfnamefont {G.~W.}\ \bibnamefont {Chern}},
  \bibinfo {author} {\bibfnamefont {Y.~B.}\ \bibnamefont {Bazaliy}}, \ and\
  \bibinfo {author} {\bibfnamefont {O.}~\bibnamefont {Tchernyshyov}}} (\bibinfo
  {year} {2008}),\ \href {\doibase 10.1103/PhysRevLett.100.127204} {\bibfield
  {journal} {\bibinfo  {journal} {Physical Review Letters}\ }\textbf {\bibinfo
  {volume} {100}}~(\bibinfo {number} {12}),\ \bibinfo {pages}
  {127204}}\BibitemShut {NoStop}%
\bibitem [{\citenamefont {Ueda}(2014)}]{Ueda2014}%
  \BibitemOpen
  \bibfield  {author} {\bibinfo {author} {\bibnamefont {Ueda}, \bibfnamefont
  {M.}}} (\bibinfo {year} {2014}),\ \href@noop {} {\bibfield  {journal}
  {\bibinfo  {journal} {Reports on Progress in Physics}\ }\textbf {\bibinfo
  {volume} {77}},\ \bibinfo {pages} {122401}}\BibitemShut {NoStop}%
\bibitem [{\citenamefont {Valenti}\ and\ \citenamefont
  {Lax}(1977)}]{Valenti1977}%
  \BibitemOpen
  \bibfield  {author} {\bibinfo {author} {\bibnamefont {Valenti}, \bibfnamefont
  {C.~F.}}, \ and\ \bibinfo {author} {\bibfnamefont {M.}~\bibnamefont {Lax}}}
  (\bibinfo {year} {1977}),\ \href@noop {} {\bibfield  {journal} {\bibinfo
  {journal} {Physical Review B}\ }\textbf {\bibinfo {volume} {16}}~(\bibinfo
  {number} {11}),\ \bibinfo {pages} {4936}}\BibitemShut {NoStop}%
\bibitem [{\citenamefont {van Vleck}(1992)}]{vanVleck}%
  \BibitemOpen
  \bibfield  {author} {\bibinfo {author} {\bibnamefont {van Vleck},
  \bibfnamefont {J.~H.}}} (\bibinfo {year} {1992}),\ \href@noop {} {\emph
  {\bibinfo {title} {Nobel Lectures in Physics (1971-1980)}}},\ edited by\
  \bibinfo {editor} {\bibfnamefont {S.}~\bibnamefont {Lundqvist}}\ (\bibinfo
  {publisher} {World Scientific},\ \bibinfo {address} {Singapore})\BibitemShut
  {NoStop}%
\bibitem [{\citenamefont {Volovik}(1986)}]{Volovik1986}%
  \BibitemOpen
  \bibfield  {author} {\bibinfo {author} {\bibnamefont {Volovik}, \bibfnamefont
  {G.~E.}}} (\bibinfo {year} {1986}),\ \href@noop {} {\enquote {\bibinfo
  {title} {{Chiral anomaly and the law of conservation of momentum in
  3He-A}},}\ }\BibitemShut {NoStop}%
\bibitem [{\citenamefont {Volovik}\ and\ \citenamefont
  {Mineev}(1977)}]{Volovik1977}%
  \BibitemOpen
  \bibfield  {author} {\bibinfo {author} {\bibnamefont {Volovik}, \bibfnamefont
  {G.~E.}}, \ and\ \bibinfo {author} {\bibfnamefont {V.~P.}\ \bibnamefont
  {Mineev}}} (\bibinfo {year} {1977}),\ \href@noop {} {\bibfield  {journal}
  {\bibinfo  {journal} {Sov. Phys. JEPT}\ }\textbf {\bibinfo {volume} {46}},\
  \bibinfo {pages} {401}}\BibitemShut {NoStop}%
\bibitem [{\citenamefont {Watanabe}\ and\ \citenamefont
  {Murayama}(2014)}]{Watanabe2014}%
  \BibitemOpen
  \bibfield  {author} {\bibinfo {author} {\bibnamefont {Watanabe},
  \bibfnamefont {H.}}, \ and\ \bibinfo {author} {\bibfnamefont
  {H.}~\bibnamefont {Murayama}}} (\bibinfo {year} {2014}),\ \href {\doibase
  10.1103/PhysRevLett.112.191804} {\bibfield  {journal} {\bibinfo  {journal}
  {Physical Review Letters}\ }\textbf {\bibinfo {volume} {112}}~(\bibinfo
  {number} {19}),\ \bibinfo {pages} {1}}\BibitemShut {NoStop}%
\bibitem [{\citenamefont {Wen}\ and\ \citenamefont {Zee}(1989)}]{Wen1989}%
  \BibitemOpen
  \bibfield  {author} {\bibinfo {author} {\bibnamefont {Wen}, \bibfnamefont
  {X.~G.}}, \ and\ \bibinfo {author} {\bibfnamefont {A.}~\bibnamefont {Zee}}}
  (\bibinfo {year} {1989}),\ \href {\doibase 10.1016/0550-3213(89)90062-X}
  {\bibfield  {journal} {\bibinfo  {journal} {Nuclear Physics B}\ }\textbf
  {\bibinfo {volume} {316}}~(\bibinfo {number} {3}),\ \bibinfo {pages}
  {641}}\BibitemShut {NoStop}%
\bibitem [{\citenamefont {Wess}\ and\ \citenamefont {Zumino}(1971)}]{Wess1971}%
  \BibitemOpen
  \bibfield  {author} {\bibinfo {author} {\bibnamefont {Wess}, \bibfnamefont
  {J.}}, \ and\ \bibinfo {author} {\bibfnamefont {B.}~\bibnamefont {Zumino}}}
  (\bibinfo {year} {1971}),\ \href@noop {} {\bibfield  {journal} {\bibinfo
  {journal} {Physics Letters B}\ }\textbf {\bibinfo {volume} {37}},\ \bibinfo
  {pages} {95}}\BibitemShut {NoStop}%
\bibitem [{\citenamefont {Wessel}\ and\ \citenamefont
  {Troyer}(2005)}]{Wessel2005}%
  \BibitemOpen
  \bibfield  {author} {\bibinfo {author} {\bibnamefont {Wessel}, \bibfnamefont
  {S.}}, \ and\ \bibinfo {author} {\bibfnamefont {M.}~\bibnamefont {Troyer}}}
  (\bibinfo {year} {2005}),\ \href@noop {} {\bibfield  {journal} {\bibinfo
  {journal} {Physical Review Letters}\ }\textbf {\bibinfo {volume} {95}},\
  \bibinfo {pages} {127205}}\BibitemShut {NoStop}%
\bibitem [{\citenamefont {Weyl}(1927)}]{Weyl1927}%
  \BibitemOpen
  \bibfield  {author} {\bibinfo {author} {\bibnamefont {Weyl}, \bibfnamefont
  {H.}}} (\bibinfo {year} {1927}),\ \href@noop {} {\bibfield  {journal}
  {\bibinfo  {journal} {Z. Phys.}\ }\textbf {\bibinfo {volume} {46}},\ \bibinfo
  {pages} {1}}\BibitemShut {NoStop}%
\bibitem [{\citenamefont {White}\ \emph {et~al.}(2014)\citenamefont {White},
  \citenamefont {Pr{\v{s}}a}, \citenamefont {Huang}, \citenamefont {Omrani},
  \citenamefont {{\v{Z}}ivkovi{\'{c}}}, \citenamefont {Bartkowiak},
  \citenamefont {Berger}, \citenamefont {Magrez}, \citenamefont {Gavilano},
  \citenamefont {Nagy}, \citenamefont {Zang},\ and\ \citenamefont
  {R{\o}nnow}}]{White2014}%
  \BibitemOpen
  \bibfield  {author} {\bibinfo {author} {\bibnamefont {White}, \bibfnamefont
  {J.~S.}}, \bibinfo {author} {\bibfnamefont {K.}~\bibnamefont {Pr{\v{s}}a}},
  \bibinfo {author} {\bibfnamefont {P.}~\bibnamefont {Huang}}, \bibinfo
  {author} {\bibfnamefont {A.~A.}\ \bibnamefont {Omrani}}, \bibinfo {author}
  {\bibfnamefont {I.}~\bibnamefont {{\v{Z}}ivkovi{\'{c}}}}, \bibinfo {author}
  {\bibfnamefont {M.}~\bibnamefont {Bartkowiak}}, \bibinfo {author}
  {\bibfnamefont {H.}~\bibnamefont {Berger}}, \bibinfo {author} {\bibfnamefont
  {A.}~\bibnamefont {Magrez}}, \bibinfo {author} {\bibfnamefont {J.~L.}\
  \bibnamefont {Gavilano}}, \bibinfo {author} {\bibfnamefont {G.}~\bibnamefont
  {Nagy}}, \bibinfo {author} {\bibfnamefont {J.}~\bibnamefont {Zang}}, \ and\
  \bibinfo {author} {\bibfnamefont {H.~M.}\ \bibnamefont {R{\o}nnow}}}
  (\bibinfo {year} {2014}),\ \href {\doibase 10.1103/PhysRevLett.113.107203}
  {\bibfield  {journal} {\bibinfo  {journal} {Physical Review Letters}\
  }\textbf {\bibinfo {volume} {113}}~(\bibinfo {number} {10}),\ \bibinfo
  {pages} {107203}}\BibitemShut {NoStop}%
\bibitem [{\citenamefont {Wilczek}(1982)}]{Wilczek1982}%
  \BibitemOpen
  \bibfield  {author} {\bibinfo {author} {\bibnamefont {Wilczek}, \bibfnamefont
  {F.}}} (\bibinfo {year} {1982}),\ \href {\doibase 10.1103/PhysRev.135.A1013}
  {\bibfield  {journal} {\bibinfo  {journal} {Physical Review}\ }\textbf
  {\bibinfo {volume} {48}}~(\bibinfo {number} {17}),\ \bibinfo {pages}
  {1144}}\BibitemShut {NoStop}%
\bibitem [{\citenamefont {Wilczek}\ and\ \citenamefont
  {Zee}(1983)}]{Wilczek1983}%
  \BibitemOpen
  \bibfield  {author} {\bibinfo {author} {\bibnamefont {Wilczek}, \bibfnamefont
  {F.}}, \ and\ \bibinfo {author} {\bibfnamefont {A.}~\bibnamefont {Zee}}}
  (\bibinfo {year} {1983}),\ \href@noop {} {\bibfield  {journal} {\bibinfo
  {journal} {Physical Review Letters}\ }\textbf {\bibinfo {volume}
  {51}}~(\bibinfo {number} {25}),\ \bibinfo {pages} {2250}}\BibitemShut
  {NoStop}%
\bibitem [{\citenamefont {Wilhelm}\ \emph {et~al.}(2011)\citenamefont
  {Wilhelm}, \citenamefont {Baenitz}, \citenamefont {Schmidt}, \citenamefont
  {R????ler}, \citenamefont {Leonov},\ and\ \citenamefont
  {Bogdanov}}]{Wilhelm2011}%
  \BibitemOpen
  \bibfield  {author} {\bibinfo {author} {\bibnamefont {Wilhelm}, \bibfnamefont
  {H.}}, \bibinfo {author} {\bibfnamefont {M.}~\bibnamefont {Baenitz}},
  \bibinfo {author} {\bibfnamefont {M.}~\bibnamefont {Schmidt}}, \bibinfo
  {author} {\bibfnamefont {U.~K.}\ \bibnamefont {R????ler}}, \bibinfo {author}
  {\bibfnamefont {A.~A.}\ \bibnamefont {Leonov}}, \ and\ \bibinfo {author}
  {\bibfnamefont {A.~N.}\ \bibnamefont {Bogdanov}}} (\bibinfo {year} {2011}),\
  \href {\doibase 10.1103/PhysRevLett.107.127203} {\bibfield  {journal}
  {\bibinfo  {journal} {Physical Review Letters}\ }\textbf {\bibinfo {volume}
  {107}}~(\bibinfo {number} {12}),\ \bibinfo {pages} {127203}}\BibitemShut
  {NoStop}%
\bibitem [{\citenamefont {Wilkinson}\ and\ \citenamefont
  {Kay}(1996)}]{Wilkinson1996}%
  \BibitemOpen
  \bibfield  {author} {\bibinfo {author} {\bibnamefont {Wilkinson},
  \bibfnamefont {M.}}, \ and\ \bibinfo {author} {\bibfnamefont {R.~J.}\
  \bibnamefont {Kay}}} (\bibinfo {year} {1996}),\ \href@noop {} {\bibfield
  {journal} {\bibinfo  {journal} {Physical Review Letters}\ }\textbf {\bibinfo
  {volume} {1}}~(\bibinfo {number} {2}),\ \bibinfo {pages} {1896}}\BibitemShut
  {NoStop}%
\bibitem [{\citenamefont {Wright}\ and\ \citenamefont
  {Mermin}(1989)}]{Wright1989}%
  \BibitemOpen
  \bibfield  {author} {\bibinfo {author} {\bibnamefont {Wright}, \bibfnamefont
  {D.~C.}}, \ and\ \bibinfo {author} {\bibfnamefont {N.~D.}\ \bibnamefont
  {Mermin}}} (\bibinfo {year} {1989}),\ \href {\doibase
  10.1103/RevModPhys.61.385} {\bibfield  {journal} {\bibinfo  {journal}
  {Reviews of Modern Physics}\ }\textbf {\bibinfo {volume} {61}}~(\bibinfo
  {number} {2}),\ \bibinfo {pages} {385}}\BibitemShut {NoStop}%
\bibitem [{\citenamefont {Wu}\ \emph {et~al.}(2007)\citenamefont {Wu},
  \citenamefont {Bergman}, \citenamefont {Balents},\ and\ \citenamefont
  {Sarma}}]{Wu2007}%
  \BibitemOpen
  \bibfield  {author} {\bibinfo {author} {\bibnamefont {Wu}, \bibfnamefont
  {C.}}, \bibinfo {author} {\bibfnamefont {D.}~\bibnamefont {Bergman}},
  \bibinfo {author} {\bibfnamefont {L.}~\bibnamefont {Balents}}, \ and\
  \bibinfo {author} {\bibfnamefont {S.~D.}\ \bibnamefont {Sarma}}} (\bibinfo
  {year} {2007}),\ \href {\doibase 10.1103/PhysRevLett.99.070401} {\bibfield
  {journal} {\bibinfo  {journal} {Physical Review Letters}\ }\textbf {\bibinfo
  {volume} {99}},\ \bibinfo {pages} {070401}}\BibitemShut {NoStop}%
\bibitem [{\citenamefont {Xiao}\ \emph {et~al.}(2010)\citenamefont {Xiao},
  \citenamefont {Chang},\ and\ \citenamefont {Niu}}]{Xiao2010}%
  \BibitemOpen
  \bibfield  {author} {\bibinfo {author} {\bibnamefont {Xiao}, \bibfnamefont
  {D.}}, \bibinfo {author} {\bibfnamefont {M.~C.}\ \bibnamefont {Chang}}, \
  and\ \bibinfo {author} {\bibfnamefont {Q.}~\bibnamefont {Niu}}} (\bibinfo
  {year} {2010}),\ \href {\doibase 10.1103/RevModPhys.82.1959} {\bibfield
  {journal} {\bibinfo  {journal} {Reviews of Modern Physics}\ }\textbf
  {\bibinfo {volume} {82}}~(\bibinfo {number} {3}),\ \bibinfo {pages}
  {1959}}\BibitemShut {NoStop}%
\bibitem [{\citenamefont {Xiao}\ \emph {et~al.}(2005)\citenamefont {Xiao},
  \citenamefont {Shi},\ and\ \citenamefont {Niu}}]{Xiao2005}%
  \BibitemOpen
  \bibfield  {author} {\bibinfo {author} {\bibnamefont {Xiao}, \bibfnamefont
  {D.}}, \bibinfo {author} {\bibfnamefont {J.}~\bibnamefont {Shi}}, \ and\
  \bibinfo {author} {\bibfnamefont {Q.}~\bibnamefont {Niu}}} (\bibinfo {year}
  {2005}),\ \href {\doibase 10.1103/PhysRevLett.95.137204} {\bibfield
  {journal} {\bibinfo  {journal} {Physical Review Letters}\ }\textbf {\bibinfo
  {volume} {95}}~(\bibinfo {number} {13}),\ \bibinfo {pages}
  {137204}}\BibitemShut {NoStop}%
\bibitem [{\citenamefont {Xiao}\ \emph {et~al.}(2006)\citenamefont {Xiao},
  \citenamefont {Yao}, \citenamefont {Fang},\ and\ \citenamefont
  {Niu}}]{Xiao2006}%
  \BibitemOpen
  \bibfield  {author} {\bibinfo {author} {\bibnamefont {Xiao}, \bibfnamefont
  {D.}}, \bibinfo {author} {\bibfnamefont {Y.}~\bibnamefont {Yao}}, \bibinfo
  {author} {\bibfnamefont {Z.}~\bibnamefont {Fang}}, \ and\ \bibinfo {author}
  {\bibfnamefont {Q.}~\bibnamefont {Niu}}} (\bibinfo {year} {2006}),\ \href
  {\doibase 10.1103/PhysRevLett.97.026603} {\bibfield  {journal} {\bibinfo
  {journal} {Physical Review Letters}\ }\textbf {\bibinfo {volume}
  {97}}~(\bibinfo {number} {2}),\ \bibinfo {pages} {026603}}\BibitemShut
  {NoStop}%
\bibitem [{\citenamefont {Xiong}\ \emph {et~al.}(2015)\citenamefont {Xiong},
  \citenamefont {Kushwaha}, \citenamefont {Liang}, \citenamefont {Krizan},
  \citenamefont {Hirschberger}, \citenamefont {Wang}, \citenamefont {Cava},\
  and\ \citenamefont {Ong}}]{Xiong2015}%
  \BibitemOpen
  \bibfield  {author} {\bibinfo {author} {\bibnamefont {Xiong}, \bibfnamefont
  {J.}}, \bibinfo {author} {\bibfnamefont {S.~K.}\ \bibnamefont {Kushwaha}},
  \bibinfo {author} {\bibfnamefont {T.}~\bibnamefont {Liang}}, \bibinfo
  {author} {\bibfnamefont {J.~W.}\ \bibnamefont {Krizan}}, \bibinfo {author}
  {\bibfnamefont {M.}~\bibnamefont {Hirschberger}}, \bibinfo {author}
  {\bibfnamefont {W.}~\bibnamefont {Wang}}, \bibinfo {author} {\bibfnamefont
  {R.~J.}\ \bibnamefont {Cava}}, \ and\ \bibinfo {author} {\bibfnamefont
  {N.~P.}\ \bibnamefont {Ong}}} (\bibinfo {year} {2015}),\ \href {\doibase
  10.1126/science.aac6089} {\bibfield  {journal} {\bibinfo  {journal}
  {Science}\ }\textbf {\bibinfo {volume} {350}}~(\bibinfo {number} {6259}),\
  \bibinfo {pages} {413}}\BibitemShut {NoStop}%
\bibitem [{\citenamefont {Yan}\ \emph {et~al.}(2013)\citenamefont {Yan},
  \citenamefont {Kamra}, \citenamefont {Cao},\ and\ \citenamefont
  {Bauer}}]{Yan2013}%
  \BibitemOpen
  \bibfield  {author} {\bibinfo {author} {\bibnamefont {Yan}, \bibfnamefont
  {P.}}, \bibinfo {author} {\bibfnamefont {A.}~\bibnamefont {Kamra}}, \bibinfo
  {author} {\bibfnamefont {Y.}~\bibnamefont {Cao}}, \ and\ \bibinfo {author}
  {\bibfnamefont {G.~E.~W.}\ \bibnamefont {Bauer}}} (\bibinfo {year} {2013}),\
  \href {\doibase 10.1103/PhysRevB.88.144413} {\bibfield  {journal} {\bibinfo
  {journal} {Physical Review B}\ }\textbf {\bibinfo {volume} {88}}~(\bibinfo
  {number} {14}),\ \bibinfo {pages} {144413}},\ \Eprint
  {http://arxiv.org/abs/1307.3432} {arXiv:1307.3432} \BibitemShut {NoStop}%
\bibitem [{\citenamefont {Yang}\ and\ \citenamefont
  {Nagaosa}(2011)}]{Yang2011}%
  \BibitemOpen
  \bibfield  {author} {\bibinfo {author} {\bibnamefont {Yang}, \bibfnamefont
  {B.-J.}}, \ and\ \bibinfo {author} {\bibfnamefont {N.}~\bibnamefont
  {Nagaosa}}} (\bibinfo {year} {2011}),\ \href {\doibase
  10.1103/PhysRevB.84.245123} {\bibfield  {journal} {\bibinfo  {journal}
  {Physical Review B}\ }\textbf {\bibinfo {volume} {84}},\ \bibinfo {pages}
  {245123}}\BibitemShut {NoStop}%
\bibitem [{\citenamefont {Yang}\ and\ \citenamefont
  {Sondhi}(1996)}]{YangSondhi}%
  \BibitemOpen
  \bibfield  {author} {\bibinfo {author} {\bibnamefont {Yang}, \bibfnamefont
  {K.}}, \ and\ \bibinfo {author} {\bibfnamefont {S.~L.}\ \bibnamefont
  {Sondhi}}} (\bibinfo {year} {1996}),\ \href@noop {} {\bibfield  {journal}
  {\bibinfo  {journal} {Physical Review B}\ }\textbf {\bibinfo {volume} {54}},\
  \bibinfo {pages} {R2331(R)}}\BibitemShut {NoStop}%
\bibitem [{\citenamefont {Yu}\ \emph {et~al.}(2012)\citenamefont {Yu},
  \citenamefont {Kanazawa}, \citenamefont {Zhang}, \citenamefont {Nagai},
  \citenamefont {Hara}, \citenamefont {Kimoto}, \citenamefont {Matsui},
  \citenamefont {Onose},\ and\ \citenamefont {Tokura}}]{Yu2012}%
  \BibitemOpen
  \bibfield  {author} {\bibinfo {author} {\bibnamefont {Yu}, \bibfnamefont
  {X.}}, \bibinfo {author} {\bibfnamefont {N.}~\bibnamefont {Kanazawa}},
  \bibinfo {author} {\bibfnamefont {W.}~\bibnamefont {Zhang}}, \bibinfo
  {author} {\bibfnamefont {T.}~\bibnamefont {Nagai}}, \bibinfo {author}
  {\bibfnamefont {T.}~\bibnamefont {Hara}}, \bibinfo {author} {\bibfnamefont
  {K.}~\bibnamefont {Kimoto}}, \bibinfo {author} {\bibfnamefont
  {Y.}~\bibnamefont {Matsui}}, \bibinfo {author} {\bibfnamefont
  {Y.}~\bibnamefont {Onose}}, \ and\ \bibinfo {author} {\bibfnamefont
  {Y.}~\bibnamefont {Tokura}}} (\bibinfo {year} {2012}),\ \href {\doibase
  10.1038/ncomms1990} {\bibfield  {journal} {\bibinfo  {journal} {Nature
  Communications}\ }\textbf {\bibinfo {volume} {3}},\ \bibinfo {pages}
  {988}}\BibitemShut {NoStop}%
\bibitem [{\citenamefont {Yu}\ \emph {et~al.}(2011)\citenamefont {Yu},
  \citenamefont {Kanazawa}, \citenamefont {Onose}, \citenamefont {Kimoto},
  \citenamefont {Zhang}, \citenamefont {Ishiwata}, \citenamefont {Matsui},\
  and\ \citenamefont {Tokura}}]{Yu2011}%
  \BibitemOpen
  \bibfield  {author} {\bibinfo {author} {\bibnamefont {Yu}, \bibfnamefont
  {X.~Z.}}, \bibinfo {author} {\bibfnamefont {N.}~\bibnamefont {Kanazawa}},
  \bibinfo {author} {\bibfnamefont {Y.}~\bibnamefont {Onose}}, \bibinfo
  {author} {\bibfnamefont {K.}~\bibnamefont {Kimoto}}, \bibinfo {author}
  {\bibfnamefont {W.~Z.}\ \bibnamefont {Zhang}}, \bibinfo {author}
  {\bibfnamefont {S.}~\bibnamefont {Ishiwata}}, \bibinfo {author}
  {\bibfnamefont {Y.}~\bibnamefont {Matsui}}, \ and\ \bibinfo {author}
  {\bibfnamefont {Y.}~\bibnamefont {Tokura}}} (\bibinfo {year} {2011}),\ \href
  {\doibase 10.1038/nmat2916} {\bibfield  {journal} {\bibinfo  {journal}
  {Nature Materials}\ }\textbf {\bibinfo {volume} {10}}~(\bibinfo {number}
  {2}),\ \bibinfo {pages} {106}}\BibitemShut {NoStop}%
\bibitem [{\citenamefont {Zak}(1964)}]{Zak1964}%
  \BibitemOpen
  \bibfield  {author} {\bibinfo {author} {\bibnamefont {Zak}, \bibfnamefont
  {J.}}} (\bibinfo {year} {1964}),\ \href@noop {} {\bibfield  {journal}
  {\bibinfo  {journal} {Physical Review}\ }\textbf {\bibinfo {volume}
  {134}}~(\bibinfo {number} {6A}),\ \bibinfo {pages} {A1602}}\BibitemShut
  {NoStop}%
\bibitem [{\citenamefont {Zaletel}\ \emph {et~al.}(2014)\citenamefont
  {Zaletel}, \citenamefont {Parameswaran}, \citenamefont {R\"uegg},\ and\
  \citenamefont {Altman}}]{Zaletel2014}%
  \BibitemOpen
  \bibfield  {author} {\bibinfo {author} {\bibnamefont {Zaletel}, \bibfnamefont
  {M.~P.}}, \bibinfo {author} {\bibfnamefont {S.~A.}\ \bibnamefont
  {Parameswaran}}, \bibinfo {author} {\bibfnamefont {A.}~\bibnamefont
  {R\"uegg}}, \ and\ \bibinfo {author} {\bibfnamefont {E.}~\bibnamefont
  {Altman}}} (\bibinfo {year} {2014}),\ \href@noop {} {\bibfield  {journal}
  {\bibinfo  {journal} {Physical Review B}\ }\textbf {\bibinfo {volume} {89}},\
  \bibinfo {pages} {155142}}\BibitemShut {NoStop}%
\bibitem [{\citenamefont {Zhu}\ \emph {et~al.}(2016)\citenamefont {Zhu},
  \citenamefont {Koch},\ and\ \citenamefont {Martin}}]{Zhu2016}%
  \BibitemOpen
  \bibfield  {author} {\bibinfo {author} {\bibnamefont {Zhu}, \bibfnamefont
  {G.}}, \bibinfo {author} {\bibfnamefont {J.}~\bibnamefont {Koch}}, \ and\
  \bibinfo {author} {\bibfnamefont {I.}~\bibnamefont {Martin}}} (\bibinfo
  {year} {2016}),\ \href {\doibase 10.1103/PhysRevB.93.144508} {\bibfield
  {journal} {\bibinfo  {journal} {Physical Review B}\ }\textbf {\bibinfo
  {volume} {93}},\ \bibinfo {pages} {144508}}\BibitemShut {NoStop}%
\end{thebibliography}%

\end{document}